\documentclass[a4paper,11pt]{article}
\usepackage{jheppub} % for details on the use of the package, please
\usepackage[T1]{fontenc} % if needed
\usepackage{amsmath,amsthm}
\usepackage{amssymb,amsfonts}
\usepackage{graphicx}
\usepackage{dsfont}
\usepackage{relsize}
\usepackage{stmaryrd}
\usepackage{lmodern}
\usepackage{slantsc}
\usepackage{scalefnt}
\usepackage{subfigure}
\usepackage{xspace}
\usepackage{boxedminipage}
\usepackage{ifpdf}
\usepackage{multirow}
\usepackage{xifthen}
\usepackage{tensor}
\usepackage{array}
\usepackage{tabu}
\usepackage{tikz}
\usepackage{color}
\usepackage{diagbox}
\usetikzlibrary{arrows,matrix,calc,scopes,decorations.markings}
\allowdisplaybreaks[1]
\usepackage[mathcal]{euscript}

\newtheorem*{ansatz*}{Ansatz}

\newcommand{\be}{\begin{equation}}
\newcommand{\ee}{\end{equation}}
\newcommand{\bse}{\begin{subequations}}
\newcommand{\ese}{\end{subequations}}
\newcommand{\ket}[1]{|{#1}\rangle}
\newcommand{\bket}[1]{\Biggl|{#1}\Biggr\rangle}

\newcommand{\Z}{\mathbb{Z}}

\newcommand{\e}{\mathrm{e}}
\newcommand{\Hil}{\mathcal{H}}

\newcommand{\F}{\mathfrak{F}}
\newcommand{\bra}[1]{\langle{#1}|}

\newcommand{\A}{\mathcal{A}}
\newcommand{\U}{\mathcal{U}}
\newcommand{\C}{\mathcal{C}}

\newcommand{\bpm}{\begin{pmatrix}}
\newcommand{\epm}{\end{pmatrix}}
\newcommand{\bmm}{\begin{matrix}}
\newcommand{\emm}{\end{matrix}}

\newcommand{\Ab}{\overline{A}}
\newcommand{\Bb}{\overline{B}}
\newcommand{\x}{\times}
\newcommand{\ox}{\otimes}

\newcommand{\rep}{\mathrm{Rep}}

\newcommand{\id}{\mathrm{id}}

\newcommand{\QD}{\mathrm{QD}}
\newcommand{\LW}{\mathrm{LW}}

\tikzset{snake it/.style={decorate, decoration={snake,amplitude=0.15mm,segment length=1mm}}}
\tikzset{->-/.style={decoration={
                        markings,
                        mark=at position .55 with {\arrow{latex}}},postaction={decorate}}}

\newtabulinestyle{dash=off 2pt}

 {\null\,\vcenter\bgroup\scriptsize
  \arraycolsep=.13885em
  \hbox\bgroup$\array{@{}#1@{}}}
 {\endarray$\egroup\egroup\,\null}

\usetikzlibrary{decorations.markings}
\usetikzlibrary{decorations.pathmorphing,positioning}
\tikzset{>=latex}
\tikzset{snake it/.style={decorate, decoration={snake,amplitude=0.15mm,segment length=1mm}}}
\tikzset{->-/.style={decoration={
			 markings,
			 mark=at position .55 with {\arrow{>}}},postaction={decorate}}}

\tikzset{-<-/.style={decoration={
			 markings,
			 mark=at position .55 with {\arrow{<}}},postaction={decorate}}}

\newcommand{\assoCondA}[6]{
\begin{tikzpicture}[baseline={([yshift=-.8ex]current bounding box.center)},scale=0.81]
\draw [->-] (2.9,12.8) -- (2.8,12.8);
\draw [->-] (1.7,10.8) -- (1.7,10.6);
\draw [->-] (2.9,8.8) -- (3.1,8.8);
\draw [->-] (3.4,10.2) -- (3.5,10.1);
\draw [->-] (4.5,10.6) -- (4.5,10.8);
\draw [->-] (4.3,11.5) -- (4.2,11.7);
\draw [->-] (3.3,11.) -- (3.3,10.8);
\draw [->-, fill] (4.4,11.3) -- (3.3,10.6);
\draw [->-, fill] (3.,9.7) -- (2.7,10.6);
\draw [->-, fill] (2.7,10.6) -- (2.2,10.8);
\draw [->-, fill] (2.7,10.6) -- (2.2,10.4);
\draw [->-, fill] (2.5,9.7) -- (3.,9.7);
\draw [->-] (3.,9.7) -- (2.9,9.3);
\draw [rotate around={0.:(3.9,10.8)}] (3.9,10.8) ++(0.:0.6 and 1.) arc(0.:360.:0.6 and 1.);
\draw [rotate around={0.:(2.9,10.8)}] (2.9,10.8) ++(31.:1.2 and 2.) arc(31.:330.:1.2 and 2.);
\draw [fill,rotate around={0.:(3.9,11.8)}] (3.9,11.8) ++(0.:0.04 and 0.04) arc(0.:360.:0.04 and 0.04);
\draw [fill,rotate around={0.:(4.4,11.3)}] (4.4,11.3) ++(0.:0.04 and 0.04) arc(0.:360.:0.04 and 0.04);
\draw [fill,rotate around={0.:(3.3,10.6)}] (3.3,10.6) ++(0.:0.04 and 0.04) arc(0.:360.:0.04 and 0.04);
\draw [fill,rotate around={0.:(2.7,10.6)}] (2.7,10.6) ++(0.:0.04 and 0.04) arc(0.:360.:0.04 and 0.04);
\draw [fill,rotate around={0.:(3.,9.7)}] (3.,9.7) ++(0.:0.04 and 0.04) arc(0.:360.:0.04 and 0.04);
\draw [fill,rotate around={0.:(3.9,9.8)}] (3.9,9.8) ++(0.:0.04 and 0.04) arc(0.:360.:0.04 and 0.04);
\node [rotate=0.] at (3.8,11.2) {\tiny $#1$};
\node [rotate=0.] at (3.4,11.7) {\tiny $#2$};
\node [rotate=0.] at (3.,10.2) {\tiny $#3$};
\node [rotate=0.] at (3.7,10.2) {\tiny $#3$};
\node [rotate=0.] at (1.6,9.8) {\tiny $#4$};
\node [rotate=0.] at (4.7,10.2) {\tiny $#5$};
\node [rotate=0.] at (4.4,11.8) {\tiny $#6$};
\node [rotate=0.] at (2.6,10.3) {\tiny $a$};
\node [rotate=0.] at (2.7,10.9) {\tiny $b$};
\node [rotate=0.] at (2.2,11.) {\tiny $\alpha_b$};
\node [rotate=0.] at (2.1,10.3) {\tiny $\alpha_a$};
\node [rotate=0.] at (2.6,9.9) {\tiny $s$};
\node [rotate=0.] at (2.2,9.7) {\tiny $\alpha_s$};
\node [rotate=0.] at (3.1,9.5) {\tiny $r$};
\node [rotate=0.] at (2.9,9.1) {\tiny $\alpha_r$};
\end{tikzpicture}
}

\newcommand{\assoCondB}[6]{
\begin{tikzpicture}[baseline={([yshift=-.8ex]current bounding box.center)},scale=0.81]
\draw [->-] (3.1,9.7) ..controls (2.847,9.773) and (2.595,9.846) .. (2.6,9.9)
			 ..controls (2.605,9.954) and (2.869,9.989) .. (3.,10.)
			 ..controls (3.131,10.011) and (3.131,9.999) .. (3.2,10.)
			 ..controls (3.269,10.001) and (3.407,10.016) .. (3.5,10.)
			 ..controls (3.593,9.984) and (3.641,9.938) .. (3.7,9.9)
			 ..controls (3.759,9.862) and (3.83,9.831) .. (3.9,9.8);
\draw [->-] (2.9,12.8) -- (2.8,12.8);
\draw [->-] (1.7,10.8) -- (1.7,10.6);
\draw [->-] (2.9,8.8) -- (3.1,8.8);
\draw [->-] (4.3,10.1) -- (4.4,10.2);
\draw [->-] (4.5,10.8) -- (4.5,11.);
\draw [->-, fill] (3.9,11.8) -- (2.4,11.);
\draw [->-, fill] (4.5,10.6) -- (2.7,10.4);
\draw [->-] (2.6,9.5) -- (3.1,9.7);
\draw [->-] (3.1,9.7) -- (3.1,9.2);
\draw [rotate around={0.:(2.9,10.8)}] (2.9,10.8) ++(31.:1.2 and 2.) arc(31.:330.:1.2 and 2.);
\draw [rotate around={0.:(3.9,10.8)}] (3.9,10.8) ++(270.:0.6 and 1.) arc(270.:450.:0.6 and 1.);
\draw [fill,rotate around={0.:(3.9,11.8)}] (3.9,11.8) ++(0.:0.04 and 0.04) arc(0.:360.:0.04 and 0.04);
\draw [fill,rotate around={0.:(4.5,10.6)}] (4.5,10.6) ++(0.:0.04 and 0.04) arc(0.:360.:0.04 and 0.04);
\draw [fill,rotate around={0.:(3.9,9.8)}] (3.9,9.8) ++(0.:0.04 and 0.04) arc(0.:360.:0.04 and 0.04);
\draw [fill,rotate around={0.:(3.1,9.7)}] (3.1,9.7) ++(0.:0.04 and 0.04) arc(0.:360.:0.04 and 0.04);
\node [rotate=0.] at (3.6,10.7) {\tiny $#1$};
\node [rotate=0.] at (3.,11.5) {\tiny $#2$};
\node [rotate=0.] at (3.7,10.1) {\tiny $#3$};
\node [rotate=0.] at (1.6,9.8) {\tiny $#4$};
\node [rotate=0.] at (4.5,11.7) {\tiny $#6$};
\node [rotate=0.] at (4.6,10.) {\tiny $#5$};
\node [rotate=0.] at (2.4,10.4) {\tiny $\alpha_a$};
\node [rotate=0.] at (2.2,10.8) {\tiny $\alpha_b$};
\node [rotate=0.] at (3.1,9.) {\tiny $\alpha_r$};
\node [rotate=0.] at (3.3,9.5) {\tiny $r$};
\node [rotate=0.] at (2.9,9.4) {\tiny $s$};
\node [rotate=0.] at (2.4,9.4) {\tiny $\alpha_s$};
\end{tikzpicture}
}

\newcommand{\assoCondBB}[6]{
\begin{tikzpicture}[baseline={([yshift=-.8ex]current bounding box.center)},scale=0.81]
\draw [->-] (3.1,9.6) ..controls (2.856,9.717) and (2.612,9.833) .. (2.6,9.9)
			 ..controls (2.588,9.967) and (2.808,9.983) .. (3.,10.)
			 ..controls (3.192,10.017) and (3.355,10.033) .. (3.5,10.)
			 ..controls (3.645,9.967) and (3.773,9.883) .. (3.9,9.8);
\draw [->-] (2.9,12.8) -- (2.8,12.8);
\draw [->-] (1.7,10.8) -- (1.7,10.6);
\draw [->-] (2.9,8.8) -- (3.1,8.8);
\draw [->-] (4.3,10.1) -- (4.4,10.2);
\draw [->-] (4.5,10.8) -- (4.5,11.);
\draw [->-] (4.5,10.6) -- (2.6,10.3);
\draw [->-] (3.9,11.8) -- (2.4,10.9);
\draw [->-] (2.7,9.3) -- (3.1,9.6);
\draw [->-] (3.1,9.6) -- (3.2,9.2);
\draw [rotate around={0.:(2.9,10.8)}] (2.9,10.8) ++(31.:1.2 and 2.) arc(31.:330.:1.2 and 2.);
\draw [rotate around={0.:(3.9,10.8)}] (3.9,10.8) ++(270.:0.6 and 1.) arc(270.:450.:0.6 and 1.);
\draw [fill,rotate around={0.:(3.9,11.8)}] (3.9,11.8) ++(0.:0.04 and 0.04) arc(0.:360.:0.04 and 0.04);
\draw [fill,rotate around={0.:(4.5,10.6)}] (4.5,10.6) ++(0.:0.04 and 0.04) arc(0.:360.:0.04 and 0.04);
\draw [fill,rotate around={0.:(3.1,9.6)}] (3.1,9.6) ++(0.:0.04 and 0.04) arc(0.:360.:0.04 and 0.04);
\draw [fill,rotate around={0.:(3.9,9.8)}] (3.9,9.8) ++(0.:0.04 and 0.04) arc(0.:360.:0.04 and 0.04);
\node [rotate=0.] at (3.6,10.7) {\tiny $#1$};
\node [rotate=0.] at (3.,11.5) {\tiny $#2$};
\node [rotate=0.] at (3.8,10.) {\tiny $#3$};
\node [rotate=0.] at (1.6,9.9) {\tiny $#4$};
\node [rotate=0.] at (4.4,11.7) {\tiny $#6$};
\node [rotate=0.] at (4.6,10.) {\tiny $#5$};
\node [circle, inner sep=1pt, fill=white,draw, inner sep=1pt,rotate=0.] at (2.6,11.) {\tiny $g$};
\node [circle, inner sep=1pt, fill=white,draw, inner sep=1pt,rotate=0.] at (3.2,10.4) {\tiny $g$};
\node [rotate=0.] at (2.2,10.8) {\tiny $\alpha_b$};
\node [rotate=0.] at (2.4,10.4) {\tiny $\alpha_a$};
\node [circle, inner sep=1pt, fill=white,draw, inner sep=1pt,rotate=0.] at (3.3,10.) {\tiny $g$};
\node [rotate=0.] at (3.,9.4) {\tiny $s$};
\node [rotate=0.] at (2.4,9.3) {\tiny $\alpha_s$};
\node [rotate=0.] at (3.3,9.5) {\tiny $r$};
\node [rotate=0.] at (3.1,9.) {\tiny $\alpha_r$};
\end{tikzpicture}
}

\newcommand{\assoCondC}[5]{
\begin{tikzpicture}[baseline={([yshift=-.8ex]current bounding box.center)},scale=0.81]
\draw [->-] (2.8,12.8) -- (2.7,12.8);
\draw [->-] (1.6,10.8) -- (1.6,10.6);
\draw [->-] (4.2,10.1) -- (4.3,10.2);
\draw [->-] (4.4,10.8) -- (4.4,11.);
\draw [->-] (4.4,10.6) -- (2.6,10.4);
\draw [->-] (3.8,11.8) -- (2.3,10.9);
\draw [rotate around={0.:(3.8,10.8)}] (3.8,10.8) ++(270.:0.6 and 1.) arc(270.:450.:0.6 and 1.);
\draw [rotate around={0.:(2.8,10.8)}] (2.8,10.8) ++(31.:1.2 and 2.) arc(31.:180.:1.2 and 2.);
\draw [fill,rotate around={0.:(3.8,11.8)}] (3.8,11.8) ++(0.:0.04 and 0.04) arc(0.:360.:0.04 and 0.04);
\draw [fill,rotate around={0.:(4.4,10.6)}] (4.4,10.6) ++(0.:0.04 and 0.04) arc(0.:360.:0.04 and 0.04);
\node [rotate=0.] at (3.5,10.7) {\tiny $#1$};
\node [rotate=0.] at (2.9,11.5) {\tiny $#2$};
\node [rotate=0.] at (4.5,11.5) {\tiny $#5$};
\node [rotate=0.] at (2.2,12.8) {\tiny $#3$};
\node [rotate=0.] at (4.5,9.9) {\tiny $#4$};
\node [rotate=0.] at (2.1,10.7) {\tiny $\alpha_b$};
\node [rotate=0.] at (2.4,10.3) {\tiny $\alpha_a$};
\node [rotate=0.] at (3.6,9.7) {\tiny $\alpha_s$};
\node [rotate=0.] at (1.6,10.4) {\tiny $\alpha_r$};
\end{tikzpicture}
}

\newcommand{\avOperatorA}{
\begin{tikzpicture}[baseline={([yshift=-.8ex]current bounding box.center)},scale=0.7]
\draw [->-, fill] (3.4,11.2) -- (4.8,11.2);
\draw [->-, fill] (3.4,11.2) -- (3.4,12.8);
\draw [->-, fill] (3.4,10.3) -- (3.4,11.2);
\draw [->-, fill] (3.4,8.1) -- (3.4,10.3);
\draw [->-, fill] (1.6,10.3) -- (3.4,10.3);
\draw [fill,rotate around={0.:(3.4,10.3)}] (3.4,10.3) ++(0.:0.04 and 0.04) arc(0.:360.:0.04 and 0.04);
\draw [fill,rotate around={0.:(3.4,11.2)}] (3.4,11.2) ++(0.:0.04 and 0.04) arc(0.:360.:0.04 and 0.04);
\node [circle, inner sep=1pt, fill=white,draw, inner sep=1pt,rotate=0.] at (3.4,12.4) {\tiny $l$};
\node [circle, inner sep=1pt, fill=white,draw, inner sep=1pt,rotate=0.] at (2.,10.3) {\tiny $h$};
\node [circle, inner sep=1pt, fill=white,draw, inner sep=1pt,rotate=0.] at (3.4,8.6) {\tiny $g$};
\node [rotate=0.] at (3.7,9.1) {\scriptsize $\mu'$};
\node [fill=white,draw, circle, inner sep=1pt, inner sep=1pt,rotate=0.] at (3.4,9.6) {\tiny $k$};
\node [fill=white,draw, circle, inner sep=1pt, inner sep=1pt,rotate=0.] at (2.9,10.3) {\tiny $k$};
\node [rotate=0.] at (2.4,10.) {\scriptsize $\nu'$};
\node [rotate=0.] at (3.7,10.7) {\scriptsize $\gamma'$};
\node [rotate=0.] at (3.7,12.) {\scriptsize $\lambda'$};
\node [rotate=0.] at (3.9,11.4) {\scriptsize $s'$};
\node [fill=white,draw, circle, inner sep=1pt, inner sep=1pt,rotate=0.] at (3.4,11.6) {\tiny $\bar k$};
\node [rotate=0.] at (4.6,11.4) {\tiny $m_{s'}$};
\node [rotate=0.] at (3.8,8.2) {\tiny $n_{\mu'}$};
\node [rotate=0.] at (3.8,12.7) {\tiny $n_{\lambda'}$};
\node [rotate=0.] at (1.7,10.6) {\tiny $n_{\nu'}$};
\end{tikzpicture}
}

\newcommand{\avOperatorB}{
\begin{tikzpicture}[baseline={([yshift=-.8ex]current bounding box.center)},scale=0.9]
\draw [->-] (4.2,8.5) -- (4.3,8.5);
\draw [->-] (4.3,8.2) -- (4.2,8.2);
\draw [->-] (4.1,12.3) -- (4.2,12.3);
\draw [->-] (4.2,12.) -- (4.1,12.);
\draw [->-, fill] (2.5,10.7) -- (3.7,10.7);
\draw [->-, fill] (3.7,11.5) -- (3.7,10.7);
\draw [->-, fill] (3.7,10.7) -- (3.7,9.9);
\draw [->-, fill] (4.7,10.7) -- (5.9,10.7);
\draw [->-, fill] (3.7,9.9) -- (3.7,9.);
\draw [->-, fill] (4.7,10.7) -- (4.7,11.5);
\draw [->-, fill] (3.7,9.9) -- (4.7,9.9);
\draw [->-, fill] (4.7,9.) -- (4.7,9.9);
\draw [->-, fill] (4.7,9.9) -- (4.7,10.7);
\draw [rotate around={0.:(4.2,11.5)}] (4.2,11.5) ++(0.:0.5 and 0.5) arc(0.:180.:0.5 and 0.5);
\draw [rotate around={0.:(4.2,9.)}] (4.2,9.) ++(180.:0.5 and 0.5) arc(180.:360.:0.5 and 0.5);
\draw [rotate around={0.:(4.2,12.8)}] (4.2,12.8) ++(180.:0.5 and 0.5) arc(180.:360.:0.5 and 0.5);
\draw [rotate around={0.:(4.2,7.7)}] (4.2,7.7) ++(0.:0.5 and 0.5) arc(0.:180.:0.5 and 0.5);
\draw [fill,rotate around={0.:(3.7,10.7)}] (3.7,10.7) ++(0.:0.04 and 0.04) arc(0.:360.:0.04 and 0.04);
\draw [fill,rotate around={0.:(3.7,9.9)}] (3.7,9.9) ++(0.:0.04 and 0.04) arc(0.:360.:0.04 and 0.04);
\draw [fill,rotate around={0.:(4.7,10.7)}] (4.7,10.7) ++(0.:0.04 and 0.04) arc(0.:360.:0.04 and 0.04);
\draw [fill,rotate around={0.:(4.7,9.9)}] (4.7,9.9) ++(0.:0.04 and 0.04) arc(0.:360.:0.04 and 0.04);
\node [rotate=0.] at (4.2,10.1) {\scriptsize $\nu$};
\node [rotate=0.] at (5.2,8.1) {\scriptsize $\mu$};
\node [rotate=0.] at (5.2,8.7) {\scriptsize $\mu$};
\node [rotate=0.] at (5.1,11.7) {\scriptsize $\lambda$};
\node [rotate=0.] at (5.4,12.3) {\scriptsize $\lambda$};
\node [rotate=0.] at (5.,10.4) {\scriptsize $\gamma$};
\node [rotate=0.] at (5.1,10.9) {\scriptsize $s$};
\node [rotate=0.] at (3.5,10.4) {\scriptsize $\gamma'$};
\node [rotate=0.] at (3.5,10.9) {\scriptsize $s'$};
\node [fill=white,draw, circle, inner sep=1pt, inner sep=1pt,rotate=0.] at (2.8,10.7) {\tiny $k$};
\node [rotate=0.] at (3.4,11.7) {\scriptsize $\lambda'$};
\node [rotate=0.] at (3.2,12.3) {\scriptsize $\lambda'$};
\node [rotate=0.] at (3.4,8.7) {\scriptsize $\mu'$};
\node [rotate=0.] at (3.3,8.1) {\scriptsize $\mu'$};
\node [rotate=0.] at (2.4,11.) {\tiny $m_{s'}$};
\node [rotate=0.] at (5.8,10.9) {\tiny $m_{s}$};
\node [rotate=0.] at (5.2,7.7) {\tiny $n_{\mu}$};
\node [rotate=0.] at (3.3,7.7) {\tiny $n_{\mu'}$};
\node [rotate=0.] at (5.4,12.7) {\tiny $n_{\lambda}$};
\node [rotate=0.] at (3.2,12.7) {\tiny $n_{\lambda'}$};
\end{tikzpicture}
}

\newcommand{\avOperatorC}{
\begin{tikzpicture}[baseline={([yshift=-.8ex]current bounding box.center)},scale=0.9]
\draw [->-] (3.6,8.4) -- (3.7,8.4);
\draw [->-] (3.6,11.9) -- (3.5,11.9);
\draw [->-, fill] (1.9,10.6) -- (3.1,10.6);
\draw [->-, fill] (3.1,11.4) -- (3.1,10.6);
\draw [->-, fill] (3.1,10.6) -- (3.1,9.8);
\draw [->-, fill] (4.1,10.6) -- (5.3,10.6);
\draw [->-, fill] (3.1,9.8) -- (3.1,8.9);
\draw [->-, fill] (4.1,10.6) -- (4.1,11.4);
\draw [->-, fill] (3.1,9.8) -- (4.1,9.8);
\draw [->-, fill] (4.1,8.9) -- (4.1,9.8);
\draw [->-, fill] (4.1,9.8) -- (4.1,10.6);
\draw [rotate around={0.:(3.6,11.4)}] (3.6,11.4) ++(0.:0.5 and 0.5) arc(0.:180.:0.5 and 0.5);
\draw [rotate around={0.:(3.6,8.9)}] (3.6,8.9) ++(180.:0.5 and 0.5) arc(180.:360.:0.5 and 0.5);
\draw [fill,rotate around={0.:(3.1,10.6)}] (3.1,10.6) ++(0.:0.04 and 0.04) arc(0.:360.:0.04 and 0.04);
\draw [fill,rotate around={0.:(3.1,9.8)}] (3.1,9.8) ++(0.:0.04 and 0.04) arc(0.:360.:0.04 and 0.04);
\draw [fill,rotate around={0.:(4.1,10.6)}] (4.1,10.6) ++(0.:0.04 and 0.04) arc(0.:360.:0.04 and 0.04);
\draw [fill,rotate around={0.:(4.1,9.8)}] (4.1,9.8) ++(0.:0.04 and 0.04) arc(0.:360.:0.04 and 0.04);
\node [rotate=0.] at (3.6,10.) {\scriptsize $\nu$};
\node [rotate=0.] at (4.6,8.6) {\scriptsize $\mu$};
\node [rotate=0.] at (4.5,11.6) {\scriptsize $\lambda$};
\node [rotate=0.] at (4.4,10.3) {\scriptsize $\gamma$};
\node [rotate=0.] at (4.5,10.8) {\scriptsize $s$};
\node [rotate=0.] at (2.9,10.3) {\scriptsize $\gamma'$};
\node [rotate=0.] at (2.9,10.8) {\scriptsize $s'$};
\node [fill=white,draw, circle, inner sep=1pt, inner sep=1pt,rotate=0.] at (2.2,10.6) {\tiny $k$};
\node [rotate=0.] at (2.8,11.6) {\scriptsize $\lambda'$};
\node [rotate=0.] at (2.8,8.6) {\scriptsize $\mu'$};
\node [rotate=0.] at (1.8,10.9) {\tiny $m_{s'}$};
\node [rotate=0.] at (5.2,10.8) {\tiny $m_{s}$};
\end{tikzpicture}
}

\newcommand{\bpOperatorA}[9]{
\begin{tikzpicture}[baseline={([yshift=-.8ex]current bounding box.center)},scale=0.8]
\draw [->-] (3.2,7.8) -- (3.2,9.2);
\draw [->-] (3.2,9.2) -- (3.2,9.8);
\draw [->-] (3.2,9.8) -- (3.2,11.);
\draw [->-] (3.2,11.) -- (3.2,11.6);
\draw [->-] (3.2,11.6) -- (3.2,12.8);
\draw [->-] (3.2,11.6) -- (4.8,11.6);
\draw [->-] (1.7,11.) -- (3.2,11.);
\draw [->-] (3.2,9.8) -- (4.8,9.8);
\draw [->-] (1.6,9.2) -- (3.2,9.2);
\draw [fill,rotate around={0.:(3.2,11.6)}] (3.2,11.6) ++(0.:0.05 and 0.05) arc(0.:360.:0.05 and 0.05);
\draw [fill,rotate around={0.:(3.2,11.)}] (3.2,11.) ++(0.:0.05 and 0.05) arc(0.:360.:0.05 and 0.05);
\draw [fill,rotate around={0.:(3.2,9.8)}] (3.2,9.8) ++(0.:0.05 and 0.05) arc(0.:360.:0.05 and 0.05);
\draw [fill,rotate around={0.:(3.2,9.2)}] (3.2,9.2) ++(0.:0.05 and 0.05) arc(0.:360.:0.05 and 0.05);
\node [rotate=0.] at (3.5,8.7) {\scriptsize $#1$};
\node [rotate=0.] at (2.6,8.9) {\scriptsize $#2$};
\node [rotate=0.] at (3.,9.6) {\scriptsize $#3$};
\node [rotate=0.] at (3.9,9.5) {\scriptsize $#4$};
\node [rotate=0.] at (3.5,10.3) {\scriptsize $#5$};
\node [rotate=0.] at (2.6,11.3) {\scriptsize $#6$};
\node [rotate=0.] at (3.5,11.2) {\scriptsize $#7$};
\node [rotate=0.] at (3.4,12.) {\scriptsize $#8$};
\node [rotate=0.] at (3.9,11.8) {\scriptsize $#9$};
\node [circle, inner sep=1pt, fill=white,draw, inner sep=1pt,rotate=0.] at (3.2,8.2) {\tiny $g$};
\node [circle, inner sep=1pt, fill=white,draw, inner sep=1pt,rotate=0.] at (1.9,9.2) {\tiny $h$};
\node [circle, inner sep=1pt, fill=white,draw, inner sep=1pt,rotate=0.] at (3.2,10.7) {\tiny $l$};
\node [circle, inner sep=1pt, fill=white,draw, inner sep=1pt,rotate=0.] at (2.1,11.) {\tiny $x$};
\node [circle, inner sep=1pt, fill=white,draw, inner sep=1pt,rotate=0.] at (3.2,12.4) {\tiny $y$};
\node [rotate=0.] at (3.6,7.9) {\tiny $n_{#1}$};
\node [rotate=0.] at (1.7,8.9) {\tiny $n_{#2}$};
\node [rotate=0.] at (1.9,11.3) {\tiny $n_{#6}$};
\node [rotate=0.] at (3.5,12.6) {\tiny $n_{#8}$};
\node [rotate=0.] at (4.7,11.8) {\tiny $\tilde{m}_{s'}$};
\node [rotate=0.] at (4.7,10.) {\tiny $\tilde{m}_{r'}$};
\end{tikzpicture}
}

\newcommand{\bpOperatorAA}[9]{
\begin{tikzpicture}[baseline={([yshift=-.8ex]current bounding box.center)},scale=0.8]
\draw [->-] (3.2,7.9) -- (3.2,9.3);
\draw [->-] (3.2,9.3) -- (3.2,9.9);
\draw [->-] (3.2,9.9) -- (3.2,11.1);
\draw [->-] (3.2,11.1) -- (3.2,11.7);
\draw [->-] (3.2,11.7) -- (3.2,13.);
\draw [->-] (3.2,11.7) -- (4.8,11.7);
\draw [->-] (1.7,11.1) -- (3.2,11.1);
\draw [->-] (3.2,9.9) -- (4.8,9.9);
\draw [->-] (1.6,9.3) -- (3.2,9.3);
\draw [fill,rotate around={0.:(3.2,11.7)}] (3.2,11.7) ++(0.:0.05 and 0.05) arc(0.:360.:0.05 and 0.05);
\draw [fill,rotate around={0.:(3.2,11.1)}] (3.2,11.1) ++(0.:0.05 and 0.05) arc(0.:360.:0.05 and 0.05);
\draw [fill,rotate around={0.:(3.2,9.9)}] (3.2,9.9) ++(0.:0.05 and 0.05) arc(0.:360.:0.05 and 0.05);
\draw [fill,rotate around={0.:(3.2,9.3)}] (3.2,9.3) ++(0.:0.05 and 0.05) arc(0.:360.:0.05 and 0.05);
\node [rotate=0.] at (3.5,8.8) {\scriptsize $#1$};
\node [rotate=0.] at (2.6,9.) {\scriptsize $#2$};
\node [rotate=0.] at (3.,9.7) {\scriptsize $#3$};
\node [rotate=0.] at (3.9,9.6) {\scriptsize $#4$};
\node [rotate=0.] at (3.5,10.4) {\scriptsize $#5$};
\node [rotate=0.] at (2.6,11.4) {\scriptsize $#6$};
\node [rotate=0.] at (3.5,11.3) {\scriptsize $#7$};
\node [rotate=0.] at (3.4,12.1) {\scriptsize $#8$};
\node [rotate=0.] at (3.9,11.9) {\scriptsize $#9$};
\node [circle, inner sep=1pt, fill=white,draw, inner sep=1pt,rotate=0.] at (3.2,8.3) {\tiny $g$};
\node [circle, inner sep=1pt, fill=white,draw, inner sep=1pt,rotate=0.] at (1.9,9.3) {\tiny $h$};
\node [circle, inner sep=1pt, fill=white,draw, inner sep=1pt,rotate=0.] at (3.2,10.8) {\tiny $l$};
\node [circle, inner sep=1pt, fill=white,draw, inner sep=1pt,rotate=0.] at (2.1,11.1) {\tiny $x$};
\node [circle, inner sep=1pt, fill=white,draw, inner sep=1pt,rotate=0.] at (3.2,12.5) {\tiny $y$};
\node [rotate=0.] at (4.7,10.1) {\tiny $\tilde{m}_{r}$};
\node [rotate=0.] at (4.7,11.9) {\tiny $\tilde{m}_{s}$};
\node [rotate=0.] at (3.6,8.) {\tiny $n_{#1}$};
\node [rotate=0.] at (3.5,12.7) {\tiny $n_{#8}$};
\node [rotate=0.] at (1.8,11.4) {\tiny $n_{#6}$};
\node [rotate=0.] at (1.7,9.) {\tiny $n_{#2}$};
\end{tikzpicture}
}

\newcommand{\bpOperatorB}[9]{
\begin{tikzpicture}[baseline={([yshift=-.8ex]current bounding box.center)},scale=0.8]
\draw [->-] (3.2,6.7) -- (3.2,8.4);
\draw [->-] (3.2,8.4) -- (3.2,9.);
\draw [->-] (3.2,9.) -- (3.2,10.5);
\draw [->-] (3.2,10.5) -- (3.2,11.4);
\draw [->-] (3.2,11.4) -- (3.2,12.8);
\draw [->-] (3.2,11.4) -- (4.8,11.4);
\draw [->-] (1.7,10.5) -- (3.2,10.5);
\draw [->-] (3.2,9.) -- (4.8,9.);
\draw [->-] (1.6,8.4) -- (3.2,8.4);
\draw [->-] (3.2,11.4) -- (3.2,11.5);
\draw [->-] (3.2,9.) -- (3.2,9.5);
\draw [->-] (3.2,10.5) -- (3.2,10.6);
\draw [fill,rotate around={0.:(3.2,11.4)}] (3.2,11.4) ++(0.:0.05 and 0.05) arc(0.:360.:0.05 and 0.05);
\draw [fill,rotate around={0.:(3.2,10.5)}] (3.2,10.5) ++(0.:0.05 and 0.05) arc(0.:360.:0.05 and 0.05);
\draw [fill,rotate around={0.:(3.2,9.)}] (3.2,9.) ++(0.:0.05 and 0.05) arc(0.:360.:0.05 and 0.05);
\draw [fill,rotate around={0.:(3.2,8.4)}] (3.2,8.4) ++(0.:0.05 and 0.05) arc(0.:360.:0.05 and 0.05);
\node [rotate=0.] at (3.4,7.5) {\scriptsize $#1$};
\node [rotate=0.] at (2.4,8.2) {\scriptsize $#2$};
\node [rotate=0.] at (3.,8.7) {\scriptsize $#3$};
\node [rotate=0.] at (3.8,8.7) {\scriptsize $#4$};
\node [rotate=0.] at (3.6,9.8) {\scriptsize $#5$};
\node [rotate=0.] at (2.4,10.8) {\scriptsize $#6$};
\node [rotate=0.] at (3.6,10.7) {\scriptsize $#7$};
\node [rotate=0.] at (3.4,11.8) {\scriptsize $#8$};
\node [rotate=0.] at (3.8,11.1) {\scriptsize $#9$};
\node [circle, inner sep=1pt, fill=white,draw, inner sep=1pt,rotate=0.] at (4.4,9.) {\tiny $l_1$};
\node [circle, inner sep=1pt, fill=white,draw, inner sep=1pt,rotate=0.] at (3.2,9.8) {\tiny $l_2$};
\node [circle, inner sep=1pt, fill=white,draw, inner sep=1pt,rotate=0.] at (3.2,10.9) {\tiny $l_3$};
\node [circle, inner sep=1pt, fill=white,draw, inner sep=1pt,rotate=0.] at (4.4,11.4) {\tiny $l_4$};
\node [circle, inner sep=1pt, fill=white,draw, inner sep=1pt,rotate=0.] at (2.,10.5) {\tiny $x'$};
\node [circle, inner sep=1pt, fill=white,draw, inner sep=1pt,rotate=0.] at (3.2,12.4) {\tiny $y'$};
\node [circle, inner sep=1pt, fill=white,draw, inner sep=1pt,rotate=0.] at (2.,8.4) {\tiny $h'$};
\node [circle, inner sep=1pt, fill=white,draw, inner sep=1pt,rotate=0.] at (3.2,8.) {\tiny $g'$};
\node [rotate=0.] at (4.7,11.8) {\tiny $\tilde{m}_{s'}$};
\node [rotate=0.] at (4.7,9.4) {\tiny $\tilde{m}_{r'}$};
\node [rotate=0.] at (3.6,12.7) {\tiny $n_{#8}$};
\node [rotate=0.] at (3.6,6.9) {\tiny $n_{#1}$};
\node [rotate=0.] at (1.7,8.8) {\tiny $n_{#2}$};
\node [rotate=0.] at (1.7,10.9) {\tiny $n_{#6}$};
\end{tikzpicture}
}

\newcommand{\bpOperatorBB}[9]{
\begin{tikzpicture}[baseline={([yshift=-.8ex]current bounding box.center)},scale=0.8]
\draw [->-] (3.2,6.7) -- (3.2,8.4);
\draw [->-] (3.2,8.4) -- (3.2,9.);
\draw [->-] (3.2,9.) -- (3.2,10.5);
\draw [->-] (3.2,10.5) -- (3.2,11.4);
\draw [->-] (3.2,11.4) -- (3.2,12.8);
\draw [->-] (3.2,11.4) -- (4.8,11.4);
\draw [->-] (1.7,10.5) -- (3.2,10.5);
\draw [->-] (3.2,9.) -- (4.8,9.);
\draw [->-] (1.6,8.4) -- (3.2,8.4);
\draw [->-] (3.2,11.4) -- (3.2,11.5);
\draw [->-] (3.2,9.) -- (3.2,9.5);
\draw [->-] (3.2,10.5) -- (3.2,10.6);
\draw [fill,rotate around={0.:(3.2,11.4)}] (3.2,11.4) ++(0.:0.05 and 0.05) arc(0.:360.:0.05 and 0.05);
\draw [fill,rotate around={0.:(3.2,10.5)}] (3.2,10.5) ++(0.:0.05 and 0.05) arc(0.:360.:0.05 and 0.05);
\draw [fill,rotate around={0.:(3.2,9.)}] (3.2,9.) ++(0.:0.05 and 0.05) arc(0.:360.:0.05 and 0.05);
\draw [fill,rotate around={0.:(3.2,8.4)}] (3.2,8.4) ++(0.:0.05 and 0.05) arc(0.:360.:0.05 and 0.05);
\node [rotate=0.] at (3.4,7.5) {\scriptsize $#1$};
\node [rotate=0.] at (2.4,8.2) {\scriptsize $#2$};
\node [rotate=0.] at (3.,8.7) {\scriptsize $#3$};
\node [rotate=0.] at (3.8,8.8) {\scriptsize $#4$};
\node [rotate=0.] at (3.6,9.8) {\scriptsize $#5$};
\node [rotate=0.] at (2.4,10.8) {\scriptsize $#6$};
\node [rotate=0.] at (3.6,10.7) {\scriptsize $#7$};
\node [rotate=0.] at (3.4,11.8) {\scriptsize $#8$};
\node [rotate=0.] at (3.8,11.2) {\scriptsize $#9$};
\node [circle, inner sep=1pt, fill=white,draw, inner sep=1pt,rotate=0.] at (4.4,9.) {\tiny $l_1$};
\node [circle, inner sep=1pt, fill=white,draw, inner sep=1pt,rotate=0.] at (3.2,9.8) {\tiny $l_2$};
\node [circle, inner sep=1pt, fill=white,draw, inner sep=1pt,rotate=0.] at (3.2,10.9) {\tiny $l_3$};
\node [circle, inner sep=1pt, fill=white,draw, inner sep=1pt,rotate=0.] at (4.4,11.4) {\tiny $l_4$};
\node [circle, inner sep=1pt, fill=white,draw, inner sep=1pt,rotate=0.] at (2.,10.5) {\tiny $x'$};
\node [circle, inner sep=1pt, fill=white,draw, inner sep=1pt,rotate=0.] at (3.2,12.4) {\tiny $y'$};
\node [circle, inner sep=1pt, fill=white,draw, inner sep=1pt,rotate=0.] at (2.,8.4) {\tiny $h'$};
\node [circle, inner sep=1pt, fill=white,draw, inner sep=1pt,rotate=0.] at (3.2,8.) {\tiny $g'$};
\node [rotate=0.] at (4.7,11.8) {\tiny $\tilde{m}_s$};
\node [rotate=0.] at (4.7,9.4) {\tiny $\tilde{m}_r$};
\node [rotate=0.] at (3.6,12.7) {\tiny $n_{#8}$};
\node [rotate=0.] at (3.6,6.9) {\tiny $n_{#1}$};
\node [rotate=0.] at (1.7,8.8) {\tiny $n_{#2}$};
\node [rotate=0.] at (1.7,10.9) {\tiny $n_{#6}$};
\end{tikzpicture}
}

\newcommand{\bpOperatorC}{
\begin{tikzpicture}[baseline={([yshift=-.8ex]current bounding box.center)},scale=0.8]
\draw [->-] (1.9,11.5) -- (2.,11.4);
\draw [->-] (2.,13.6) -- (1.9,13.5);
\draw [rotate around={0.:(2.5,12.5)}] (2.5,12.5) ++(10.:0.9 and 1.3) arc(10.:350.:0.9 and 1.3);
\node [circle, inner sep=1pt, fill=white,draw, inner sep=1pt,rotate=0.] at (1.6,12.5) {\tiny $l$};
\node [rotate=0.] at (3.2,13.6) {\scriptsize $t$};
\node [rotate=0.] at (3.6,12.1) {\scriptsize $\alpha_t$};
\node [rotate=0.] at (3.6,12.8) {\scriptsize $\alpha_t$};
\end{tikzpicture}
}

\newcommand{\bpOperatorD}{
\begin{tikzpicture}[baseline={([yshift=-.8ex]current bounding box.center)},scale=0.8]
\draw [->-] (1.8,13.1) -- (1.7,12.9);
\draw [->-] (2.5,13.6) -- (2.4,13.6);
\draw [->-] (2.5,11.) -- (2.6,11.);
\draw [rotate around={0.:(2.5,12.3)}] (2.5,12.3) ++(10.:0.9 and 1.3) arc(10.:350.:0.9 and 1.3);
\node [circle, inner sep=1pt, fill=white,draw, inner sep=1pt,rotate=0.] at (2.,11.2) {\tiny $l_1$};
\node [circle, inner sep=1pt, fill=white,draw, inner sep=1pt,rotate=0.] at (1.7,11.8) {\tiny $l_2$};
\node [circle, inner sep=1pt, fill=white,draw, inner sep=1pt,rotate=0.] at (1.6,12.5) {\tiny $l_3$};
\node [circle, inner sep=1pt, fill=white,draw, inner sep=1pt,rotate=0.] at (2.1,13.5) {\tiny $l_4$};
\node [rotate=0.] at (3.3,13.5) {\scriptsize $t$};
\node [rotate=0.] at (3.6,12.) {\scriptsize $\alpha_t$};
\node [rotate=0.] at (3.6,12.7) {\scriptsize $\alpha_t$};
\end{tikzpicture}
}

\newcommand{\bpOperatorE}{
\begin{tikzpicture}[baseline={([yshift=-.8ex]current bounding box.center)},scale=0.7]
\draw [->-] (8.4,12.1) -- (8.4,12.4);
\draw [->-] (8.4,12.1) -- (10.,12.1);
\draw [->-] (4.5,12.1) -- (6.,12.1);
\draw [->-] (6.,12.1) -- (6.,10.6);
\draw [->-] (8.4,10.6) -- (8.4,12.1);
\draw [->-] (4.5,8.7) -- (6.,8.7);
\draw [->-] (8.4,8.7) -- (8.4,10.6);
\draw [->-] (6.,10.6) -- (8.4,10.6);
\draw [->-] (8.4,8.7) -- (10.,8.7);
\draw [->-] (6.,8.7) -- (6.,7.8);
\draw [->-] (8.4,7.8) -- (8.4,8.7);
\draw [->-] (6.,10.6) -- (6.,8.7);
\draw [->-] (10.6,11.2) -- (10.5,11.3);
\draw [->-] (8.4,7.1) -- (8.4,7.8);
\draw [->-] (6.,7.8) -- (6.,7.1);
\draw [->-] (10.5,9.7) -- (10.6,9.8);
\draw [->-] (6.,12.4) -- (6.,12.1);
\draw [->-] (7.2,13.) -- (7.1,13.);
\draw [->-] (7.2,6.5) -- (7.3,6.5);
\draw [->-] (8.4,10.6) -- (8.4,11.4);
\draw [->-] (6.,11.4) -- (6.,10.6);
\draw [->-] (6.,9.7) -- (6.,8.7);
\draw [->-] (8.4,8.7) -- (8.4,9.7);
\draw [->-] (9.6,10.6) -- (9.6,10.5);
\draw [->-] (6.,7.8) -- (8.4,7.8);
\draw [fill,rotate around={0.:(6.,12.1)}] (6.,12.1) ++(0.:0.05 and 0.05) arc(0.:360.:0.05 and 0.05);
\draw [fill,rotate around={0.:(8.4,12.1)}] (8.4,12.1) ++(0.:0.05 and 0.05) arc(0.:360.:0.05 and 0.05);
\draw [fill,rotate around={0.:(6.,10.6)}] (6.,10.6) ++(0.:0.05 and 0.05) arc(0.:360.:0.05 and 0.05);
\draw [fill,rotate around={0.:(8.4,10.6)}] (8.4,10.6) ++(0.:0.05 and 0.05) arc(0.:360.:0.05 and 0.05);
\draw [fill,rotate around={0.:(6.,8.7)}] (6.,8.7) ++(0.:0.05 and 0.05) arc(0.:360.:0.05 and 0.05);
\draw [fill,rotate around={0.:(8.4,8.7)}] (8.4,8.7) ++(0.:0.05 and 0.05) arc(0.:360.:0.05 and 0.05);
\draw [fill,rotate around={0.:(6.,7.8)}] (6.,7.8) ++(0.:0.05 and 0.05) arc(0.:360.:0.05 and 0.05);
\draw [fill,rotate around={0.:(8.4,7.8)}] (8.4,7.8) ++(0.:0.05 and 0.05) arc(0.:360.:0.05 and 0.05);
\draw [rotate around={0.:(10.2,10.5)}] (10.2,10.5) ++(10.:0.6 and 0.9) arc(10.:350.:0.6 and 0.9);
\draw [rotate around={0.:(7.2,12.4)}] (7.2,12.4) ++(0.:1.2 and 0.6) arc(0.:180.:1.2 and 0.6);
\draw [rotate around={0.:(7.2,7.1)}] (7.2,7.1) ++(180.:1.2 and 0.6) arc(180.:360.:1.2 and 0.6);
\node [fill=white,draw, circle, inner sep=1pt, inner sep=1pt,rotate=0.] at (5.6,12.1) {\tiny $l_4$};
\node [fill=white,draw, circle, inner sep=1pt, inner sep=1pt,rotate=0.] at (8.7,12.1) {\tiny $l_4$};
\node [fill=white,draw, circle, inner sep=1pt, inner sep=1pt,rotate=0.] at (10.1,11.4) {\tiny $l_4$};
\node [fill=white,draw, circle, inner sep=1pt, inner sep=1pt,rotate=0.] at (9.7,10.9) {\tiny $l_3$};
\node [fill=white,draw, circle, inner sep=1pt, inner sep=1pt,rotate=0.] at (9.6,10.2) {\tiny $l_2$};
\node [fill=white,draw, circle, inner sep=1pt, inner sep=1pt,rotate=0.] at (10.,9.7) {\tiny $l_1$};
\node [fill=white,draw, circle, inner sep=1pt, inner sep=1pt,rotate=0.] at (5.6,8.7) {\tiny $l_1$};
\node [fill=white,draw, circle, inner sep=1pt, inner sep=1pt,rotate=0.] at (8.7,8.7) {\tiny $l_1$};
\node [fill=white,draw, circle, inner sep=1pt, inner sep=1pt,rotate=0.] at (8.4,11.4) {\tiny $l_3$};
\node [fill=white,draw, circle, inner sep=1pt, inner sep=1pt,rotate=0.] at (6.,11.4) {\tiny $l_3$};
\node [fill=white,draw, circle, inner sep=1pt, inner sep=1pt,rotate=0.] at (6.,9.7) {\tiny $l_2$};
\node [fill=white,draw, circle, inner sep=1pt, inner sep=1pt,rotate=0.] at (8.4,9.7) {\tiny $l_2$};
\node [rotate=0.] at (9.2,12.3) {\tiny $s$};
\node [rotate=0.] at (5.2,12.4) {\tiny $s'$};
\node [rotate=0.] at (9.8,12.3) {\tiny $\tilde{m}_s$};
\node [rotate=0.] at (8.6,11.1) {\tiny $\eta$};
\node [rotate=0.] at (6.3,11.1) {\tiny $\eta'$};
\node [rotate=0.] at (6.2,10.) {\tiny $\lambda'$};
\node [rotate=0.] at (8.6,10.) {\tiny $\lambda$};
\node [rotate=0.] at (10.8,11.2) {\tiny $t$};
\node [rotate=0.] at (11.,10.8) {\tiny $\alpha_t$};
\node [rotate=0.] at (11.,10.2) {\tiny $\alpha_t$};
\node [rotate=0.] at (9.8,9.) {\tiny $\tilde{m}_r$};
\node [rotate=0.] at (5.2,9.) {\tiny $r'$};
\node [rotate=0.] at (9.2,9.) {\tiny $r$};
\node [rotate=0.] at (5.7,8.2) {\tiny $\gamma'$};
\node [rotate=0.] at (8.7,8.2) {\tiny $\gamma$};
\node [rotate=0.] at (4.6,12.3) {\tiny $\tilde{m}_{s'}$};
\node [rotate=0.] at (4.6,8.9) {\tiny $\tilde{m}_{r'}$};
\node [rotate=0.] at (7.1,10.9) {\tiny $\kappa$};
\node [rotate=0.] at (7.1,8.) {\tiny $\nu$};
\node [rotate=0.] at (7.2,6.2) {\tiny $\mu$};
\node [rotate=0.] at (7.2,13.2) {\tiny $\rho$};
\end{tikzpicture}
}

\newcommand{\bpOperatorF}{
\begin{tikzpicture}[baseline={([yshift=-.8ex]current bounding box.center)},scale=0.6]
\draw [->-] (2.8,11.8) -- (2.8,10.7);
\draw [->-] (2.8,10.3) -- (2.8,9.3);
\draw [->-] (2.8,9.1) -- (2.8,8.2);
\draw [->-] (1.6,5.6) -- (4.,5.6);
\draw [->-] (2.8,7.5) -- (2.8,6.3);
\draw [->-] (4.,5.6) -- (2.8,6.3);
\draw [->-] (5.,12.4) -- (5.6,11.8);
\draw [->-] (5.6,11.8) -- (6.2,12.4);
\draw [->-] (5.6,6.6) -- (6.2,5.8);
\draw [->-] (5.,5.8) -- (5.6,6.6);
\draw [->-] (5.6,11.) -- (5.6,11.8);
\draw [->-] (5.6,6.6) -- (5.6,7.5);
\draw [->-] (4.,12.4) -- (1.6,12.4);
\draw [->-] (4.,12.4) -- (2.8,11.8);
\draw [->-] (2.8,10.7) -- (4.,12.4);
\draw [->-] (2.8,11.8) -- (1.6,12.4);
\draw [->-] (1.6,12.4) -- (2.8,10.7);
\draw [->-] (2.8,8.2) -- (4.,9.2);
\draw [->-] (4.,9.2) -- (2.8,10.3);
\draw [->-] (2.8,10.3) -- (1.6,9.2);
\draw [->-] (1.6,9.2) -- (2.8,8.2);
\draw [->-] (1.6,9.2) -- (4.,9.2);
\draw [->-] (2.8,6.3) -- (1.6,5.6);
\draw [->-] (4.,5.6) -- (2.8,7.5);
\draw [->-] (2.8,7.5) -- (1.6,5.6);
\draw [fill,rotate around={0.:(1.6,12.4)}] (1.6,12.4) ++(0.:0.05 and 0.05) arc(0.:360.:0.05 and 0.05);
\draw [fill,rotate around={0.:(4.,12.4)}] (4.,12.4) ++(0.:0.05 and 0.05) arc(0.:360.:0.05 and 0.05);
\draw [fill,rotate around={0.:(2.8,10.7)}] (2.8,10.7) ++(0.:0.05 and 0.05) arc(0.:360.:0.05 and 0.05);
\draw [fill,rotate around={0.:(2.8,11.8)}] (2.8,11.8) ++(0.:0.05 and 0.05) arc(0.:360.:0.05 and 0.05);
\draw [fill,rotate around={0.:(5.6,11.8)}] (5.6,11.8) ++(0.:0.05 and 0.05) arc(0.:360.:0.05 and 0.05);
\draw [fill,rotate around={0.:(5.6,6.6)}] (5.6,6.6) ++(0.:0.05 and 0.05) arc(0.:360.:0.05 and 0.05);
\draw [fill,rotate around={0.:(2.8,10.3)}] (2.8,10.3) ++(0.:0.05 and 0.05) arc(0.:360.:0.05 and 0.05);
\draw [fill,rotate around={0.:(4.,9.2)}] (4.,9.2) ++(0.:0.05 and 0.05) arc(0.:360.:0.05 and 0.05);
\draw [fill,rotate around={0.:(1.6,9.2)}] (1.6,9.2) ++(0.:0.05 and 0.05) arc(0.:360.:0.05 and 0.05);
\draw [fill,rotate around={0.:(2.8,8.2)}] (2.8,8.2) ++(0.:0.05 and 0.05) arc(0.:360.:0.05 and 0.05);
\draw [fill,rotate around={0.:(2.8,7.5)}] (2.8,7.5) ++(0.:0.05 and 0.05) arc(0.:360.:0.05 and 0.05);
\draw [fill,rotate around={0.:(2.8,6.3)}] (2.8,6.3) ++(0.:0.05 and 0.05) arc(0.:360.:0.05 and 0.05);
\draw [fill,rotate around={0.:(4.,5.6)}] (4.,5.6) ++(0.:0.05 and 0.05) arc(0.:360.:0.05 and 0.05);
\draw [fill,rotate around={0.:(1.6,5.6)}] (1.6,5.6) ++(0.:0.05 and 0.05) arc(0.:360.:0.05 and 0.05);
\node [rotate=0.] at (2.5,12.2) {\tiny $s'$};
\node [rotate=0.] at (3.2,12.2) {\tiny $s$};
\node [rotate=0.] at (3.,11.4) {\tiny $t$};
\node [rotate=0.] at (3.6,11.3) {\tiny $\eta$};
\node [rotate=0.] at (2.,11.3) {\tiny $\eta'$};
\node [rotate=0.] at (3.5,10.1) {\tiny $\eta$};
\node [rotate=0.] at (3.,8.8) {\tiny $t$};
\node [rotate=0.] at (3.,6.6) {\tiny $t$};
\node [rotate=0.] at (2.4,9.4) {\tiny $\kappa$};
\node [rotate=0.] at (3.5,8.3) {\tiny $\lambda$};
\node [rotate=0.] at (2.1,8.3) {\tiny $\lambda'$};
\node [rotate=0.] at (2.,6.9) {\tiny $\lambda'$};
\node [rotate=0.] at (3.5,6.9) {\tiny $\lambda$};
\node [rotate=0.] at (3.3,6.2) {\tiny $r$};
\node [rotate=0.] at (2.3,6.2) {\tiny $r'$};
\node [rotate=0.] at (2.8,5.4) {\tiny $\gamma$};
\node [rotate=0.] at (5.1,12.) {\tiny $s'$};
\node [rotate=0.] at (5.7,12.2) {\tiny $s$};
\node [rotate=0.] at (6.4,12.6) {\tiny $\tilde{m}_s$};
\node [rotate=0.] at (5.8,11.4) {\tiny $t$};
\node [rotate=0.] at (5.6,10.7) {\tiny $\alpha_t$};
\node [rotate=0.] at (5.6,7.8) {\tiny $\alpha_t$};
\node [rotate=0.] at (5.1,6.3) {\tiny $r'$};
\node [rotate=0.] at (6.4,5.6) {\tiny $\tilde{m}_r$};
\node [rotate=0.] at (6.1,6.3) {\tiny $r$};
\node [rotate=0.] at (2.1,10.1) {\tiny $\eta'$};
\node [rotate=0.] at (5.8,7.) {\tiny $t$};
\node [rotate=0.] at (4.8,5.6) {\tiny $\tilde{m}_{r'}$};
\node [rotate=0.] at (4.8,12.6) {\tiny $\tilde{m}_{s'}$};
\node [rotate=0.] at (2.8,12.6) {\tiny $\rho$};
\end{tikzpicture}
}

\newcommand{\bpOperatorG}[9]{
\begin{tikzpicture}[baseline={([yshift=-.8ex]current bounding box.center)},scale=0.8]
\draw [->-] (3.2,5.6) -- (3.2,7.2);
\draw [->-] (3.2,7.2) -- (3.2,7.8);
\draw [->-] (3.2,7.8) -- (3.2,10.8);
\draw [->-] (3.2,10.8) -- (3.2,11.4);
\draw [->-] (3.2,11.4) -- (3.2,12.8);
\draw [->-] (3.2,11.4) -- (4.8,11.4);
\draw [->-] (1.7,10.8) -- (3.2,10.8);
\draw [->-] (3.2,7.8) -- (4.8,7.8);
\draw [->-] (1.6,7.2) -- (3.2,7.2);
\draw [fill,rotate around={0.:(3.2,11.4)}] (3.2,11.4) ++(0.:0.05 and 0.05) arc(0.:360.:0.05 and 0.05);
\draw [fill,rotate around={0.:(3.2,10.8)}] (3.2,10.8) ++(0.:0.05 and 0.05) arc(0.:360.:0.05 and 0.05);
\draw [fill,rotate around={0.:(3.2,7.8)}] (3.2,7.8) ++(0.:0.05 and 0.05) arc(0.:360.:0.05 and 0.05);
\draw [fill,rotate around={0.:(3.2,7.2)}] (3.2,7.2) ++(0.:0.05 and 0.05) arc(0.:360.:0.05 and 0.05);
\node [rotate=0.] at (3.6,6.4) {\scriptsize $#1$};
\node [rotate=0.] at (2.6,7.) {\scriptsize $#2$};
\node [rotate=0.] at (3.,7.6) {\scriptsize $#3$};
\node [rotate=0.] at (3.7,7.6) {\scriptsize $#4$};
\node [rotate=0.] at (3.7,9.9) {\scriptsize $#5$};
\node [rotate=0.] at (2.6,11.) {\scriptsize $#6$};
\node [rotate=0.] at (3.5,11.) {\scriptsize $#7$};
\node [rotate=0.] at (3.6,12.2) {\scriptsize $#8$};
\node [rotate=0.] at (3.7,11.6) {\scriptsize $#9$};
\node [circle, inner sep=1pt, fill=white,draw, inner sep=1pt,rotate=0.] at (3.2,6.) {\tiny $g$};
\node [circle, inner sep=1pt, fill=white,draw, inner sep=1pt,rotate=0.] at (1.9,7.2) {\tiny $h$};
\node [circle, inner sep=1pt, fill=white,draw, inner sep=1pt,rotate=0.] at (2.1,10.8) {\tiny $x$};
\node [circle, inner sep=1pt, fill=white,draw, inner sep=1pt,rotate=0.] at (3.2,12.4) {\tiny $y$};
\node [circle, inner sep=1pt, fill=white,draw, inner sep=1pt,rotate=0.] at (3.2,8.3) {\tiny $l_1$};
\node [circle, inner sep=1pt, fill=white,draw, inner sep=1pt,rotate=0.] at (3.2,8.9) {\tiny $l_2$};
\node [circle, inner sep=1pt, fill=white,draw, inner sep=1pt,rotate=0.] at (3.2,9.9) {\tiny $l_3$};
\node [circle, inner sep=1pt, fill=white,draw, inner sep=1pt,rotate=0.] at (3.2,10.4) {\tiny $l_4$};
\node [rotate=0.] at (4.7,8.) {\tiny $\tilde{m}_{r'}$};
\node [rotate=0.] at (4.7,11.6) {\tiny $\tilde{m}_{s'}$};
\node [rotate=0.] at (3.6,12.7) {\tiny $n_{#8}$};
\node [rotate=0.] at (3.6,5.7) {\tiny $n_{#1}$};
\node [rotate=0.] at (1.7,11.2) {\tiny $n_{#6}$};
\node [rotate=0.] at (1.7,7.6) {\tiny $n_{#2}$};
\end{tikzpicture}
}

\newcommand{\bpOperatorGG}[9]{
\begin{tikzpicture}[baseline={([yshift=-.8ex]current bounding box.center)},scale=0.8]
\draw [->-] (3.2,5.6) -- (3.2,7.2);
\draw [->-] (3.2,7.2) -- (3.2,7.8);
\draw [->-] (3.2,7.8) -- (3.2,10.8);
\draw [->-] (3.2,10.8) -- (3.2,11.4);
\draw [->-] (3.2,11.4) -- (3.2,12.8);
\draw [->-] (3.2,11.4) -- (4.8,11.4);
\draw [->-] (1.7,10.8) -- (3.2,10.8);
\draw [->-] (3.2,7.8) -- (4.8,7.8);
\draw [->-] (1.6,7.2) -- (3.2,7.2);
\draw [fill,rotate around={0.:(3.2,11.4)}] (3.2,11.4) ++(0.:0.05 and 0.05) arc(0.:360.:0.05 and 0.05);
\draw [fill,rotate around={0.:(3.2,10.8)}] (3.2,10.8) ++(0.:0.05 and 0.05) arc(0.:360.:0.05 and 0.05);
\draw [fill,rotate around={0.:(3.2,7.8)}] (3.2,7.8) ++(0.:0.05 and 0.05) arc(0.:360.:0.05 and 0.05);
\draw [fill,rotate around={0.:(3.2,7.2)}] (3.2,7.2) ++(0.:0.05 and 0.05) arc(0.:360.:0.05 and 0.05);
\node [rotate=0.] at (3.6,6.4) {\scriptsize $#1$};
\node [rotate=0.] at (2.6,7.) {\scriptsize $#2$};
\node [rotate=0.] at (3.,7.6) {\scriptsize $#3$};
\node [rotate=0.] at (3.7,7.6) {\scriptsize $#4$};
\node [rotate=0.] at (3.7,9.8) {\scriptsize $#5$};
\node [rotate=0.] at (2.6,11.) {\scriptsize $#6$};
\node [rotate=0.] at (3.5,11.) {\scriptsize $#7$};
\node [rotate=0.] at (3.7,12.2) {\scriptsize $#8$};
\node [rotate=0.] at (3.7,11.6) {\scriptsize $#9$};
\node [circle, inner sep=1pt, fill=white,draw, inner sep=1pt,rotate=0.] at (3.2,6.) {\tiny $g$};
\node [circle, inner sep=1pt, fill=white,draw, inner sep=1pt,rotate=0.] at (1.9,7.2) {\tiny $h$};
\node [circle, inner sep=1pt, fill=white,draw, inner sep=1pt,rotate=0.] at (2.1,10.8) {\tiny $x$};
\node [circle, inner sep=1pt, fill=white,draw, inner sep=1pt,rotate=0.] at (3.2,12.4) {\tiny $y$};
\node [circle, inner sep=1pt, fill=white,draw, inner sep=1pt,rotate=0.] at (3.2,8.3) {\tiny $l_1$};
\node [circle, inner sep=1pt, fill=white,draw, inner sep=1pt,rotate=0.] at (3.2,8.9) {\tiny $l_2$};
\node [circle, inner sep=1pt, fill=white,draw, inner sep=1pt,rotate=0.] at (3.2,9.9) {\tiny $l_3$};
\node [circle, inner sep=1pt, fill=white,draw, inner sep=1pt,rotate=0.] at (3.2,10.4) {\tiny $l_4$};
\node [rotate=0.] at (4.7,11.6) {\tiny $\tilde{m}_s$};
\node [rotate=0.] at (4.7,8.) {\tiny $\tilde{m}_r$};
\node [rotate=0.] at (3.6,12.7) {\tiny $n_{#8}$};
\node [rotate=0.] at (1.7,11.2) {\tiny $n_{#6}$};
\node [rotate=0.] at (1.7,7.6) {\tiny $n_{#2}$};
\node [rotate=0.] at (3.6,5.7) {\tiny $n_{#1}$};
\end{tikzpicture}
}

\newcommand{\bpTensorBasis}[9]{
\begin{tikzpicture}[baseline={([yshift=-.8ex]current bounding box.center)},scale=0.8]
\path [lightgray, fill] (3.4,12.8) -- (1.6,12.8) -- (1.6,8.) -- (3.4,8.) -- cycle;
\draw [->-] (3.4,8.) -- (3.4,9.3);
\draw [->-] (3.4,9.3) -- (3.4,9.9);
\draw [->-] (3.4,9.9) -- (3.4,11.1);
\draw [->-] (3.4,11.1) -- (3.4,11.7);
\draw [->-] (3.4,11.7) -- (3.4,12.8);
\draw [->-] (3.4,11.7) -- (5.,11.7);
\draw [->-] (1.8,11.1) -- (3.4,11.1);
\draw [->-] (3.4,9.9) -- (5.,9.9);
\draw [->-] (1.8,9.3) -- (3.4,9.3);
\draw [fill,rotate around={0.:(3.4,11.7)}] (3.4,11.7) ++(0.:0.05 and 0.05) arc(0.:360.:0.05 and 0.05);
\draw [fill,rotate around={0.:(3.4,11.1)}] (3.4,11.1) ++(0.:0.05 and 0.05) arc(0.:360.:0.05 and 0.05);
\draw [fill,rotate around={0.:(3.4,9.9)}] (3.4,9.9) ++(0.:0.05 and 0.05) arc(0.:360.:0.05 and 0.05);
\draw [fill,rotate around={0.:(3.4,9.3)}] (3.4,9.3) ++(0.:0.05 and 0.05) arc(0.:360.:0.05 and 0.05);
\node [rotate=0.] at (3.6,8.6) {\scriptsize $#1$};
\node [rotate=0.] at (2.5,9.) {\scriptsize $#2$};
\node [rotate=0.] at (3.2,9.6) {\scriptsize $#3$};
\node [rotate=0.] at (4.1,10.1) {\scriptsize $#4$};
\node [rotate=0.] at (3.2,10.4) {\scriptsize $#5$};
\node [rotate=0.] at (2.6,11.4) {\scriptsize $#6$};
\node [rotate=0.] at (3.2,11.4) {\scriptsize $#7$};
\node [rotate=0.] at (3.6,12.1) {\scriptsize $#8$};
\node [rotate=0.] at (4.1,11.9) {\scriptsize $#9$};
\node [rotate=0.] at (3.8,9.3) {\scriptsize $v_1$};
\node [rotate=0.] at (3.8,11.1) {\scriptsize $v_2$};
\node [rotate=0.] at (4.5,10.8) {\large $p$};
\node [rotate=0.] at (4.9,11.9) {\tiny $\tilde{m}_s$};
\node [rotate=0.] at (4.9,10.1) {\tiny $\tilde{m}_r$};
\node [rotate=0.] at (3.7,8.1) {\tiny $n_{#1}$};
\node [rotate=0.] at (2.,9.5) {\tiny $n_{#2}$};
\node [rotate=0.] at (2.,11.3) {\tiny $n_{#6}$};
\node [rotate=0.] at (3.7,12.7) {\tiny $n_{#8}$};
\end{tikzpicture}
}

\newcommand{\bulkBpOperator}[6]{
\begin{tikzpicture}[baseline={([yshift=-.8ex]current bounding box.center)},scale=1]
\draw [->-] (3.1,13.2) -- (2.2,13.2);
\draw [->-] (2.2,13.2) -- (2.2,12.6);
\draw [->-] (2.2,12.6) -- (2.2,12.);
\draw [->-] (2.2,12.) -- (3.1,12.);
\draw [->-] (2.2,13.2) -- (2.2,13.6);
\draw [->-] (1.6,12.6) -- (2.2,12.6);
\draw [->-] (2.2,12.) -- (2.2,11.6);
\draw [->-] (3.1,11.6) -- (3.1,12.);
\draw [->-] (3.1,12.) -- (3.1,12.6);
\draw [->-] (3.1,12.6) -- (3.6,12.6);
\draw [->-, fill] (3.1,12.6) -- (3.1,13.2);
\draw [->-, fill] (3.1,13.6) -- (3.1,13.2);
\node [rotate=0.] at (2.7,12.6) {\scriptsize $p$};
\node [rotate=0.] at (2.7,13.4) {\scriptsize $#1$};
\node [rotate=0.] at (2.,12.9) {\scriptsize $#2$};
\node [rotate=0.] at (2.,12.2) {\scriptsize $#3$};
\node [rotate=0.] at (2.7,11.8) {\scriptsize $#4$};
\node [rotate=0.] at (3.3,12.2) {\scriptsize $#5$};
\node [rotate=0.] at (3.3,12.9) {\scriptsize $#6$};
\end{tikzpicture}
}

\newcommand{\bulkHamiltonian}[4]{
\begin{tikzpicture}[baseline={([yshift=-.8ex]current bounding box.center)},scale=0.6]
\draw [->-] (3.4,10.2) -- (2.6,11.4);
\draw [->-] (1.8,10.2) -- (2.6,11.4);
\draw [->-] (2.6,11.4) -- (2.6,12.6);
\node [rotate=0.] at (3.6,10.6) {\scriptsize $#1$};
\node [rotate=0.] at (1.6,10.6) {\scriptsize $#2$};
\node [rotate=0.] at (3.2,12.2) {\scriptsize $#3$};
\node [rotate=0.] at (3.,11.4) {\scriptsize $#4$};
\end{tikzpicture}
}

\newcommand{\cGCoefficientA}{
\begin{tikzpicture}[baseline={([yshift=-.8ex]current bounding box.center)},scale=1]
\draw [->-] (2.2,12.) -- (2.2,12.8);
\draw [->-] (1.6,11.3) -- (2.2,12.);
\draw [->-] (2.8,11.3) -- (2.2,12.);
\draw [fill,rotate around={0.:(2.2,12.)}] (2.2,12.) ++(0.:0.04 and 0.04) arc(0.:360.:0.04 and 0.04);
\node [rotate=0.] at (2.1,11.7) {\tiny $\mu$};
\node [rotate=0.] at (2.7,11.7) {\tiny $\nu$};
\node [rotate=0.] at (2.4,12.3) {\tiny $\lambda$};
\node [rotate=0.] at (2.,11.4) {\tiny $m_\mu$};
\node [rotate=0.] at (3.,11.4) {\tiny $m_\nu$};
\node [rotate=0.] at (2.4,12.7) {\tiny $m_\lambda$};
\end{tikzpicture}
}

\newcommand{\cGCoefficientB}{
\begin{tikzpicture}[baseline={([yshift=-.8ex]current bounding box.center)},scale=1]
\draw [->-] (1.6,11.6) -- (2.4,12.7);
\draw [->-] (2.4,11.6) -- (2.4,12.7);
\draw [->-] (3.2,11.6) -- (2.4,12.7);
\draw [->-] (3.2,11.6) -- (3.2,12.8);
\draw [fill,rotate around={0.:(2.4,12.7)}] (2.4,12.7) ++(0.:0.04 and 0.04) arc(0.:360.:0.04 and 0.04);
\draw [fill,rotate around={0.:(3.2,11.6)}] (3.2,11.6) ++(0.:0.04 and 0.04) arc(0.:360.:0.04 and 0.04);
\node [rotate=0.] at (2.2,12.2) {\tiny $\mu$};
\node [rotate=0.] at (2.6,12.2) {\tiny $\nu$};
\node [rotate=0.] at (3.4,12.) {\tiny $\lambda$};
\node [rotate=0.] at (1.9,11.7) {\tiny $m_\mu$};
\node [rotate=0.] at (2.6,11.7) {\tiny $m_\nu$};
\node [rotate=0.] at (3.4,12.7) {\tiny $m_\lambda$};
\node [rotate=0.] at (2.8,12.4) {\tiny $\lambda^*$};
\end{tikzpicture}
}

\newcommand{\cGCoefficientC}{
\begin{tikzpicture}[baseline={([yshift=-.8ex]current bounding box.center)},scale=1]
\draw [->-] (2.2,11.2) -- (2.2,12.);
\draw [->-] (2.2,12.) -- (1.6,12.8);
\draw [->-] (2.2,12.) -- (2.8,12.8);
\draw [fill,rotate around={0.:(2.2,12.)}] (2.2,12.) ++(0.:0.04 and 0.04) arc(0.:360.:0.04 and 0.04);
\node [rotate=0.] at (2.1,12.4) {\tiny $\mu$};
\node [rotate=0.] at (2.7,12.4) {\tiny $\nu$};
\node [rotate=0.] at (2.4,11.7) {\tiny $\lambda$};
\node [rotate=0.] at (1.9,12.7) {\tiny $m_\mu$};
\node [rotate=0.] at (3.,12.7) {\tiny $m_\nu$};
\node [rotate=0.] at (2.4,11.3) {\tiny $m_\lambda$};
\end{tikzpicture}
}

\newcommand{\cGCoefficientD}{
\begin{tikzpicture}[baseline={([yshift=-.8ex]current bounding box.center)},scale=1]
\draw [->-] (4.8,11.7) -- (4.,12.8);
\draw [->-] (4.8,11.7) -- (4.8,12.8);
\draw [->-] (4.8,11.7) -- (5.6,12.8);
\draw [->-] (5.6,11.6) -- (5.6,12.8);
\draw [fill,rotate around={0.:(4.8,11.7)}] (4.8,11.7) ++(0.:0.04 and 0.04) arc(0.:360.:0.04 and 0.04);
\draw [fill,rotate around={0.:(5.6,12.8)}] (5.6,12.8) ++(0.:0.04 and 0.04) arc(0.:360.:0.04 and 0.04);
\node [rotate=0.] at (5.8,12.4) {\tiny $\lambda$};
\node [rotate=0.] at (4.3,12.7) {\tiny $m_\mu$};
\node [rotate=0.] at (5.,12.7) {\tiny $m_\nu$};
\node [rotate=0.] at (5.8,11.7) {\tiny $m_\lambda$};
\node [rotate=0.] at (4.6,12.4) {\tiny $\mu$};
\node [rotate=0.] at (5.,12.4) {\tiny $\nu$};
\node [rotate=0.] at (5.3,12.1) {\tiny $\lambda^*$};
\end{tikzpicture}
}

\newcommand{\conjugate}{
\begin{tikzpicture}[baseline={([yshift=-.8ex]current bounding box.center)},scale=1]
\draw [->-] (3.,12.4) ..controls (2.953,12.167) and (2.907,11.935) .. (2.9,11.9)
			 ..controls (2.893,11.865) and (2.927,12.028) .. (2.9,11.9)
			 ..controls (2.873,11.772) and (2.785,11.354) .. (2.7,11.2)
			 ..controls (2.615,11.046) and (2.532,11.155) .. (2.5,11.2)
			 ..controls (2.468,11.245) and (2.487,11.226) .. (2.5,11.2)
			 ..controls (2.513,11.174) and (2.521,11.141) .. (2.5,11.2)
			 ..controls (2.479,11.259) and (2.428,11.411) .. (2.4,11.5)
			 ..controls (2.372,11.589) and (2.368,11.615) .. (2.3,11.8)
			 ..controls (2.232,11.985) and (2.102,12.328) .. (2.,12.4)
			 ..controls (1.898,12.472) and (1.826,12.272) .. (1.8,12.2)
			 ..controls (1.774,12.128) and (1.794,12.185) .. (1.8,12.2)
			 ..controls (1.806,12.215) and (1.797,12.19) .. (1.8,12.2)
			 ..controls (1.803,12.21) and (1.816,12.257) .. (1.8,12.2)
			 ..controls (1.784,12.143) and (1.738,11.984) .. (1.7,11.8)
			 ..controls (1.662,11.616) and (1.631,11.408) .. (1.6,11.2);
\draw [->-] (3.,12.4) -- (2.9,11.9);
\draw [->-] (1.7,11.8) -- (1.6,11.2);
\draw [->-] (2.5,11.2) -- (2.4,11.5);
\node [rotate=0.] at (1.7,10.9) {\scriptsize $n_{\mu}$};
\node [rotate=0.] at (3.1,12.8) {\scriptsize $m_{\mu}$};
\node [rotate=0.] at (2.4,12.3) {\scriptsize $\mu$};
\node [fill=white,draw, circle, inner sep=1pt, inner sep=1pt,rotate=0.] at (2.3,11.8) {\scriptsize $\bar g$};
\end{tikzpicture}
}

\newcommand{\conjugateRep}{
\begin{tikzpicture}[baseline={([yshift=-.8ex]current bounding box.center)},scale=1]
\draw [->-] (1.6,10.) -- (1.6,9.2);
\draw [->-] (1.6,9.2) -- (1.6,8.4);
\node [circle, inner sep=1pt, fill=white,draw, inner sep=1pt,rotate=0.] at (1.6,9.2) {\scriptsize $g$};
\node [rotate=0.] at (1.8,8.9) {\scriptsize $\mu$};
\node [rotate=0.] at (1.9,8.5) {\scriptsize $n_{\mu}$};
\node [rotate=0.] at (2.,9.9) {\scriptsize $m_{\mu}$};
\end{tikzpicture}
}

\newcommand{\contractA}{
\begin{tikzpicture}[baseline={([yshift=-.8ex]current bounding box.center)},scale=0.8]
\draw [->-] (3.,11.4) -- (3.,12.8);
\draw [->-] (3.,11.4) -- (4.4,11.4);
\draw [->-] (1.6,10.8) -- (3.,10.8);
\draw [->-] (3.,9.2) -- (3.,10.8);
\draw [->-] (3.,10.8) -- (3.,11.4);
\draw [fill,rotate around={0.:(3.,10.8)}] (3.,10.8) ++(0.:0.04 and 0.04) arc(0.:360.:0.04 and 0.04);
\draw [fill,rotate around={0.:(3.,11.4)}] (3.,11.4) ++(0.:0.04 and 0.04) arc(0.:360.:0.04 and 0.04);
\node [rotate=0.] at (3.3,10.) {\scriptsize $\mu$};
\node [rotate=0.] at (2.5,11.) {\scriptsize $\nu$};
\node [rotate=0.] at (3.2,11.1) {\scriptsize $\gamma$};
\node [rotate=0.] at (3.3,12.1) {\scriptsize $\lambda$};
\node [rotate=0.] at (3.4,11.6) {\scriptsize $s$};
\node [circle, inner sep=1pt, fill=white,draw, inner sep=1pt,rotate=0.] at (3.,12.4) {\tiny $l$};
\node [circle, inner sep=1pt, fill=white,draw, inner sep=1pt,rotate=0.] at (1.9,10.8) {\tiny $h$};
\node [circle, inner sep=1pt, fill=white,draw, inner sep=1pt,rotate=0.] at (3.,9.5) {\tiny $g$};
\node [rotate=0.] at (4.2,11.6) {\tiny $m_{s}$};
\node [rotate=0.] at (1.7,11.2) {\tiny $n_{\nu}$};
\node [rotate=0.] at (3.4,9.3) {\tiny $n_{\mu}$};
\node [rotate=0.] at (3.4,12.7) {\tiny $n_{\lambda}$};
\end{tikzpicture}
}

\newcommand{\contractAvOperator}{
\begin{tikzpicture}[baseline={([yshift=-.8ex]current bounding box.center)},scale=0.8]
\draw [->-] (3.,12.8) -- (3.,11.3);
\draw [->-] (3.,11.8) -- (3.,11.5);
\draw [->-] (6.4,8.8) -- (6.4,10.6);
\draw [->-, fill] (6.4,11.3) -- (7.8,11.3);
\draw [->-, fill] (6.4,11.3) -- (6.4,12.8);
\draw [->-, fill] (1.6,11.3) -- (3.,11.3);
\draw [->-, fill] (3.,10.6) -- (4.4,10.6);
\draw [->-, fill] (6.4,10.6) -- (6.4,11.3);
\draw [->-] (3.,11.3) -- (3.,10.6);
\draw [->-] (3.,10.6) -- (3.,8.8);
\draw [->-] (5.2,10.6) -- (6.4,10.6);
\draw [fill,rotate around={0.:(3.,10.6)}] (3.,10.6) ++(0.:0.04 and 0.04) arc(0.:360.:0.04 and 0.04);
\draw [fill,rotate around={0.:(3.,11.3)}] (3.,11.3) ++(0.:0.04 and 0.04) arc(0.:360.:0.04 and 0.04);
\draw [fill,rotate around={0.:(6.4,11.3)}] (6.4,11.3) ++(0.:0.04 and 0.04) arc(0.:360.:0.04 and 0.04);
\draw [fill,rotate around={0.:(6.4,10.6)}] (6.4,10.6) ++(0.:0.04 and 0.04) arc(0.:360.:0.04 and 0.04);
\node [rotate=0.] at (6.7,9.8) {\scriptsize $\mu$};
\node [rotate=0.] at (5.9,10.3) {\scriptsize $\nu$};
\node [rotate=0.] at (6.1,10.9) {\scriptsize $\gamma$};
\node [rotate=0.] at (6.8,12.2) {\scriptsize $\lambda$};
\node [rotate=0.] at (6.8,11.5) {\scriptsize $s$};
\node [circle, inner sep=1pt, fill=white,draw, inner sep=1pt,rotate=0.] at (6.4,11.8) {\tiny $l$};
\node [circle, inner sep=1pt, fill=white,draw, inner sep=1pt,rotate=0.] at (5.5,10.6) {\tiny $h$};
\node [circle, inner sep=1pt, fill=white,draw, inner sep=1pt,rotate=0.] at (6.4,9.4) {\tiny $g$};
\node [circle, inner sep=1pt, fill=white,draw, inner sep=1pt,rotate=0.] at (4.,10.6) {\tiny $h$};
\node [circle, inner sep=1pt, fill=white,draw, inner sep=1pt,rotate=0.] at (3.,11.8) {\tiny $l$};
\node [circle, inner sep=1pt, fill=white,draw, inner sep=1pt,rotate=0.] at (3.,9.4) {\tiny $g$};
\node [rotate=0.] at (3.3,12.2) {\scriptsize $\lambda'$};
\node [rotate=0.] at (2.6,11.5) {\scriptsize $s'$};
\node [rotate=0.] at (2.7,10.9) {\scriptsize $\gamma'$};
\node [rotate=0.] at (3.5,10.3) {\scriptsize $\nu'$};
\node [rotate=0.] at (2.7,9.9) {\scriptsize $\mu'$};
\node [fill=white,draw, circle, inner sep=1pt, inner sep=1pt,rotate=0.] at (2.1,11.3) {\tiny $k$};
\node [rotate=0.] at (7.6,11.5) {\tiny $m_{s}$};
\node [rotate=0.] at (1.8,11.6) {\tiny $m_{s'}$};
\node [rotate=0.] at (6.7,12.7) {\tiny $n_{\lambda}$};
\node [rotate=0.] at (3.3,12.7) {\tiny $n_{\lambda'}$};
\node [rotate=0.] at (6.7,8.9) {\tiny $n_{\mu}$};
\node [rotate=0.] at (3.3,8.9) {\tiny $n_{\mu'}$};
\node [rotate=0.] at (5.2,10.9) {\tiny $n_{\nu}$};
\node [rotate=0.] at (4.4,10.9) {\tiny $n_{\nu'}$};
\end{tikzpicture}
}

\newcommand{\dualGraph}{
\begin{tikzpicture}[baseline={([yshift=-.8ex]current bounding box.center)},scale=1]
\draw [] (2.4,9.6) -- (2.4,10.4);
\draw [] (2.4,10.4) -- (2.4,11.2);
\draw [] (2.4,11.2) -- (2.4,12.);
\draw [] (2.4,12.) -- (2.4,12.8);
\draw [] (3.2,9.6) -- (3.2,10.4);
\draw [] (3.2,10.4) -- (3.2,11.2);
\draw [] (4.,9.6) -- (4.,10.4);
\draw [] (4.,10.4) -- (4.,11.2);
\draw [] (4.,11.2) -- (4.,12.);
\draw [] (4.,12.) -- (4.,12.8);
\draw [] (4.8,9.6) -- (4.8,10.4);
\draw [] (4.8,10.4) -- (4.8,11.2);
\draw [] (4.8,11.2) -- (4.8,12.);
\draw [] (4.8,12.) -- (4.8,12.8);
\draw [] (1.6,11.6) -- (2.4,11.6);
\draw [] (1.6,10.8) -- (2.4,10.8);
\draw [] (1.6,12.4) -- (2.4,12.4);
\draw [] (1.6,10.) -- (2.4,10.);
\draw [] (3.2,11.6) -- (4.,11.6);
\draw [] (3.2,12.4) -- (4.,12.4);
\draw [] (2.4,11.2) -- (3.2,11.2);
\draw [] (2.4,10.4) -- (3.2,10.4);
\draw [] (4.,11.2) -- (4.8,11.2);
\draw [] (4.,10.4) -- (4.8,10.4);
\draw [] (4.,12.) -- (4.8,12.);
\draw [red, very thick] (2.4,12.) -- (3.2,12.);
\draw [] (3.2,11.2) -- (3.2,11.6);
\draw [] (3.2,12.4) -- (3.2,12.8);
\draw [red, ultra thick, very thick] (3.2,11.6) -- (3.2,12.);
\draw [red, ultra thick, very thick] (3.2,12.) -- (3.2,12.4);
\draw [blue, very thick] (3.2,10.8) -- (4.,10.8);
\draw [blue, very thick] (4.,10.8) -- (4.,10.);
\draw [blue, very thick] (4.,10.) -- (3.2,10.);
\node [rotate=0.] at (3.4,12.) {\scriptsize $1$};
\node [rotate=0.] at (3.6,10.4) {\scriptsize $2$};
\draw [blue, very thick] (3.2,10.) -- (3.2,10.8);
\end{tikzpicture}
}

\newcommand{\dualGraphAA}{
\begin{tikzpicture}[baseline={([yshift=-.8ex]current bounding box.center)},scale=1]
\draw [dashed] (2.4,9.6) -- (2.4,9.7);
\draw [dashed] (2.4,9.9) -- (2.4,10.1);
\draw [dashed] (2.4,10.3) -- (2.4,10.4);
\draw [dashed] (2.4,10.4) -- (2.4,10.5);
\draw [dashed] (2.4,10.7) -- (2.4,10.9);
\draw [dashed] (2.4,11.1) -- (2.4,11.2);
\draw [dashed] (2.4,11.2) -- (2.4,11.3);
\draw [dashed] (2.4,11.5) -- (2.4,11.7);
\draw [dashed] (2.4,11.9) -- (2.4,12.);
\draw [dashed] (2.4,12.) -- (2.4,12.1);
\draw [dashed] (2.4,12.3) -- (2.4,12.5);
\draw [dashed] (2.4,12.7) -- (2.4,12.8);
\draw [dashed] (3.2,9.6) -- (3.2,9.7);
\draw [dashed] (3.2,9.9) -- (3.2,10.1);
\draw [dashed] (3.2,10.3) -- (3.2,10.4);
\draw [dashed] (3.2,10.4) -- (3.2,10.5);
\draw [dashed] (3.2,10.7) -- (3.2,10.9);
\draw [dashed] (3.2,11.1) -- (3.2,11.2);
\draw [dashed] (3.2,11.2) -- (3.2,11.3);
\draw [dashed] (3.2,11.5) -- (3.2,11.7);
\draw [dashed] (3.2,11.9) -- (3.2,12.);
\draw [dashed] (3.2,12.) -- (3.2,12.1);
\draw [dashed] (3.2,12.3) -- (3.2,12.5);
\draw [dashed] (3.2,12.7) -- (3.2,12.8);
\draw [dashed] (4.,9.6) -- (4.,9.7);
\draw [dashed] (4.,9.9) -- (4.,10.1);
\draw [dashed] (4.,10.3) -- (4.,10.4);
\draw [dashed] (4.,10.4) -- (4.,10.5);
\draw [dashed] (4.,10.7) -- (4.,10.9);
\draw [dashed] (4.,11.1) -- (4.,11.2);
\draw [dashed] (4.,11.2) -- (4.,11.3);
\draw [dashed] (4.,11.5) -- (4.,11.7);
\draw [dashed] (4.,11.9) -- (4.,12.);
\draw [dashed] (4.,12.) -- (4.,12.1);
\draw [dashed] (4.,12.3) -- (4.,12.5);
\draw [dashed] (4.,12.7) -- (4.,12.8);
\draw [dashed] (4.8,9.6) -- (4.8,10.4);
\draw [dashed] (4.8,10.4) -- (4.8,11.2);
\draw [dashed] (4.8,11.2) -- (4.8,12.);
\draw [dashed] (4.8,12.) -- (4.8,12.8);
\draw [dashed] (1.6,11.6) -- (1.9,11.6);
\draw [dashed] (2.1,11.6) -- (2.4,11.6);
\draw [dashed] (1.6,10.8) -- (1.9,10.8);
\draw [dashed] (2.1,10.8) -- (2.4,10.8);
\draw [dashed] (1.6,12.4) -- (1.9,12.4);
\draw [dashed] (2.1,12.4) -- (2.4,12.4);
\draw [dashed] (1.6,10.) -- (1.9,10.);
\draw [dashed] (2.1,10.) -- (2.4,10.);
\draw [dashed] (3.2,11.6) -- (3.5,11.6);
\draw [dashed] (3.7,11.6) -- (4.,11.6);
\draw [dashed] (3.2,10.8) -- (3.5,10.8);
\draw [dashed] (3.7,10.8) -- (4.,10.8);
\draw [dashed] (3.2,12.4) -- (3.5,12.4);
\draw [dashed] (3.7,12.4) -- (4.,12.4);
\draw [dashed] (3.2,10.) -- (3.5,10.);
\draw [dashed] (3.7,10.) -- (4.,10.);
\draw [dashed] (2.4,11.2) -- (2.7,11.2);
\draw [dashed] (2.9,11.2) -- (3.2,11.2);
\draw [dashed] (2.4,10.4) -- (2.7,10.4);
\draw [dashed] (2.9,10.4) -- (3.2,10.4);
\draw [dashed] (2.4,12.) -- (2.7,12.);
\draw [dashed] (2.9,12.) -- (3.2,12.);
\draw [dashed] (4.,11.2) -- (4.3,11.2);
\draw [dashed] (4.5,11.2) -- (4.8,11.2);
\draw [dashed] (4.,10.4) -- (4.3,10.4);
\draw [dashed] (4.5,10.4) -- (4.8,10.4);
\draw [dashed] (4.,12.) -- (4.3,12.);
\draw [dashed] (4.5,12.) -- (4.8,12.);
\draw [] (2.8,11.6) -- (2.8,10.8);
\draw [] (2.8,10.8) -- (2.8,10.);
\draw [] (2.8,11.6) -- (3.6,11.2);
\draw [] (3.6,11.2) -- (2.8,10.8);
\draw [] (3.6,12.) -- (3.6,11.2);
\draw [] (3.6,12.) -- (4.4,11.6);
\draw [] (4.4,11.6) -- (3.6,11.2);
\draw [] (3.6,11.2) -- (4.4,10.8);
\draw [] (4.4,10.) -- (3.6,9.6);
\draw [] (2.,12.) -- (2.,11.2);
\draw [] (2.,11.2) -- (2.,10.4);
\draw [] (2.,10.4) -- (2.,9.6);
\draw [] (2.,12.) -- (2.8,11.6);
\draw [] (2.8,11.6) -- (2.,11.2);
\draw [] (2.,11.2) -- (2.8,10.8);
\draw [] (2.8,10.8) -- (2.,10.4);
\draw [] (2.,10.4) -- (2.8,10.);
\draw [] (2.8,10.) -- (2.,9.6);
\draw [] (2.8,10.) -- (3.6,9.6);
\draw [] (2.,12.) -- (2.8,12.4);
\draw [] (4.4,10.) -- (4.4,10.8);
\draw [] (4.4,10.8) -- (4.4,11.6);
\draw [] (4.4,11.6) -- (4.4,12.4);
\draw [] (4.4,12.4) -- (3.6,12.);
\draw [] (2.8,12.4) -- (3.6,12.8);
\draw [] (3.6,12.8) -- (3.6,12.);
\draw [] (3.6,12.8) -- (4.4,12.4);
\draw [] (2.,12.) -- (2.,12.8);
\draw [] (2.,12.8) -- (2.8,12.4);
\draw [red, very thick] (2.8,12.4) -- (2.8,11.6);
\draw [red, very thick] (2.8,12.4) -- (3.6,12.);
\draw [red, very thick] (3.6,12.) -- (2.8,11.6);
\draw [blue, very thick] (2.8,10.8) -- (3.6,10.4);
\draw [blue, very thick] (3.6,10.4) -- (4.4,10.);
\draw [blue, very thick] (3.6,11.2) -- (3.6,10.4);
\draw [blue, very thick] (3.6,10.4) -- (3.6,9.6);
\draw [blue, very thick] (3.6,10.4) -- (2.8,10.);
\draw [blue, very thick] (4.4,10.8) -- (3.6,10.4);
\node [rotate=0.] at (3.,11.9) {\scriptsize $1$};
\node [rotate=0.] at (3.8,10.7) {\scriptsize $2$};
\end{tikzpicture}
}

\newcommand{\dualGraphBoundaryAA}{
\begin{tikzpicture}[baseline={([yshift=-.8ex]current bounding box.center)},scale=1]
\path [lightgray, fill] (1.6,12.8) -- (3.2,12.8) -- (3.2,8.8) -- (1.6,8.8) -- cycle;
\draw [] (3.2,9.8) -- (3.2,10.4);
\draw [] (3.2,10.4) -- (3.2,10.6);
\draw [] (3.2,10.6) -- (3.2,11.2);
\draw [] (2.4,10.4) -- (3.2,10.4);
\draw [] (2.4,11.2) -- (3.2,11.2);
\draw [] (3.2,11.2) -- (3.2,11.4);
\draw [] (3.2,11.4) -- (3.2,12.);
\draw [] (3.2,12.) -- (3.2,12.2);
\draw [] (3.2,12.2) -- (3.2,12.8);
\draw [] (2.4,12.8) -- (3.2,12.8);
\draw [] (2.4,8.8) -- (2.4,10.4);
\draw [] (2.4,10.4) -- (2.4,11.2);
\draw [] (2.4,11.2) -- (2.4,12.);
\draw [] (2.4,12.) -- (2.4,12.8);
\draw [] (1.6,12.4) -- (2.4,12.4);
\draw [] (1.6,11.6) -- (2.4,11.6);
\draw [] (1.6,10.8) -- (2.4,10.8);
\draw [] (1.6,10.) -- (2.4,10.);
\draw [red, very thick] (2.4,12.) -- (3.2,12.);
\draw [red, very thick] (3.2,12.) -- (3.2,12.8);
\draw [red, very thick] (3.2,11.2) -- (3.2,12.);
\draw [] (3.2,8.8) -- (3.2,9.);
\draw [] (3.2,9.) -- (3.2,9.6);
\draw [] (3.2,9.6) -- (3.2,9.8);
\draw [] (2.4,9.6) -- (3.2,9.6);
\draw [] (1.6,9.2) -- (2.4,9.2);
\draw [blue, very thick] (3.2,9.6) -- (3.2,10.4);
\end{tikzpicture}
}

\newcommand{\dualGraphBoundaryAB}{
\begin{tikzpicture}[baseline={([yshift=-.8ex]current bounding box.center)},scale=1]
\path [lightgray, fill] (1.6,12.8) -- (3.2,12.8) -- (3.2,8.8) -- (1.6,8.8) -- cycle;
\draw [] (3.2,10.6) -- (3.2,11.2);
\draw [] (2.4,10.4) -- (3.2,10.4);
\draw [] (2.4,11.2) -- (3.2,11.2);
\draw [] (3.2,11.2) -- (3.2,11.4);
\draw [] (3.2,11.4) -- (3.2,12.);
\draw [] (3.2,12.) -- (3.2,12.2);
\draw [] (3.2,12.2) -- (3.2,12.8);
\draw [] (2.4,12.) -- (3.2,12.);
\draw [] (2.4,12.8) -- (3.2,12.8);
\draw [] (2.4,8.8) -- (2.4,10.4);
\draw [] (2.4,10.4) -- (2.4,11.2);
\draw [] (2.4,11.2) -- (2.4,12.);
\draw [] (2.4,12.) -- (2.4,12.8);
\draw [] (1.6,12.4) -- (2.4,12.4);
\draw [] (1.6,11.6) -- (2.4,11.6);
\draw [] (1.6,10.8) -- (2.4,10.8);
\draw [] (1.6,10.) -- (2.4,10.);
\draw [] (3.2,11.4) -- (3.8,11.4);
\draw [] (3.2,8.8) -- (3.2,9.);
\draw [] (3.2,9.) -- (3.2,9.6);
\draw [] (3.2,9.6) -- (3.2,9.8);
\draw [] (2.4,9.6) -- (3.2,9.6);
\draw [] (1.6,9.2) -- (2.4,9.2);
\draw [] (3.2,9.) -- (3.8,9.);
\draw [red, very thick] (3.2,12.2) -- (3.8,12.2);
\draw [blue, very thick] (3.8,10.6) -- (3.2,10.6);
\draw [blue, very thick] (3.2,10.6) -- (3.2,10.4);
\draw [blue, very thick] (3.2,10.4) -- (3.2,9.8);
\draw [blue, very thick] (3.2,9.8) -- (3.8,9.8);
\end{tikzpicture}
}

\newcommand{\dualGraphBoundaryAC}{
\begin{tikzpicture}[baseline={([yshift=-.8ex]current bounding box.center)},scale=1]
\path [lightgray, fill] (1.6,12.8) -- (3.2,12.8) -- (3.2,8.8) -- (1.6,8.8) -- cycle;
\draw [dashed] (3.2,10.8) -- (3.2,10.9);
\draw [dashed] (3.2,11.1) -- (3.2,11.2);
\draw [dashed] (2.4,10.4) -- (2.7,10.4);
\draw [dashed] (2.9,10.4) -- (3.2,10.4);
\draw [dashed] (2.4,11.2) -- (2.7,11.2);
\draw [dashed] (2.9,11.2) -- (3.2,11.2);
\draw [dashed] (3.2,11.2) -- (3.2,11.3);
\draw [dashed] (3.2,11.5) -- (3.2,11.6);
\draw [dashed] (3.2,11.6) -- (3.2,11.7);
\draw [dashed] (3.2,11.9) -- (3.2,12.);
\draw [dashed] (3.2,12.) -- (3.2,12.1);
\draw [dashed] (3.2,12.3) -- (3.2,12.4);
\draw [dashed] (3.2,12.4) -- (3.2,12.5);
\draw [dashed] (3.2,12.7) -- (3.2,12.8);
\draw [dashed] (2.4,12.) -- (2.7,12.);
\draw [dashed] (2.9,12.) -- (3.2,12.);
\draw [dashed] (2.4,12.8) -- (3.2,12.8);
\draw [dashed] (2.4,8.8) -- (2.4,8.9);
\draw [dashed] (2.4,9.1) -- (2.4,9.3);
\draw [dashed] (2.4,9.5) -- (2.4,9.7);
\draw [dashed] (2.4,9.9) -- (2.4,10.1);
\draw [dashed] (2.4,10.3) -- (2.4,10.4);
\draw [dashed] (2.4,10.4) -- (2.4,10.5);
\draw [dashed] (2.4,10.7) -- (2.4,10.9);
\draw [dashed] (2.4,11.1) -- (2.4,11.2);
\draw [dashed] (2.4,11.2) -- (2.4,11.3);
\draw [dashed] (2.4,11.5) -- (2.4,11.7);
\draw [dashed] (2.4,11.9) -- (2.4,12.);
\draw [dashed] (2.4,12.) -- (2.4,12.1);
\draw [dashed] (2.4,12.3) -- (2.4,12.5);
\draw [dashed] (2.4,12.7) -- (2.4,12.8);
\draw [dashed] (1.6,12.4) -- (1.9,12.4);
\draw [dashed] (2.1,12.4) -- (2.4,12.4);
\draw [dashed] (1.6,11.6) -- (1.9,11.6);
\draw [dashed] (2.1,11.6) -- (2.4,11.6);
\draw [dashed] (1.6,10.8) -- (1.9,10.8);
\draw [dashed] (2.1,10.8) -- (2.4,10.8);
\draw [dashed] (1.6,10.) -- (1.9,10.);
\draw [dashed] (2.1,10.) -- (2.4,10.);
\draw [dashed] (3.2,11.6) -- (3.5,11.6);
\draw [dashed] (3.7,11.6) -- (3.8,11.6);
\draw [dashed] (3.2,8.8) -- (3.2,8.9);
\draw [dashed] (3.2,9.1) -- (3.2,9.2);
\draw [dashed] (3.2,9.2) -- (3.2,9.3);
\draw [dashed] (3.2,9.5) -- (3.2,9.6);
\draw [dashed] (3.2,9.6) -- (3.2,9.7);
\draw [dashed] (3.2,9.9) -- (3.2,10.);
\draw [dashed] (2.4,9.6) -- (2.7,9.6);
\draw [dashed] (2.9,9.6) -- (3.2,9.6);
\draw [dashed] (1.6,9.2) -- (1.9,9.2);
\draw [dashed] (2.1,9.2) -- (2.4,9.2);
\draw [dashed] (3.2,9.2) -- (3.5,9.2);
\draw [dashed] (3.7,9.2) -- (3.8,9.2);
\draw [dashed] (3.2,12.4) -- (3.5,12.4);
\draw [dashed] (3.7,12.4) -- (3.8,12.4);
\draw [dashed] (3.8,10.8) -- (3.7,10.8);
\draw [dashed] (3.5,10.8) -- (3.2,10.8);
\draw [dashed] (3.2,10.8) -- (3.2,10.7);
\draw [dashed] (3.2,10.5) -- (3.2,10.4);
\draw [dashed] (3.2,10.4) -- (3.2,10.3);
\draw [dashed] (3.2,10.1) -- (3.2,10.);
\draw [dashed] (3.2,10.) -- (3.5,10.);
\draw [dashed] (3.7,10.) -- (3.8,10.);
\draw [] (2.8,12.4) -- (3.6,12.8);
\draw [] (3.6,12.) -- (3.6,11.2);
\draw [] (3.6,9.6) -- (3.6,8.8);
\draw [] (3.6,8.8) -- (2.8,9.2);
\draw [] (2.8,9.2) -- (3.6,9.6);
\draw [] (3.6,9.6) -- (2.8,10.);
\draw [] (2.8,10.8) -- (3.6,11.2);
\draw [] (3.6,11.2) -- (2.8,11.6);
\draw [] (2.8,11.6) -- (3.6,12.);
\draw [] (3.6,12.) -- (2.8,12.4);
\draw [] (2.8,12.4) -- (2.,12.8);
\draw [] (2.,12.8) -- (2.,12.);
\draw [] (2.,12.) -- (2.,11.2);
\draw [] (2.,11.2) -- (2.,10.4);
\draw [] (2.,10.4) -- (2.,9.6);
\draw [] (2.,9.6) -- (2.,8.8);
\draw [] (2.,8.8) -- (2.8,9.2);
\draw [] (2.8,9.2) -- (2.,9.6);
\draw [] (2.,9.6) -- (2.8,10.);
\draw [] (2.8,10.) -- (2.,10.4);
\draw [] (2.,10.4) -- (2.8,10.8);
\draw [] (2.8,10.8) -- (2.,11.2);
\draw [] (2.,11.2) -- (2.8,11.6);
\draw [] (2.8,11.6) -- (2.,12.);
\draw [] (2.,12.) -- (2.8,12.4);
\draw [] (2.8,9.2) -- (2.8,10.);
\draw [] (2.8,10.) -- (2.8,10.8);
\draw [] (2.8,10.8) -- (2.8,11.6);
\draw [] (2.8,11.6) -- (2.8,12.4);
\draw [red, very thick] (3.6,12.) -- (3.6,12.8);
\draw [blue, very thick] (3.6,9.6) -- (3.6,10.4);
\draw [blue, very thick] (3.6,10.4) -- (2.8,10.8);
\draw [blue, very thick] (2.8,10.) -- (3.6,10.4);
\draw [blue, very thick] (3.6,10.4) -- (3.6,11.2);
\end{tikzpicture}
}

\newcommand{\dualMap}{
\begin{tikzpicture}[baseline={([yshift=-.8ex]current bounding box.center)},scale=1]
\draw [->-] (5.6,11.2) -- (5.2,12.);
\draw [->-] (5.6,11.2) -- (6.,12.);
\draw [fill,rotate around={0.:(5.6,11.2)}] (5.6,11.2) ++(0.:0.04 and 0.04) arc(0.:360.:0.04 and 0.04);
\node [rotate=0.] at (5.3,11.6) {\tiny $\mu$};
\node [rotate=0.] at (5.5,11.9) {\tiny $m_\mu$};
\node [rotate=0.] at (6.,11.6) {\tiny $\mu^*$};
\node [rotate=0.] at (6.3,11.9) {\tiny $n_{\mu^*}$};
\end{tikzpicture}
}

\newcommand{\dualMapInverse}{
\begin{tikzpicture}[baseline={([yshift=-.8ex]current bounding box.center)},scale=1]
\draw [->-] (5.2,10.4) -- (5.6,11.2);
\draw [->-] (6.,10.4) -- (5.6,11.2);
\draw [fill,rotate around={0.:(5.6,11.2)}] (5.6,11.2) ++(0.:0.04 and 0.04) arc(0.:360.:0.04 and 0.04);
\node [rotate=0.] at (6.,10.9) {\tiny $\mu$};
\node [rotate=0.] at (6.2,10.5) {\tiny $m_\mu$};
\node [rotate=0.] at (5.2,10.9) {\tiny $\mu^*$};
\node [rotate=0.] at (5.6,10.5) {\tiny $n_{\mu^*}$};
\end{tikzpicture}
}

\newcommand{\ftBasisA}{
\begin{tikzpicture}[baseline={([yshift=-.8ex]current bounding box.center)},scale=1]
\path [lightgray, fill] (3.1,12.8) -- (1.6,12.8) -- (1.6,10.3) -- (3.1,10.3) -- cycle;
\draw [->-] (3.1,10.3) -- (3.1,11.4);
\draw [->-] (1.7,11.5) -- (3.,11.5);
\draw [->-] (3.1,11.6) -- (3.1,12.8);
\node [rotate=0.] at (3.3,10.9) {\scriptsize $\mu$};
\node [rotate=0.] at (2.3,11.7) {\scriptsize $\nu$};
\node [rotate=0.] at (3.3,12.2) {\scriptsize $\lambda$};
\node [rotate=0.] at (3.3,11.3) {\tiny $m_{\mu}$};
\node [rotate=0.] at (3.3,11.7) {\tiny $m_{\lambda}$};
\node [rotate=0.] at (2.8,11.6) {\tiny $m_{\nu}$};
\node [rotate=0.] at (1.8,11.6) {\tiny $n_{\nu}$};
\node [rotate=0.] at (3.3,12.7) {\tiny $n_{\lambda}$};
\node [rotate=0.] at (3.3,10.4) {\tiny $n_{\mu}$};
\end{tikzpicture}
}

\newcommand{\ftBasisB}{
\begin{tikzpicture}[baseline={([yshift=-.8ex]current bounding box.center)},scale=0.8]
\path [lightgray, fill] (3.1,12.8) -- (3.1,12.8) -- (3.1,9.6) -- (3.1,9.6) -- cycle;
\draw [->-, fill] (1.6,11.2) -- (3.,11.2);
\draw [->-, fill] (3.1,9.6) -- (3.1,11.1);
\draw [->-] (3.1,11.3) -- (3.1,12.8);
\node [rotate=0.] at (3.3,10.3) {\scriptsize $\mu$};
\node [rotate=0.] at (2.4,11.4) {\scriptsize $\nu$};
\node [rotate=0.] at (3.3,12.) {\scriptsize $\lambda$};
\node [fill=white,draw, circle, inner sep=1pt, inner sep=1pt,rotate=0.] at (3.1,10.) {\tiny $g$};
\node [fill=white,draw, circle, inner sep=1pt, inner sep=1pt,rotate=0.] at (2.,11.2) {\tiny $h$};
\node [fill=white,draw, circle, inner sep=1pt, inner sep=1pt,rotate=0.] at (3.1,12.4) {\tiny $l$};
\node [rotate=0.] at (3.4,11.) {\tiny $m_{\mu}$};
\node [rotate=0.] at (2.9,11.3) {\tiny $m_{\nu}$};
\node [rotate=0.] at (3.4,11.4) {\tiny $m_{\lambda}$};
\node [rotate=0.] at (3.4,9.7) {\tiny $n_{\mu}$};
\node [rotate=0.] at (3.4,12.7) {\tiny $n_{\lambda}$};
\node [rotate=0.] at (1.7,11.4) {\tiny $n_{\nu}$};
\end{tikzpicture}
}

\newcommand{\ftBasisC}{
\begin{tikzpicture}[baseline={([yshift=-.8ex]current bounding box.center)},scale=0.8]
\draw [->-] (2.5,12.6) -- (2.4,12.6);
\draw [->-] (2.4,8.8) -- (2.5,8.8);
\draw [->-, fill] (2.9,11.2) -- (3.7,11.2);
\draw [->-, fill] (2.9,11.2) -- (2.9,12.1);
\draw [->-, fill] (2.1,10.6) -- (2.9,10.6);
\draw [->-, fill] (2.9,10.6) -- (2.9,11.2);
\draw [->-] (1.9,12.1) -- (1.9,10.8);
\draw [->-] (2.9,9.3) -- (2.9,10.6);
\draw [->-] (1.9,10.4) -- (1.9,9.3);
\draw [rotate around={0.:(2.4,12.1)}] (2.4,12.1) ++(0.:0.5 and 0.5) arc(0.:180.:0.5 and 0.5);
\draw [rotate around={0.:(2.4,9.3)}] (2.4,9.3) ++(180.:0.5 and 0.5) arc(180.:360.:0.5 and 0.5);
\draw [fill,rotate around={0.:(2.9,11.2)}] (2.9,11.2) ++(0.:0.04 and 0.04) arc(0.:360.:0.04 and 0.04);
\draw [fill,rotate around={0.:(2.9,10.6)}] (2.9,10.6) ++(0.:0.04 and 0.04) arc(0.:360.:0.04 and 0.04);
\node [rotate=0.] at (2.4,10.4) {\scriptsize $\nu$};
\node [rotate=0.] at (2.5,8.6) {\scriptsize $\mu$};
\node [rotate=0.] at (2.5,12.8) {\scriptsize $\lambda$};
\node [rotate=0.] at (3.1,10.9) {\scriptsize $\gamma$};
\node [rotate=0.] at (3.6,11.4) {\scriptsize $m_s$};
\node [rotate=0.] at (3.1,11.4) {\scriptsize $s$};
\node [rotate=0.] at (1.6,10.9) {\tiny $m_{\lambda}$};
\node [rotate=0.] at (2.2,10.8) {\tiny $m_{\nu}$};
\node [rotate=0.] at (1.6,10.3) {\tiny $m_{\mu}$};
\end{tikzpicture}
}

\newcommand{\ftBasisD}{
\begin{tikzpicture}[baseline={([yshift=-.8ex]current bounding box.center)},scale=0.8]
\draw [->-] (2.7,12.5) -- (2.6,12.5);
\draw [->-] (2.6,8.7) -- (2.7,8.7);
\draw [->-, fill] (1.2,11.2) -- (2.2,11.2);
\draw [->-, fill] (2.2,11.2) -- (2.2,10.7);
\draw [->-, fill] (2.2,10.7) -- (3.,10.7);
\draw [->-] (3.2,9.2) -- (3.2,10.5);
\draw [->-] (2.2,12.) -- (2.2,11.2);
\draw [->-] (2.2,10.7) -- (2.2,9.2);
\draw [->-] (3.2,10.9) -- (3.2,12.);
\draw [rotate around={0.:(2.7,12.)}] (2.7,12.) ++(0.:0.5 and 0.5) arc(0.:180.:0.5 and 0.5);
\draw [rotate around={0.:(2.7,9.2)}] (2.7,9.2) ++(180.:0.5 and 0.5) arc(180.:360.:0.5 and 0.5);
\draw [fill,rotate around={0.:(2.2,11.2)}] (2.2,11.2) ++(0.:0.04 and 0.04) arc(0.:360.:0.04 and 0.04);
\draw [fill,rotate around={0.:(2.2,10.7)}] (2.2,10.7) ++(0.:0.04 and 0.04) arc(0.:360.:0.04 and 0.04);
\node [rotate=0.] at (2.7,10.8) {\scriptsize $\nu$};
\node [rotate=0.] at (2.7,8.4) {\scriptsize $\mu$};
\node [rotate=0.] at (2.7,12.8) {\scriptsize $\lambda$};
\node [rotate=0.] at (3.5,10.3) {\tiny $m_{\mu}$};
\node [rotate=0.] at (2.9,10.5) {\tiny $m_{\nu}$};
\node [rotate=0.] at (3.5,11.1) {\tiny $m_{\lambda}$};
\node [rotate=0.] at (1.9,11.4) {\scriptsize $s$};
\node [rotate=0.] at (1.3,11.4) {\scriptsize $m_s$};
\node [rotate=0.] at (2.,10.9) {\scriptsize $\gamma$};
\end{tikzpicture}
}

\newcommand{\ftBasisE}[9]{
\begin{tikzpicture}[baseline={([yshift=-.8ex]current bounding box.center)},scale=1]
\path [lightgray, fill] (3.1,12.8) -- (1.6,12.8) -- (1.6,9.8) -- (3.1,9.8) -- cycle;
\draw [->-] (3.1,9.8) -- (3.1,11.2);
\draw [->-] (2.,11.2) -- (3.1,11.2);
\draw [->-] (3.1,11.2) -- (3.1,11.8);
\draw [->-] (3.1,11.8) -- (4.6,11.8);
\draw [->-] (3.1,11.8) -- (3.1,12.8);
\draw [fill,rotate around={0.:(3.1,11.2)}] (3.1,11.2) ++(0.:0.025 and 0.025) arc(0.:360.:0.025 and 0.025);
\draw [fill,rotate around={0.:(3.1,11.8)}] (3.1,11.8) ++(0.:0.025 and 0.025) arc(0.:360.:0.025 and 0.025);
\draw [fill,rotate around={0.:(3.1,11.2)}] (3.1,11.2) ++(0.:0.03 and 0.03) arc(0.:360.:0.03 and 0.03);
\draw [fill,rotate around={0.:(3.1,11.8)}] (3.1,11.8) ++(0.:0.03 and 0.03) arc(0.:360.:0.03 and 0.03);
\draw [fill,rotate around={0.:(3.1,11.2)}] (3.1,11.2) ++(0.:0.04 and 0.04) arc(0.:360.:0.04 and 0.04);
\draw [fill,rotate around={0.:(3.1,11.8)}] (3.1,11.8) ++(0.:0.04 and 0.04) arc(0.:360.:0.04 and 0.04);
\node [rotate=0.] at (3.3,10.6) {\scriptsize $#1$};
\node [rotate=0.] at (2.3,11.) {\scriptsize $#2$};
\node [rotate=0.] at (3.3,11.6) {\scriptsize $#3$};
\node [rotate=0.] at (3.9,12.) {\scriptsize $#4$};
\node [rotate=0.] at (3.3,12.2) {\scriptsize $#5$};
\node [rotate=0.] at (3.4,11.2) {\scriptsize $v$};
\node [rotate=0.] at (3.3,9.9) {\tiny $#6$};
\node [rotate=0.] at (1.9,11.4) {\tiny $#7$};
\node [rotate=0.] at (3.3,12.7) {\tiny $#8$};
\node [rotate=0.] at (4.5,12.) {\tiny $#9$};
\end{tikzpicture}
}

\newcommand{\ftLocalBasis}[9]{
\begin{tikzpicture}[baseline={([yshift=-.8ex]current bounding box.center)},scale=0.8]
\path [lightgray, fill] (3.1,12.8) -- (1.6,12.8) -- (1.6,9.8) -- (3.1,9.8) -- cycle;
\draw [->-] (3.1,9.8) -- (3.1,11.2);
\draw [->-] (2.,11.2) -- (3.1,11.2);
\draw [->-] (3.1,11.2) -- (3.1,11.8);
\draw [->-] (3.1,11.8) -- (4.6,11.8);
\draw [->-] (3.1,11.8) -- (3.1,12.8);
\draw [fill,rotate around={0.:(3.1,11.2)}] (3.1,11.2) ++(0.:0.025 and 0.025) arc(0.:360.:0.025 and 0.025);
\draw [fill,rotate around={0.:(3.1,11.8)}] (3.1,11.8) ++(0.:0.025 and 0.025) arc(0.:360.:0.025 and 0.025);
\draw [fill,rotate around={0.:(3.1,11.2)}] (3.1,11.2) ++(0.:0.03 and 0.03) arc(0.:360.:0.03 and 0.03);
\draw [fill,rotate around={0.:(3.1,11.8)}] (3.1,11.8) ++(0.:0.03 and 0.03) arc(0.:360.:0.03 and 0.03);
\draw [fill,rotate around={0.:(3.1,11.2)}] (3.1,11.2) ++(0.:0.04 and 0.04) arc(0.:360.:0.04 and 0.04);
\draw [fill,rotate around={0.:(3.1,11.8)}] (3.1,11.8) ++(0.:0.04 and 0.04) arc(0.:360.:0.04 and 0.04);
\node [rotate=0.] at (3.3,10.6) {\scriptsize $#1$};
\node [rotate=0.] at (2.3,11.) {\scriptsize $#2$};
\node [rotate=0.] at (3.3,11.6) {\scriptsize $#3$};
\node [rotate=0.] at (3.9,12.) {\scriptsize $#4$};
\node [rotate=0.] at (3.3,12.2) {\scriptsize $#5$};
\node [rotate=0.] at (3.4,11.2) {\scriptsize $v$};
\node [rotate=0.] at (3.4,9.9) {\tiny $#6$};
\node [rotate=0.] at (1.9,11.4) {\tiny $#7$};
\node [rotate=0.] at (3.4,12.7) {\tiny $#8$};
\node [rotate=0.] at (4.5,12.) {\tiny $#9$};
\end{tikzpicture}
}

\newcommand{\greatOrTheomA}{
\begin{tikzpicture}[baseline={([yshift=-.8ex]current bounding box.center)},scale=1]
\draw [->-] (1.6,10.8) -- (1.6,11.8);
\draw [->-] (1.6,11.8) -- (1.6,12.8);
\draw [->-] (2.5,12.8) -- (2.5,11.8);
\draw [->-] (2.5,11.8) -- (2.5,10.8);
\node [fill=white,draw, circle, inner sep=1pt, inner sep=1pt,rotate=0.] at (1.6,11.8) {\scriptsize $g$};
\node [rotate=0.] at (1.8,11.4) {\scriptsize $\mu$};
\node [rotate=0.] at (1.8,10.9) {\scriptsize $m$};
\node [rotate=0.] at (1.7,12.7) {\scriptsize $n$};
\node [circle, inner sep=1pt, fill=white,draw, inner sep=1pt,rotate=0.] at (2.5,11.8) {\scriptsize $g$};
\node [rotate=0.] at (2.6,12.7) {\scriptsize $b$};
\node [rotate=0.] at (2.6,10.9) {\scriptsize $a$};
\node [rotate=0.] at (2.7,11.4) {\scriptsize $\nu$};
\end{tikzpicture}
}

\newcommand{\greatOrTheomB}{
\begin{tikzpicture}[baseline={([yshift=-.8ex]current bounding box.center)},scale=0.8]
\draw [->-] (3.2,12.8) ..controls (3.117,12.571) and (3.033,12.343) .. (2.9,12.1)
			 ..controls (2.767,11.857) and (2.583,11.6) .. (2.4,11.6)
			 ..controls (2.217,11.6) and (2.033,11.857) .. (1.9,12.1)
			 ..controls (1.767,12.343) and (1.683,12.571) .. (1.6,12.8);
\draw [->-] (1.6,10.) ..controls (1.683,10.271) and (1.767,10.543) .. (1.9,10.8)
			 ..controls (2.033,11.057) and (2.217,11.3) .. (2.4,11.3)
			 ..controls (2.583,11.3) and (2.767,11.057) .. (2.9,10.8)
			 ..controls (3.033,10.543) and (3.117,10.271) .. (3.2,10.);
\node [rotate=0.] at (1.9,10.2) {\scriptsize $m$};
\node [rotate=0.] at (3.5,10.2) {\scriptsize $a$};
\node [rotate=0.] at (1.9,12.7) {\scriptsize $n$};
\node [rotate=0.] at (3.5,12.7) {\scriptsize $b$};
\node [rotate=0.] at (1.7,11.) {\scriptsize $\mu$};
\node [rotate=0.] at (3.1,11.) {\scriptsize $\nu$};
\node [rotate=0.] at (3.1,11.9) {\scriptsize $\nu$};
\node [rotate=0.] at (1.7,11.9) {\scriptsize $\mu$};
\end{tikzpicture}
}

\newcommand{\greatOrTheomC}{
\begin{tikzpicture}[baseline={([yshift=-.8ex]current bounding box.center)},scale=1]
\draw [->-] (1.6,12.) -- (1.6,12.4);
\draw [->-] (1.6,12.4) -- (1.6,12.8);
\draw [->-] (2.4,12.8) -- (2.4,12.4);
\draw [->-] (2.4,12.4) -- (2.4,12.);
\draw [rotate around={0.:(2.,12.8)}] (2.,12.8) ++(0.:0.4 and 0.4) arc(0.:180.:0.4 and 0.4);
\draw [rotate around={0.:(2.,12.)}] (2.,12.) ++(180.:0.4 and 0.4) arc(180.:360.:0.4 and 0.4);
\node [fill=white,draw, circle, inner sep=1pt, inner sep=1pt,rotate=0.] at (1.6,12.4) {\scriptsize $g$};
\node [rotate=0.] at (2.6,12.) {\scriptsize $\mu$};
\node [fill=white,draw, circle, inner sep=1pt, inner sep=1pt,rotate=0.] at (2.4,12.4) {\scriptsize $h$};
\end{tikzpicture}
}

\newcommand{\greatOrTheomD}{
\begin{tikzpicture}[baseline={([yshift=-.8ex]current bounding box.center)},scale=1]
\draw [->-] (1.6,12.) -- (1.6,12.4);
\draw [->-] (1.6,12.4) -- (1.6,13.1);
\draw [->-] (2.4,13.1) -- (2.4,12.);
\draw [rotate around={0.:(2.,13.1)}] (2.,13.1) ++(0.:0.4 and 0.4) arc(0.:180.:0.4 and 0.4);
\draw [rotate around={0.:(2.,12.)}] (2.,12.) ++(180.:0.4 and 0.4) arc(180.:360.:0.4 and 0.4);
\node [fill=white,draw, circle, inner sep=1pt, inner sep=1pt,rotate=0.] at (1.6,12.4) {\scriptsize $g$};
\node [rotate=0.] at (2.6,12.) {\scriptsize $\mu$};
\node [fill=white,draw, circle, inner sep=1pt, inner sep=1pt,rotate=0.] at (1.6,12.9) {\scriptsize $\bar h$};
\end{tikzpicture}
}

\newcommand{\hilbertBasisA}{
\begin{tikzpicture}[baseline={([yshift=-.8ex]current bounding box.center)},scale=0.7]
\path [lightgray, fill] (-4.9,13.) -- (-7.9,13.) -- (-7.9,7.6) -- (-4.9,7.6) -- cycle;
\path [lightgray, fill] (0.5,12.9) -- (-2.8,12.9) -- (-2.8,7.6) -- (0.5,7.6) -- cycle;
\path [lightgray, fill] (5.9,12.9) -- (2.6,12.9) -- (2.6,7.6) -- (5.9,7.6) -- cycle;
\path [lightgray, fill] (11.3,12.9) -- (8.,12.9) -- (8.,7.6) -- (11.3,7.6) -- cycle;
\draw [->-] (-4.9,7.6) -- (-4.9,9.);
\draw [->-] (-6.2,7.6) -- (-6.2,9.);
\draw [->-] (-4.9,9.) -- (-4.9,11.7);
\draw [->-] (-6.2,9.) -- (-6.2,10.4);
\draw [->-] (-6.2,10.4) -- (-6.2,11.7);
\draw [->-] (-7.8,10.4) -- (-6.2,10.4);
\draw [->-] (-6.2,9.) -- (-4.9,9.);
\draw [->-] (-6.2,11.7) -- (-4.9,11.7);
\draw [->-] (-4.9,11.7) -- (-4.9,13.);
\draw [->-] (-6.2,11.7) -- (-6.2,13.);
\draw [->-] (0.5,7.6) -- (0.5,8.8);
\draw [->-] (-0.8,7.6) -- (-0.8,8.8);
\draw [->-] (0.5,9.) -- (0.5,11.5);
\draw [->-] (-0.8,9.) -- (-0.8,10.2);
\draw [->-] (-0.8,10.4) -- (-0.8,11.5);
\draw [->-] (-2.7,10.3) -- (-1.1,10.3);
\draw [->-] (-0.7,8.9) -- (0.4,8.9);
\draw [->-] (-0.7,11.6) -- (0.4,11.6);
\draw [->-] (0.5,11.7) -- (0.5,12.9);
\draw [->-] (-0.8,11.7) -- (-0.8,12.9);
\draw [->-] (5.9,7.6) -- (5.9,8.8);
\draw [->-] (4.6,7.6) -- (4.6,8.8);
\draw [->-] (5.9,9.) -- (5.9,11.6);
\draw [->-] (4.6,9.) -- (4.6,10.2);
\draw [->-] (4.6,10.4) -- (4.6,11.5);
\draw [->-] (2.7,10.3) -- (4.3,10.3);
\draw [->-] (4.7,8.9) -- (5.8,8.9);
\draw [->-] (4.7,11.6) -- (5.9,11.6);
\draw [->-] (5.9,12.) -- (5.9,12.9);
\draw [->-] (4.6,11.7) -- (4.6,12.9);
\draw [->-] (5.9,11.6) -- (5.9,12.);
\draw [->-] (5.9,12.) -- (6.9,12.);
\draw [->-] (11.3,7.6) -- (11.3,8.9);
\draw [->-] (10.,7.6) -- (10.,8.9);
\draw [->-] (11.3,9.3) -- (11.3,11.6);
\draw [->-] (8.1,10.3) -- (10.,10.3);
\draw [->-] (10.,8.9) -- (11.3,8.9);
\draw [->-] (10.,11.6) -- (11.3,11.6);
\draw [->-] (11.3,12.) -- (11.3,12.9);
\draw [->-] (11.3,11.6) -- (11.3,12.);
\draw [->-] (11.3,12.) -- (12.3,12.);
\draw [->-] (11.3,8.9) -- (11.3,9.3);
\draw [->-] (11.3,9.3) -- (12.3,9.3);
\draw [->-] (-6.2,13.6) -- (-1.,13.6);
\draw [->-] (-0.7,13.6) -- (4.5,13.6);
\draw [->-] (4.8,13.6) -- (10.,13.6);
\draw [] (-6.2,13.3) -- (-6.2,13.6);
\draw [] (-1.,13.6) -- (-1.,13.3);
\draw [] (-0.7,13.3) -- (-0.7,13.6);
\draw [] (4.5,13.6) -- (4.5,13.3);
\draw [] (4.8,13.6) -- (4.8,13.3);
\draw [] (10.,13.6) -- (10.,13.3);
\draw [->-] (10.,12.) -- (10.5,12.);
\draw [->-] (10.,10.7) -- (9.4,10.7);
\draw [->-] (10.,9.3) -- (10.5,9.3);
\draw [->-] (10.,11.6) -- (10.,12.9);
\draw [->-] (10.,10.3) -- (10.,10.7);
\draw [->-] (10.,10.7) -- (10.,11.6);
\draw [->-] (10.,8.9) -- (10.,9.3);
\draw [->-] (10.,9.3) -- (10.,10.3);
\draw [->-] (10.,11.6) -- (10.,12.);
\draw [fill,rotate around={0.:(5.9,11.6)}] (5.9,11.6) ++(0.:0.04 and 0.04) arc(0.:360.:0.04 and 0.04);
\draw [fill,rotate around={0.:(5.9,12.)}] (5.9,12.) ++(0.:0.04 and 0.04) arc(0.:360.:0.04 and 0.04);
\draw [fill,rotate around={0.:(11.3,8.9)}] (11.3,8.9) ++(0.:0.04 and 0.04) arc(0.:360.:0.04 and 0.04);
\draw [fill,rotate around={0.:(11.3,9.3)}] (11.3,9.3) ++(0.:0.04 and 0.04) arc(0.:360.:0.04 and 0.04);
\draw [fill,rotate around={0.:(11.3,11.6)}] (11.3,11.6) ++(0.:0.04 and 0.04) arc(0.:360.:0.04 and 0.04);
\draw [fill,rotate around={0.:(11.3,12.)}] (11.3,12.) ++(0.:0.04 and 0.04) arc(0.:360.:0.04 and 0.04);
\draw [fill,rotate around={0.:(10.,8.9)}] (10.,8.9) ++(0.:0.04 and 0.04) arc(0.:360.:0.04 and 0.04);
\draw [fill,rotate around={0.:(10.,9.3)}] (10.,9.3) ++(0.:0.04 and 0.04) arc(0.:360.:0.04 and 0.04);
\draw [fill,rotate around={0.:(10.,10.3)}] (10.,10.3) ++(0.:0.04 and 0.04) arc(0.:360.:0.04 and 0.04);
\draw [fill,rotate around={0.:(10.,10.7)}] (10.,10.7) ++(0.:0.04 and 0.04) arc(0.:360.:0.04 and 0.04);
\draw [fill,rotate around={0.:(10.,11.6)}] (10.,11.6) ++(0.:0.04 and 0.04) arc(0.:360.:0.04 and 0.04);
\draw [fill,rotate around={0.:(10.,12.)}] (10.,12.) ++(0.:0.04 and 0.04) arc(0.:360.:0.04 and 0.04);
\node [rotate=0.] at (-4.5,9.) {\scriptsize $v_1$};
\node [rotate=0.] at (-4.5,11.7) {\scriptsize $v_2$};
\node [rotate=0.] at (0.8,8.7) {\tiny $n_1$};
\node [rotate=0.] at (0.8,9.1) {\tiny $m_2$};
\node [rotate=0.] at (0.8,11.4) {\tiny $n_2$};
\node [rotate=0.] at (0.8,11.8) {\tiny $m_3$};
\node [rotate=0.] at (-0.5,9.1) {\tiny $m_4$};
\node [rotate=0.] at (0.3,9.1) {\tiny $n_4$};
\node [rotate=0.] at (-0.5,11.8) {\tiny $m_5$};
\node [rotate=0.] at (0.3,11.8) {\tiny $n_5$};
\node [rotate=0.] at (-1.1,8.7) {\tiny $n_6$};
\node [rotate=0.] at (-1.1,11.8) {\tiny $m_{9}$};
\node [rotate=0.] at (-0.5,10.5) {\tiny $m_8$};
\node [rotate=0.] at (-1.1,11.4) {\tiny $n_8$};
\node [rotate=0.] at (0.8,12.8) {\tiny $n_3$};
\node [rotate=0.] at (-1.1,7.7) {\tiny $m_6$};
\node [rotate=0.] at (-1.1,12.8) {\tiny $n_9$};
\node [rotate=0.] at (-1.1,9.1) {\tiny $m_7$};
\node [rotate=0.] at (-1.1,10.1) {\tiny $n_7$};
\node [rotate=0.] at (0.8,7.7) {\tiny $m_1$};
\node [rotate=0.] at (1.3,8.9) {\scriptsize $v_1$};
\node [rotate=0.] at (1.3,11.6) {\scriptsize $v_2$};
\node [rotate=0.] at (-2.5,10.1) {\tiny $m_0$};
\node [rotate=0.] at (-1.3,10.5) {\tiny $n_0$};
\node [rotate=0.] at (6.2,8.7) {\tiny $n_1$};
\node [rotate=0.] at (6.2,9.1) {\tiny $m_2$};
\node [rotate=0.] at (4.9,9.1) {\tiny $m_4$};
\node [rotate=0.] at (5.7,9.1) {\tiny $n_4$};
\node [rotate=0.] at (4.9,11.8) {\tiny $m_5$};
\node [rotate=0.] at (4.3,8.7) {\tiny $n_6$};
\node [rotate=0.] at (4.3,11.8) {\tiny $m_{9}$};
\node [rotate=0.] at (4.9,10.5) {\tiny $m_8$};
\node [rotate=0.] at (4.3,11.4) {\tiny $n_8$};
\node [rotate=0.] at (6.2,12.8) {\tiny $n_3$};
\node [rotate=0.] at (4.3,7.7) {\tiny $m_6$};
\node [rotate=0.] at (4.3,12.8) {\tiny $n_9$};
\node [rotate=0.] at (4.3,9.1) {\tiny $m_7$};
\node [rotate=0.] at (4.3,10.1) {\tiny $n_7$};
\node [rotate=0.] at (6.2,7.7) {\tiny $m_1$};
\node [rotate=0.] at (6.2,11.5) {\scriptsize $v_2$};
\node [rotate=0.] at (6.7,8.9) {\scriptsize $v_1$};
\node [rotate=0.] at (4.,10.5) {\tiny $n_0$};
\node [rotate=0.] at (2.9,10.1) {\tiny $m_0$};
\node [rotate=0.] at (6.9,12.1) {\tiny $m_s$};
\node [rotate=0.] at (6.2,12.2) {\scriptsize $s$};
\node [rotate=0.] at (11.6,8.7) {\scriptsize $v_1$};
\node [rotate=0.] at (11.6,11.5) {\scriptsize $v_2$};
\node [rotate=0.] at (12.3,9.4) {\tiny $m_r$};
\node [rotate=0.] at (12.3,12.1) {\tiny $m_s$};
\node [rotate=0.] at (-6.3,7.1) {\normalsize $(a)$};
\node [rotate=0.] at (-0.9,7.1) {\normalsize $(b)$};
\node [rotate=0.] at (4.4,7.1) {\normalsize $(c)$};
\node [rotate=0.] at (9.8,7.1) {\normalsize $(d)$};
\node [rotate=0.] at (-3.5,13.3) {\small $\text{FT}$};
\node [rotate=0.] at (2.,13.3) {\small $\text{Rewrite $v_2$}$};
\node [rotate=0.] at (10.7,12.) {\tiny $q_1$};
\node [rotate=0.] at (9.1,10.7) {\tiny $q_2$};
\node [rotate=0.] at (10.7,9.3) {\tiny $q_3$};
\node [rotate=0.] at (9.7,11.8) {\tiny $\nu_1$};
\node [rotate=0.] at (10.3,10.5) {\tiny $\nu_2$};
\node [rotate=0.] at (9.7,9.1) {\tiny $\nu_3$};
\node [rotate=0.] at (10.3,12.2) {\tiny $\lambda_1$};
\node [rotate=0.] at (9.6,10.9) {\tiny $\lambda_2$};
\node [rotate=0.] at (10.3,9.5) {\tiny $\lambda_3$};
\node [rotate=0.] at (8.5,10.5) {\tiny $\mu_0$};
\node [rotate=0.] at (10.3,11.2) {\tiny $\mu_8$};
\node [rotate=0.] at (10.7,11.4) {\tiny $\mu_5$};
\node [rotate=0.] at (10.3,12.7) {\tiny $\mu_9$};
\node [rotate=0.] at (11.,12.5) {\tiny $\mu_3$};
\node [rotate=0.] at (11.5,11.8) {\tiny $\eta$};
\node [rotate=0.] at (9.7,10.) {\tiny $\mu_7$};
\node [rotate=0.] at (11.6,10.3) {\tiny $\mu_2$};
\node [rotate=0.] at (11.5,9.1) {\tiny $\gamma$};
\node [rotate=0.] at (10.7,8.7) {\tiny $\mu_4$};
\node [rotate=0.] at (11.,8.3) {\tiny $\mu_1$};
\node [rotate=0.] at (10.3,8.1) {\tiny $\mu_6$};
\node [rotate=0.] at (11.6,9.5) {\tiny $r$};
\node [rotate=0.] at (11.6,12.2) {\tiny $s$};
\node [rotate=0.] at (3.3,10.5) {\tiny $\mu_0$};
\node [rotate=0.] at (4.9,12.3) {\tiny $\mu_9$};
\node [rotate=0.] at (4.9,11.) {\tiny $\mu_8$};
\node [rotate=0.] at (4.9,9.7) {\tiny $\mu_7$};
\node [rotate=0.] at (5.6,12.5) {\tiny $\mu_3$};
\node [rotate=0.] at (5.3,8.7) {\tiny $\mu_4$};
\node [rotate=0.] at (4.9,8.2) {\tiny $\mu_6$};
\node [rotate=0.] at (5.6,8.2) {\tiny $\mu_1$};
\node [rotate=0.] at (6.1,11.8) {\tiny $\eta$};
\node [rotate=0.] at (6.3,10.3) {\tiny $\mu_2$};
\node [rotate=0.] at (5.3,11.4) {\tiny $\mu_5$};
\node [rotate=0.] at (-2.,10.5) {\tiny $\mu_0$};
\node [rotate=0.] at (-0.5,9.7) {\tiny $\mu_7$};
\node [rotate=0.] at (0.9,10.3) {\tiny $\mu_2$};
\node [rotate=0.] at (0.2,8.2) {\tiny $\mu_1$};
\node [rotate=0.] at (-0.5,8.2) {\tiny $\mu_6$};
\node [rotate=0.] at (-0.1,8.7) {\tiny $\mu_4$};
\node [rotate=0.] at (-0.1,11.4) {\tiny $\mu_5$};
\node [rotate=0.] at (0.2,12.3) {\tiny $\mu_3$};
\node [rotate=0.] at (-0.5,10.9) {\tiny $\mu_8$};
\node [rotate=0.] at (-0.5,12.3) {\tiny $\mu_9$};
\node [rotate=0.] at (7.5,13.3) {\small $\text{Rewrite all vertices}$};
\node [rotate=0.] at (-4.6,8.1) {\tiny $g_1$};
\node [rotate=0.] at (-4.6,10.2) {\tiny $g_2$};
\node [rotate=0.] at (-4.6,12.2) {\tiny $g_3$};
\node [rotate=0.] at (-5.6,8.7) {\tiny $g_4$};
\node [rotate=0.] at (-5.6,11.4) {\tiny $g_5$};
\node [rotate=0.] at (-5.9,8.1) {\tiny $g_6$};
\node [rotate=0.] at (-5.9,9.6) {\tiny $g_7$};
\node [rotate=0.] at (-5.9,10.9) {\tiny $g_8$};
\node [rotate=0.] at (-5.9,12.2) {\tiny $g_9$};
\node [rotate=0.] at (-7.1,10.1) {\tiny $g_0$};
\end{tikzpicture}
}

\newcommand{\identityB}{
\begin{tikzpicture}[baseline={([yshift=-.8ex]current bounding box.center)},scale=0.8]
\draw [-<-, fill] (2.,10.) -- (2.,8.2);
\draw [fill, ->-] (2.6,8.2) -- (2.6,10.);
\draw [fill, ->-] (3.2,8.2) -- (3.2,10.);
\draw [dashed] (1.6,10.) -- (3.6,10.);
\draw [dashed,rotate around={0.:(2.6,10.)}] (2.6,10.) ++(0.:1. and 1.) arc(0.:180.:1. and 1.);
\node [rotate=0.] at (2.2,9.1) {\tiny $a$};
\node [rotate=0.] at (2.8,9.1) {\tiny $b$};
\node [rotate=0.] at (2.3,8.3) {\tiny $m_a$};
\node [rotate=0.] at (2.9,8.3) {\tiny $m_b$};
\node [rotate=0.] at (3.5,8.3) {\tiny $m_c$};
\node [rotate=0.] at (3.4,9.1) {\tiny $c$};
\node [circle, inner sep=1pt, fill=white,draw, inner sep=1pt,rotate=0.] at (2.,9.4) {\tiny $g$};
\node [circle, inner sep=1pt, fill=white,draw, inner sep=1pt,rotate=0.] at (2.6,9.4) {\tiny $g$};
\node [circle, inner sep=1pt, fill=white,draw, inner sep=1pt,rotate=0.] at (3.2,9.4) {\tiny $g$};
\node [rotate=0.] at (2.6,10.4) {\scriptsize $\Gamma_{closed}$};
\end{tikzpicture}
}

\newcommand{\identityC}{
\begin{tikzpicture}[baseline={([yshift=-.8ex]current bounding box.center)},scale=0.8]
\draw [-<-, fill] (2.,10.) -- (2.,8.2);
\draw [fill, ->-] (2.6,8.2) -- (2.6,10.);
\draw [fill, ->-] (3.2,8.2) -- (3.2,10.);
\draw [dashed] (1.6,10.) -- (3.6,10.);
\draw [dashed,rotate around={0.:(2.6,10.)}] (2.6,10.) ++(0.:1. and 1.) arc(0.:180.:1. and 1.);
\node [rotate=0.] at (2.2,9.1) {\tiny $a$};
\node [rotate=0.] at (2.8,9.1) {\tiny $b$};
\node [rotate=0.] at (2.3,8.3) {\tiny $m_a$};
\node [rotate=0.] at (2.9,8.3) {\tiny $m_b$};
\node [rotate=0.] at (3.5,8.3) {\tiny $m_c$};
\node [rotate=0.] at (3.4,9.1) {\tiny $c$};
\node [rotate=0.] at (2.6,10.4) {\scriptsize $\Gamma_{closed}$};
\end{tikzpicture}
}

\newcommand{\identityD}{
\begin{tikzpicture}[baseline={([yshift=-.8ex]current bounding box.center)},scale=0.6]
\draw [->-, fill] (1.6,7.9) -- (2.4,8.8);
\draw [->-, fill] (3.2,7.9) -- (2.4,8.8);
\draw [->-, fill] (2.4,8.8) -- (2.4,9.6);
\draw [->-, fill] (2.4,9.6) -- (1.6,10.4);
\draw [->-, fill] (2.4,9.6) -- (3.2,10.4);
\draw [fill,rotate around={0.:(2.4,9.6)}] (2.4,9.6) ++(0.:0.05 and 0.05) arc(0.:360.:0.05 and 0.05);
\draw [fill,rotate around={0.:(2.4,8.8)}] (2.4,8.8) ++(0.:0.05 and 0.05) arc(0.:360.:0.05 and 0.05);
\node [rotate=0.] at (1.8,8.5) {\scriptsize $a$};
\node [rotate=0.] at (3.,8.5) {\scriptsize $b$};
\node [rotate=0.] at (3.,9.9) {\scriptsize $b$};
\node [rotate=0.] at (1.8,9.9) {\scriptsize $a$};
\node [rotate=0.] at (2.6,9.2) {\scriptsize $c$};
\node [rotate=0.] at (2.1,7.9) {\scriptsize $m_a$};
\node [rotate=0.] at (3.6,7.9) {\scriptsize $m_b$};
\node [rotate=0.] at (3.6,10.3) {\scriptsize $m_b'$};
\node [rotate=0.] at (2.1,10.3) {\scriptsize $m_a'$};
\end{tikzpicture}
}

\newcommand{\identityE}{
\begin{tikzpicture}[baseline={([yshift=-.8ex]current bounding box.center)},scale=0.6]
\draw [->-, fill] (2.4,7.9) -- (2.4,10.4);
\draw [->-, fill] (1.6,7.9) -- (1.6,10.4);
\node [rotate=0.] at (1.8,9.1) {\scriptsize $a$};
\node [rotate=0.] at (2.6,9.1) {\scriptsize $b$};
\node [rotate=0.] at (2.,8.) {\scriptsize $m_a$};
\node [rotate=0.] at (2.,10.3) {\scriptsize $m_a'$};
\node [rotate=0.] at (2.8,10.3) {\scriptsize $m_b'$};
\node [rotate=0.] at (2.8,8.) {\scriptsize $m_b$};
\end{tikzpicture}
}

\newcommand{\idnetityA}{
\begin{tikzpicture}[baseline={([yshift=-.8ex]current bounding box.center)},scale=0.8]
\draw [->-] (1.8,8.2) -- (2.6,9.);
\draw [->-] (2.6,8.1) -- (2.6,9.);
\draw [->-] (3.4,8.2) -- (2.6,9.);
\draw [->-, fill] (2.6,9.2) -- (1.8,10.);
\draw [->-, fill] (2.6,9.2) -- (2.6,10.);
\draw [->-, fill] (2.6,9.2) -- (3.4,10.);
\draw [dashed] (1.6,10.) -- (3.6,10.);
\draw [fill,rotate around={0.:(2.6,9.2)}] (2.6,9.2) ++(0.:0.05 and 0.05) arc(0.:360.:0.05 and 0.05);
\draw [fill,rotate around={0.:(2.6,9.)}] (2.6,9.) ++(0.:0.05 and 0.05) arc(0.:360.:0.05 and 0.05);
\draw [dashed,rotate around={0.:(2.6,10.)}] (2.6,10.) ++(0.:1. and 1.) arc(0.:180.:1. and 1.);
\node [rotate=0.] at (2.2,8.8) {\tiny $a$};
\node [rotate=0.] at (2.7,8.6) {\tiny $b$};
\node [rotate=0.] at (3.,8.8) {\tiny $c$};
\node [rotate=0.] at (2.1,8.2) {\tiny $m_a$};
\node [rotate=0.] at (2.9,8.2) {\tiny $m_b$};
\node [rotate=0.] at (3.7,8.2) {\tiny $m_c$};
\node [rotate=0.] at (2.1,9.5) {\tiny $a$};
\node [rotate=0.] at (2.7,9.6) {\tiny $b$};
\node [rotate=0.] at (2.6,10.4) {\scriptsize $\Gamma_{closed}$};
\node [rotate=0.] at (3.2,9.5) {\tiny $c$};
\end{tikzpicture}
}

\newcommand{\intertwiner}{
\begin{tikzpicture}[baseline={([yshift=-.8ex]current bounding box.center)},scale=0.6]
\draw [->-] (4.8,10.4) -- (3.2,12.8);
\draw [->-] (3.2,10.4) -- (3.2,12.8);
\draw [->-] (1.6,10.4) -- (3.2,12.8);
\draw [fill,rotate around={0.:(3.2,12.8)}] (3.2,12.8) ++(0.:0.08 and 0.08) arc(0.:360.:0.08 and 0.08);
\node [rotate=0.] at (3.,11.6) {\scriptsize $b$};
\node [rotate=0.] at (2.,11.6) {\scriptsize $a$};
\node [rotate=0.] at (4.4,11.6) {\scriptsize $c$};
\node [rotate=0.] at (5.1,10.5) {\scriptsize $m_c$};
\node [rotate=0.] at (3.6,10.5) {\scriptsize $m_b$};
\node [rotate=0.] at (2.1,10.5) {\scriptsize $m_a$};
\node [circle, inner sep=1pt, fill=white,draw, inner sep=1pt,rotate=0.] at (2.,11.) {\tiny $g$};
\node [circle, inner sep=1pt, fill=white,draw, inner sep=1pt,rotate=0.] at (3.2,11.) {\tiny $g$};
\node [circle, inner sep=1pt, fill=white,draw, inner sep=1pt,rotate=0.] at (4.4,11.) {\tiny $g$};
\end{tikzpicture}
}

\newcommand{\lwAvop}[3]{
\begin{tikzpicture}[baseline={([yshift=-.8ex]current bounding box.center)},scale=1]
\draw [->-] (1.6,12.) -- (2.4,12.8);
\draw [->-] (3.2,12.) -- (2.4,12.8);
\draw [->-] (2.4,13.6) -- (2.4,12.8);
\node [rotate=0.] at (1.7,12.4) {\tiny $#1$};
\node [rotate=0.] at (3.,12.5) {\tiny $#2$};
\node [rotate=0.] at (2.6,13.2) {\tiny $#3$};
\node [rotate=0.] at (2.4,12.5) {\tiny $v$};
\end{tikzpicture}
}

\newcommand{\lwbdAvop}[3]{
\begin{tikzpicture}[baseline={([yshift=-.8ex]current bounding box.center)},scale=1]
\path [lightgray, fill] (2.4,13.6) -- (1.6,13.6) -- (1.6,12.) -- (2.4,12.) -- cycle;
\draw [->-] (2.4,12.) -- (2.4,12.8);
\draw [->-] (2.4,12.8) -- (2.4,13.6);
\draw [->-] (2.4,12.8) -- (3.,12.8);
\node [rotate=0.] at (2.6,12.2) {\tiny $#1$};
\node [rotate=0.] at (2.6,13.4) {\tiny $#2$};
\node [rotate=0.] at (2.8,13.) {\tiny $#3$};
\end{tikzpicture}
}

\newcommand{\lwbdBpop}[9]{
\begin{tikzpicture}[baseline={([yshift=-.8ex]current bounding box.center)},scale=0.6]
\path [lightgray, fill] (3.6,13.6) -- (1.6,13.6) -- (1.6,9.2) -- (3.6,9.2) -- cycle;
\draw [->-] (3.6,9.2) -- (3.6,10.8);
\draw [->-] (3.6,10.8) -- (3.6,12.4);
\draw [->-] (3.6,12.4) -- (3.6,13.6);
\draw [->-] (2.4,10.) -- (3.6,10.);
\draw [->-] (3.6,10.6) -- (4.6,10.6);
\draw [->-] (2.4,12.) -- (3.6,12.);
\draw [->-] (3.6,12.6) -- (4.6,12.6);
\draw [->-] (3.6,12.) -- (3.6,12.6);
\draw [->-] (3.6,10.) -- (3.6,10.6);
\node [rotate=0.] at (3.8,9.6) {\tiny $#1$};
\node [rotate=0.] at (3.,9.7) {\tiny $#2$};
\node [rotate=0.] at (3.8,10.3) {\tiny $#3$};
\node [rotate=0.] at (4.2,10.8) {\tiny $#4$};
\node [rotate=0.] at (3.9,11.3) {\tiny $#5$};
\node [rotate=0.] at (3.,11.7) {\tiny $#6$};
\node [rotate=0.] at (3.9,12.2) {\tiny $#7$};
\node [rotate=0.] at (4.2,12.8) {\tiny $#8$};
\node [rotate=0.] at (3.8,13.2) {\tiny $#9$};
\node [rotate=0.] at (4.4,11.6) {\scriptsize $p$};
\end{tikzpicture}
}

\newcommand{\lwBpop}[6]{
\begin{tikzpicture}[baseline={([yshift=-.8ex]current bounding box.center)},scale=1]
\draw [->-] (4.7,11.7) -- (3.8,11.7);
\draw [->-] (3.8,11.7) -- (3.8,11.1);
\draw [->-] (3.8,11.1) -- (3.8,10.5);
\draw [->-] (3.8,10.5) -- (4.7,10.5);
\draw [->-] (3.2,11.1) -- (3.8,11.1);
\draw [->-] (4.7,10.1) -- (4.7,10.5);
\draw [->-] (4.7,10.5) -- (4.7,11.1);
\draw [->-, fill] (4.7,11.1) -- (4.7,11.7);
\draw [->-, fill] (4.7,12.1) -- (4.7,11.7);
\draw [->-, fill] (3.8,12.1) -- (3.8,11.7);
\draw [->-, fill] (3.8,10.1) -- (3.8,10.5);
\draw [->-, fill] (5.2,11.1) -- (4.7,11.1);
\node [rotate=0.] at (4.3,11.1) {\scriptsize $p$};
\node [rotate=0.] at (4.3,11.9) {\scriptsize $#1$};
\node [rotate=0.] at (3.6,11.4) {\scriptsize $#2$};
\node [rotate=0.] at (3.6,10.7) {\scriptsize $#3$};
\node [rotate=0.] at (4.3,10.3) {\scriptsize $#4$};
\node [rotate=0.] at (4.9,10.7) {\scriptsize $#5$};
\node [rotate=0.] at (4.9,11.4) {\scriptsize $#6$};
\node [rotate=0.] at (4.9,12.2) {\scriptsize $\nu_1$};
\node [rotate=0.] at (3.6,12.2) {\scriptsize $\nu_2$};
\node [rotate=0.] at (3.,10.9) {\scriptsize $\nu_3$};
\node [rotate=0.] at (3.7,9.9) {\scriptsize $\nu_4$};
\node [rotate=0.] at (4.8,9.9) {\scriptsize $\nu_5$};
\node [rotate=0.] at (5.4,11.3) {\scriptsize $\nu_6$};
\end{tikzpicture}
}

\newcommand{\lwReview}{
\begin{tikzpicture}[baseline={([yshift=-.8ex]current bounding box.center)},scale=0.9]
\path [lightgray, fill] (5.4,13.2) -- (1.6,13.2) -- (1.6,8.2) -- (5.4,8.2) -- cycle;
\draw [->-] (1.8,12.4) -- (3.,12.4);
\draw [->-] (1.8,10.8) -- (3.,10.8);
\draw [->-] (1.8,9.2) -- (3.,9.2);
\draw [->-] (3.,12.4) -- (3.,13.2);
\draw [->-] (3.,10.8) -- (3.,11.6);
\draw [->-] (3.,9.2) -- (3.,10.);
\draw [->-] (3.,10.) -- (4.2,10.);
\draw [->-] (4.2,10.) -- (4.2,10.8);
\draw [->-] (4.2,8.2) -- (4.2,9.2);
\draw [->-] (4.2,9.2) -- (4.2,10.);
\draw [->-] (4.2,9.2) -- (5.4,9.2);
\draw [->-] (3.,11.6) -- (4.2,11.6);
\draw [->-] (4.2,10.8) -- (4.2,11.6);
\draw [->-] (4.2,12.4) -- (5.4,12.4);
\draw [->-] (4.2,12.4) -- (4.2,13.2);
\draw [->-] (3.,8.2) -- (3.,9.2);
\draw [->-] (3.,10.) -- (3.,10.8);
\draw [->-] (3.,11.6) -- (3.,12.4);
\draw [->-] (4.2,11.6) -- (4.2,12.4);
\draw [->-, fill] (5.4,8.2) -- (5.4,9.2);
\draw [->-, fill] (5.4,9.2) -- (5.4,9.6);
\draw [->-, fill] (5.4,9.6) -- (6.2,9.6);
\draw [->-, fill] (5.4,9.6) -- (5.4,10.8);
\draw [->-, fill] (4.2,10.8) -- (5.4,10.8);
\draw [->-, fill] (5.4,10.8) -- (5.4,11.2);
\draw [->-, fill] (5.4,11.2) -- (6.2,11.2);
\draw [->-, fill] (5.4,11.2) -- (5.4,12.4);
\draw [->-, fill] (5.4,12.4) -- (5.4,12.8);
\draw [->-, fill] (5.4,12.8) -- (6.2,12.8);
\draw [->-, fill] (5.4,12.8) -- (5.4,13.2);
\node [rotate=0.] at (5.4,13.6) {\small $\text{Boundary}$};
\node [rotate=0.] at (5.9,9.8) {\tiny $s_1$};
\node [rotate=0.] at (5.9,11.4) {\tiny $s_2$};
\node [rotate=0.] at (5.9,13.) {\tiny $s_3$};
\node [rotate=0.] at (5.6,8.7) {\tiny $\mu_1$};
\node [rotate=0.] at (5.6,9.4) {\tiny $\lambda_1$};
\node [rotate=0.] at (5.6,10.2) {\tiny $\mu_2$};
\node [rotate=0.] at (5.6,12.) {\tiny $\mu_3$};
\node [rotate=0.] at (5.6,11.) {\tiny $\lambda_2$};
\node [rotate=0.] at (5.6,12.6) {\tiny $\lambda_3$};
\node [rotate=0.] at (5.2,13.1) {\tiny $\mu_4$};
\node [rotate=0.] at (4.8,9.) {\tiny $\nu_1$};
\node [rotate=0.] at (4.8,10.6) {\tiny $\nu_2$};
\node [rotate=0.] at (4.8,12.2) {\tiny $\nu_3$};
\node [rotate=0.] at (4.4,12.8) {\tiny $\nu_4$};
\node [rotate=0.] at (4.,12.) {\tiny $\nu_5$};
\node [rotate=0.] at (4.4,11.2) {\tiny $\nu_6$};
\node [rotate=0.] at (4.,10.4) {\tiny $\nu_7$};
\node [rotate=0.] at (4.4,9.6) {\tiny $\nu_8$};
\node [rotate=0.] at (4.,8.6) {\tiny $\nu_9$};
\node [rotate=0.] at (3.6,11.3) {\tiny $\nu_{10}$};
\node [rotate=0.] at (3.6,9.7) {\tiny $\nu_{11}$};
\node [rotate=0.] at (3.3,8.6) {\tiny $\nu_{12}$};
\node [rotate=0.] at (2.7,9.6) {\tiny $\nu_{13}$};
\node [rotate=0.] at (3.3,10.4) {\tiny $\nu_{14}$};
\node [rotate=0.] at (2.7,11.2) {\tiny $\nu_{15}$};
\node [rotate=0.] at (3.3,12.) {\tiny $\nu_{16}$};
\node [rotate=0.] at (2.7,12.9) {\tiny $\nu_{16}$};
\node [rotate=0.] at (2.,12.2) {\tiny $\nu_{17}$};
\node [rotate=0.] at (2.,10.6) {\tiny $\nu_{18}$};
\node [rotate=0.] at (2.,9.) {\tiny $\nu_{19}$};
\node [rotate=0.] at (6.2,9.8) {\tiny $\alpha_1$};
\node [rotate=0.] at (6.2,11.4) {\tiny $\alpha_2$};
\node [rotate=0.] at (6.2,13.) {\tiny $\alpha_3$};
\end{tikzpicture}
}

\newcommand{\lwReviewTail}{
\begin{tikzpicture}[baseline={([yshift=-.8ex]current bounding box.center)},scale=1.1]
\path [lightgray, fill] (5.4,13.2) -- (1.6,13.2) -- (1.6,8.2) -- (5.4,8.2) -- cycle;
\draw [->-] (1.8,12.4) -- (3.,12.4);
\draw [->-] (1.8,10.8) -- (3.,10.8);
\draw [->-] (1.8,9.1) -- (3.,9.1);
\draw [->-] (3.,12.4) -- (3.,13.2);
\draw [->-] (3.,10.8) -- (3.,11.6);
\draw [->-] (3.,9.1) -- (3.,10.);
\draw [->-] (3.,10.) -- (4.2,10.);
\draw [->-] (4.2,10.) -- (4.2,10.8);
\draw [->-] (4.2,8.2) -- (4.2,9.1);
\draw [->-] (4.2,9.1) -- (4.2,10.);
\draw [->-] (4.2,9.1) -- (5.4,9.1);
\draw [->-] (3.,11.6) -- (4.2,11.6);
\draw [->-] (4.2,10.8) -- (4.2,11.6);
\draw [->-] (4.2,12.4) -- (5.4,12.4);
\draw [->-] (4.2,12.4) -- (4.2,13.2);
\draw [->-] (3.,8.2) -- (3.,9.1);
\draw [->-] (3.,10.) -- (3.,10.8);
\draw [->-] (3.,11.6) -- (3.,12.4);
\draw [->-] (4.2,11.6) -- (4.2,12.4);
\draw [->-, fill] (5.4,8.2) -- (5.4,9.1);
\draw [->-, fill] (5.4,9.1) -- (5.4,9.6);
\draw [->-, fill] (5.4,9.6) -- (6.2,9.6);
\draw [->-, fill] (5.4,9.6) -- (5.4,10.8);
\draw [->-, fill] (4.2,10.8) -- (5.4,10.8);
\draw [->-, fill] (5.4,10.8) -- (5.4,11.2);
\draw [->-, fill] (5.4,11.2) -- (6.2,11.2);
\draw [->-, fill] (5.4,11.2) -- (5.4,12.4);
\draw [->-, fill] (5.4,12.4) -- (5.4,12.8);
\draw [->-, fill] (5.4,12.8) -- (6.2,12.8);
\draw [->-, fill] (5.4,12.8) -- (5.4,13.2);
\draw [->-] (4.2,9.5) -- (4.7,9.5);
\draw [->-] (3.,9.5) -- (3.5,9.5);
\draw [->-] (3.,11.2) -- (3.5,11.2);
\draw [->-] (4.2,11.2) -- (4.7,11.2);
\draw [->-] (3.,12.7) -- (3.5,12.7);
\draw [->-] (4.2,12.7) -- (4.7,12.7);
\draw [->-] (3.,11.9) -- (2.5,11.9);
\draw [->-] (4.2,11.9) -- (3.7,11.9);
\draw [->-] (3.,10.4) -- (2.5,10.4);
\draw [->-] (4.2,10.4) -- (3.7,10.4);
\node [rotate=0.] at (5.4,13.6) {\small $\text{Boundary}$};
\node [rotate=0.] at (5.7,9.8) {\tiny $s_1$};
\node [rotate=0.] at (5.7,11.4) {\tiny $s_2$};
\node [rotate=0.] at (5.7,13.) {\tiny $s_3$};
\node [rotate=0.] at (5.6,8.7) {\tiny $\mu_1$};
\node [rotate=0.] at (5.6,9.4) {\tiny $\lambda_1$};
\node [rotate=0.] at (5.6,10.2) {\tiny $\mu_2$};
\node [rotate=0.] at (5.6,12.) {\tiny $\mu_3$};
\node [rotate=0.] at (5.6,11.) {\tiny $\lambda_2$};
\node [rotate=0.] at (5.6,12.6) {\tiny $\lambda_3$};
\node [rotate=0.] at (5.2,13.1) {\tiny $\mu_4$};
\node [rotate=0.] at (4.8,8.9) {\tiny $\nu_1$};
\node [rotate=0.] at (4.8,10.6) {\tiny $\nu_2$};
\node [rotate=0.] at (4.8,12.2) {\tiny $\nu_3$};
\node [rotate=0.] at (4.4,13.1) {\tiny $\nu_4$};
\node [rotate=0.] at (4.4,12.1) {\tiny $\nu_5$};
\node [rotate=0.] at (4.4,11.4) {\tiny $\nu_6$};
\node [rotate=0.] at (4.4,10.6) {\tiny $\nu_7$};
\node [rotate=0.] at (4.4,9.8) {\tiny $\nu_8$};
\node [rotate=0.] at (4.,8.6) {\tiny $\nu_9$};
\node [rotate=0.] at (3.6,11.4) {\tiny $\nu_{10}$};
\node [rotate=0.] at (3.6,9.8) {\tiny $\nu_{11}$};
\node [rotate=0.] at (3.3,8.6) {\tiny $\nu_{12}$};
\node [rotate=0.] at (2.8,9.8) {\tiny $\nu_{13}$};
\node [rotate=0.] at (3.2,10.7) {\tiny $\nu_{14}$};
\node [rotate=0.] at (2.8,11.4) {\tiny $\nu_{15}$};
\node [rotate=0.] at (3.2,12.2) {\tiny $\nu_{16}$};
\node [rotate=0.] at (2.8,13.1) {\tiny $\nu_{16}$};
\node [rotate=0.] at (2.,12.2) {\tiny $\nu_{17}$};
\node [rotate=0.] at (1.9,11.) {\tiny $\nu_{18}$};
\node [rotate=0.] at (2.,8.9) {\tiny $\nu_{19}$};
\node [rotate=0.] at (3.2,9.7) {\tiny $q_1$};
\node [rotate=0.] at (6.5,9.6) {\tiny $\tilde m_{s_1}$};
\node [rotate=0.] at (6.5,11.2) {\tiny $\tilde m_{s_1}$};
\node [rotate=0.] at (6.5,12.8) {\tiny $\tilde m_{s_1}$};
\node [rotate=0.] at (3.8,9.5) {\tiny $\tilde m_{q_1}$};
\node [rotate=0.] at (5.,9.5) {\tiny $\tilde m_{q_2}$};
\node [rotate=0.] at (4.4,9.3) {\tiny $q_2$};
\node [rotate=0.] at (3.2,9.2) {\tiny $\gamma_1$};
\node [rotate=0.] at (4.,9.2) {\tiny $\gamma_2$};
\node [rotate=0.] at (2.8,10.6) {\tiny $q_3$};
\node [rotate=0.] at (2.1,10.4) {\tiny $\tilde m_{q_3}$};
\node [rotate=0.] at (4.,10.6) {\tiny $q_4$};
\node [rotate=0.] at (3.5,10.4) {\tiny $\tilde m_{q_4}$};
\node [rotate=0.] at (2.8,10.2) {\tiny $\gamma_3$};
\node [rotate=0.] at (4.4,10.2) {\tiny $\gamma_4$};
\node [rotate=0.] at (2.8,11.) {\tiny $\gamma_5$};
\node [rotate=0.] at (4.,10.9) {\tiny $\gamma_6$};
\node [rotate=0.] at (3.2,11.) {\tiny $q_5$};
\node [rotate=0.] at (3.8,11.2) {\tiny $\tilde m_{q_5}$};
\node [rotate=0.] at (4.4,11.) {\tiny $q_6$};
\node [rotate=0.] at (5.,11.2) {\tiny $\tilde m_{q_6}$};
\node [rotate=0.] at (3.2,11.7) {\tiny $\gamma_7$};
\node [rotate=0.] at (4.4,11.7) {\tiny $\gamma_8$};
\node [rotate=0.] at (2.8,11.7) {\tiny $q_7$};
\node [rotate=0.] at (2.2,11.9) {\tiny $\tilde m_{q_7}$};
\node [rotate=0.] at (4.,12.1) {\tiny $q_8$};
\node [rotate=0.] at (3.5,11.9) {\tiny $\tilde m_{q_8}$};
\node [rotate=0.] at (3.2,12.5) {\tiny $\gamma_9$};
\node [rotate=0.] at (4.3,12.5) {\tiny $\gamma_0$};
\node [rotate=0.] at (3.2,12.9) {\tiny $q_9$};
\node [rotate=0.] at (3.8,12.7) {\tiny $\tilde m_{q_9}$};
\node [rotate=0.] at (4.4,12.9) {\tiny $q_0$};
\node [rotate=0.] at (5.,12.7) {\tiny $\tilde m_{q_0}$};
\end{tikzpicture}
}

\newcommand{\normalizeThreej}{
\begin{tikzpicture}[baseline={([yshift=-.8ex]current bounding box.center)},scale=0.8]
\draw [->-, fill] (2.5,11.6) -- (2.5,12.8);
\draw [->-, fill] (1.9,12.2) -- (1.9,12.3);
\draw [->-, fill] (3.1,12.2) -- (3.1,12.3);
\draw [rotate around={0.:(2.5,12.2)}] (2.5,12.2) ++(0.:0.6 and 0.6) arc(0.:360.:0.6 and 0.6);
\draw [fill,rotate around={0.:(2.5,12.8)}] (2.5,12.8) ++(0.:0.05 and 0.05) arc(0.:360.:0.05 and 0.05);
\draw [fill,rotate around={0.:(2.5,11.6)}] (2.5,11.6) ++(0.:0.05 and 0.05) arc(0.:360.:0.05 and 0.05);
\node [rotate=0.] at (2.7,12.2) {\scriptsize $b$};
\node [rotate=0.] at (3.4,12.2) {\scriptsize $c$};
\node [rotate=0.] at (1.6,12.2) {\scriptsize $a$};
\end{tikzpicture}
}

\newcommand{\qdLocalBasisA}[4]{
\begin{tikzpicture}[baseline={([yshift=-.8ex]current bounding box.center)},scale=0.6]
\path [lightgray, fill] (3.6,10.4) -- (1.6,10.4) -- (1.6,7.2) -- (3.6,7.2) -- cycle;
\draw [->-] (3.6,7.2) -- (3.6,8.8);
\draw [->-] (3.6,8.8) -- (3.6,10.4);
\draw [->-] (1.8,8.8) -- (3.6,8.8);
\node [rotate=0.] at (4.,7.8) {\scriptsize $#1$};
\node [rotate=0.] at (2.6,8.5) {\scriptsize $#2$};
\node [rotate=0.] at (4.,9.6) {\scriptsize $#3$};
\node [rotate=0.] at (3.9,8.8) {\scriptsize $#4$};
\end{tikzpicture}
}

\newcommand{\qdLocalBasisB}[5]{
\begin{tikzpicture}[baseline={([yshift=-.8ex]current bounding box.center)},scale=0.6]
\path [lightgray, fill] (3.8,13.6) -- (1.6,13.6) -- (1.6,8.8) -- (3.8,8.8) -- cycle;
\draw [->-] (3.8,8.8) -- (3.8,10.4);
\draw [->-] (3.8,10.4) -- (3.8,12.);
\draw [->-] (2.2,12.) -- (3.8,12.);
\draw [->-] (2.2,10.4) -- (3.8,10.4);
\draw [fill, ->-] (3.8,12.) -- (3.8,13.6);
\draw [dashed] (3.8,12.) -- (5.,11.2);
\draw [dashed] (5.,11.2) -- (3.8,10.4);
\node [rotate=0.] at (4.,9.8) {\scriptsize $#1$};
\node [rotate=0.] at (3.,10.1) {\scriptsize $#2$};
\node [rotate=0.] at (3.6,11.3) {\scriptsize $#3$};
\node [rotate=0.] at (3.,11.7) {\scriptsize $#4$};
\node [rotate=0.] at (4.,13.) {\scriptsize $#5$};
\node [rotate=0.] at (4.3,11.2) {\small $p$};
\end{tikzpicture}
}

\newcommand{\qdLocalBasisPlus}[9]{
\begin{tikzpicture}[baseline={([yshift=-.8ex]current bounding box.center)},scale=1]
\path [lightgray, fill] (2.9,12.8) -- (1.6,12.8) -- (1.6,10.4) -- (2.9,10.4) -- cycle;
\draw [->-] (2.9,10.4) -- (2.9,11.6);
\draw [->-] (2.9,11.6) -- (2.9,12.8);
\draw [->-] (1.8,11.6) -- (2.9,11.6);
\node [rotate=0.] at (3.1,11.6) {\scriptsize $v$};
\node [rotate=0.] at (3.1,11.) {\scriptsize $#1$};
\node [rotate=0.] at (2.2,11.4) {\scriptsize $#2$};
\node [rotate=0.] at (3.1,12.1) {\scriptsize $#3$};
\node [rotate=0.] at (3.1,10.5) {\scriptsize $#4$};
\node [rotate=0.] at (3.,11.4) {\scriptsize $#5$};
\node [rotate=0.] at (1.8,11.8) {\scriptsize $#6$};
\node [rotate=0.] at (2.7,11.8) {\scriptsize $#7$};
\node [rotate=0.] at (3.1,11.8) {\scriptsize $#8$};
\node [rotate=0.] at (3.1,12.7) {\scriptsize $#9$};
\end{tikzpicture}
}

\newcommand{\qdReview}{
\begin{tikzpicture}[baseline={([yshift=-.8ex]current bounding box.center)},scale=0.9]
\path [lightgray, fill] (7.2,14.4) -- (3.4,14.4) -- (3.4,9.4) -- (7.2,9.4) -- cycle;
\draw [->-] (6.,12.) -- (7.2,12.);
\draw [->-] (3.6,13.6) -- (4.8,13.6);
\draw [->-] (3.6,12.) -- (4.8,12.);
\draw [->-] (3.6,10.4) -- (4.8,10.4);
\draw [->-] (4.8,13.6) -- (4.8,14.4);
\draw [->-] (4.8,12.) -- (4.8,12.8);
\draw [->-] (4.8,10.4) -- (4.8,11.2);
\draw [->-] (4.8,11.2) -- (6.,11.2);
\draw [->-] (6.,11.2) -- (6.,12.);
\draw [->-] (6.,9.4) -- (6.,10.4);
\draw [->-] (6.,10.4) -- (6.,11.2);
\draw [->-] (6.,10.4) -- (7.2,10.4);
\draw [->-] (4.8,12.8) -- (6.,12.8);
\draw [->-] (6.,12.) -- (6.,12.8);
\draw [->-] (6.,13.6) -- (7.2,13.6);
\draw [->-] (6.,13.6) -- (6.,14.4);
\draw [->-] (4.8,9.4) -- (4.8,10.4);
\draw [->-] (4.8,11.2) -- (4.8,12.);
\draw [->-] (4.8,12.8) -- (4.8,13.6);
\draw [->-] (6.,12.8) -- (6.,13.6);
\draw [->-, fill, thick] (7.2,9.4) -- (7.2,10.4);
\draw [->-, fill, thick] (7.2,10.4) -- (7.2,12.);
\draw [->-, fill, thick] (7.2,12.) -- (7.2,13.6);
\draw [->-, fill, thick] (7.2,13.6) -- (7.2,14.4);
\node [rotate=0.] at (7.5,14.1) {\tiny $g_0$};
\node [rotate=0.] at (7.5,12.8) {\tiny $g_1$};
\node [rotate=0.] at (7.5,11.2) {\tiny $g_2$};
\node [rotate=0.] at (7.5,9.9) {\tiny $g_3$};
\node [rotate=0.] at (6.6,13.4) {\tiny $g_4$};
\node [rotate=0.] at (6.6,11.8) {\tiny $g_5$};
\node [rotate=0.] at (6.6,10.2) {\tiny $g_6$};
\node [rotate=0.] at (5.7,9.9) {\tiny $g_7$};
\node [rotate=0.] at (6.3,10.8) {\tiny $g_8$};
\node [rotate=0.] at (5.7,11.6) {\tiny $g_9$};
\node [rotate=0.] at (6.3,12.4) {\tiny $g_{10}$};
\node [rotate=0.] at (5.7,13.2) {\tiny $g_{11}$};
\node [rotate=0.] at (6.3,14.1) {\tiny $g_{12}$};
\node [rotate=0.] at (5.4,10.9) {\tiny $g_{13}$};
\node [rotate=0.] at (5.4,12.6) {\tiny $g_{14}$};
\node [rotate=0.] at (5.1,9.9) {\tiny $g_{15}$};
\node [rotate=0.] at (4.5,10.8) {\tiny $g_{16}$};
\node [rotate=0.] at (5.1,11.6) {\tiny $g_{17}$};
\node [rotate=0.] at (4.5,12.4) {\tiny $g_{18}$};
\node [rotate=0.] at (5.1,13.2) {\tiny $g_{19}$};
\node [rotate=0.] at (5.1,14.1) {\tiny $g_{20}$};
\node [rotate=0.] at (3.9,13.4) {\tiny $g_{21}$};
\node [rotate=0.] at (3.9,11.8) {\tiny $g_{22}$};
\node [rotate=0.] at (3.9,10.2) {\tiny $g_{23}$};
\node [rotate=0.] at (7.2,14.8) {\small $\text{Boundary}$};
\end{tikzpicture}
}

\newcommand{\repBasisA}{
\begin{tikzpicture}[baseline={([yshift=-.8ex]current bounding box.center)},scale=0.8]
\draw [->-] (3.,11.4) -- (3.,12.8);
\draw [->-] (3.,11.4) -- (4.4,11.4);
\draw [->-] (1.6,10.8) -- (3.,10.8);
\draw [->-] (3.,9.2) -- (3.,10.8);
\draw [->-] (3.,10.8) -- (3.,11.4);
\draw [fill,rotate around={0.:(3.,10.8)}] (3.,10.8) ++(0.:0.04 and 0.04) arc(0.:360.:0.04 and 0.04);
\draw [fill,rotate around={0.:(3.,11.4)}] (3.,11.4) ++(0.:0.04 and 0.04) arc(0.:360.:0.04 and 0.04);
\node [rotate=0.] at (3.2,10.2) {\scriptsize $\mu$};
\node [rotate=0.] at (2.6,10.9) {\scriptsize $\nu$};
\node [rotate=0.] at (3.2,11.1) {\scriptsize $\gamma$};
\node [rotate=0.] at (3.2,11.8) {\scriptsize $\lambda$};
\node [rotate=0.] at (3.4,11.5) {\scriptsize $s$};
\node [circle, inner sep=1pt, fill=white,draw, inner sep=1pt,rotate=0.] at (3.,12.5) {\tiny $l$};
\node [circle, inner sep=1pt, fill=white,draw, inner sep=1pt,rotate=0.] at (1.9,10.8) {\tiny $h$};
\node [circle, inner sep=1pt, fill=white,draw, inner sep=1pt,rotate=0.] at (3.,9.6) {\tiny $g$};
\node [rotate=0.] at (4.3,11.5) {\tiny $m_s$};
\node [rotate=0.] at (3.3,9.3) {\tiny $n_\mu$};
\node [rotate=0.] at (1.7,11.1) {\tiny $n_\nu$};
\node [rotate=0.] at (3.3,12.7) {\tiny $n_\lambda$};
\end{tikzpicture}
}

\newcommand{\repBasisAA}{
\begin{tikzpicture}[baseline={([yshift=-.8ex]current bounding box.center)},scale=0.8]
\draw [->-] (3.,11.4) -- (3.,12.8);
\draw [->-] (3.,11.4) -- (4.4,11.4);
\draw [->-] (1.6,10.8) -- (3.,10.8);
\draw [->-] (3.,9.2) -- (3.,10.8);
\draw [->-] (3.,10.8) -- (3.,11.4);
\draw [fill,rotate around={0.:(3.,10.8)}] (3.,10.8) ++(0.:0.04 and 0.04) arc(0.:360.:0.04 and 0.04);
\draw [fill,rotate around={0.:(3.,11.4)}] (3.,11.4) ++(0.:0.04 and 0.04) arc(0.:360.:0.04 and 0.04);
\node [circle, inner sep=1pt, fill=white,draw, inner sep=1pt,rotate=0.] at (3.,12.4) {\tiny $l$};
\node [circle, inner sep=1pt, fill=white,draw, inner sep=1pt,rotate=0.] at (1.9,10.8) {\tiny $h$};
\node [circle, inner sep=1pt, fill=white,draw, inner sep=1pt,rotate=0.] at (3.,9.6) {\tiny $g$};
\node [rotate=0.] at (3.4,11.6) {\scriptsize $s'$};
\node [rotate=0.] at (3.2,11.1) {\scriptsize $\gamma'$};
\node [rotate=0.] at (4.2,11.5) {\tiny $m_{s'}$};
\node [rotate=0.] at (1.7,11.1) {\tiny $n_{\nu'}$};
\node [rotate=0.] at (3.4,12.7) {\tiny $n_{\lambda'}$};
\node [rotate=0.] at (3.4,9.3) {\tiny $n_{\mu'}$};
\node [rotate=0.] at (3.2,12.) {\scriptsize $\lambda'$};
\node [rotate=0.] at (3.2,10.1) {\scriptsize $\mu'$};
\node [rotate=0.] at (2.4,11.) {\scriptsize $\nu'$};
\end{tikzpicture}
}

\newcommand{\repBasisB}{
\begin{tikzpicture}[baseline={([yshift=-.8ex]current bounding box.center)},scale=0.8]
\draw [->-] (3.,11.4) -- (3.,12.8);
\draw [->-] (3.,11.4) -- (4.4,11.4);
\draw [->-] (1.6,10.8) -- (3.,10.8);
\draw [->-] (3.,9.2) -- (3.,10.8);
\draw [->-] (3.,10.8) -- (3.,11.4);
\draw [fill,rotate around={0.:(3.,10.8)}] (3.,10.8) ++(0.:0.04 and 0.04) arc(0.:360.:0.04 and 0.04);
\draw [fill,rotate around={0.:(3.,11.4)}] (3.,11.4) ++(0.:0.04 and 0.04) arc(0.:360.:0.04 and 0.04);
\node [rotate=0.] at (3.2,10.3) {\scriptsize $\mu$};
\node [rotate=0.] at (2.5,11.) {\scriptsize $\nu$};
\node [rotate=0.] at (3.2,11.1) {\scriptsize $\gamma$};
\node [rotate=0.] at (3.2,12.) {\scriptsize $\lambda$};
\node [rotate=0.] at (3.4,11.5) {\scriptsize $s$};
\node [fill=white,draw, circle, inner sep=1pt, inner sep=1pt,rotate=0.] at (3.,9.7) {\tiny $x$};
\node [fill=white,draw, circle, inner sep=1pt, inner sep=1pt,rotate=0.] at (2.,10.8) {\tiny $y$};
\node [fill=white,draw, circle, inner sep=1pt, inner sep=1pt,rotate=0.] at (3.,12.3) {\tiny $z$};
\node [rotate=0.] at (4.2,11.5) {\tiny $m_s$};
\node [rotate=0.] at (3.2,12.7) {\tiny $n_{\lambda}$};
\node [rotate=0.] at (3.2,9.3) {\tiny $n_{\mu}$};
\node [rotate=0.] at (1.7,11.1) {\tiny $n_{\nu}$};
\end{tikzpicture}
}

\newcommand{\repBasisC}{
\begin{tikzpicture}[baseline={([yshift=-.8ex]current bounding box.center)},scale=0.7]
\draw [->-] (7.,12.8) -- (7.,11.6);
\draw [->-, fill] (10.3,11.2) -- (11.7,11.2);
\draw [->-] (10.3,8.6) -- (10.3,10.5);
\draw [->-, fill] (8.2,10.5) -- (8.4,10.5);
\draw [->-, fill] (10.3,12.1) -- (10.3,12.8);
\draw [->-, fill] (7.,10.5) -- (7.,8.6);
\draw [->-, fill] (7.,11.2) -- (7.,10.5);
\draw [->-, fill] (5.6,11.2) -- (7.,11.2);
\draw [->-, fill] (10.3,11.2) -- (10.3,12.1);
\draw [->-, fill] (9.,10.5) -- (10.3,10.5);
\draw [->-, fill] (10.3,10.5) -- (10.3,11.2);
\draw [->-, fill] (7.,10.5) -- (8.4,10.5);
\draw [->-] (7.,11.6) -- (7.,11.2);
\draw [fill,rotate around={0.:(7.,10.5)}] (7.,10.5) ++(0.:0.04 and 0.04) arc(0.:360.:0.04 and 0.04);
\draw [fill,rotate around={0.:(7.,11.2)}] (7.,11.2) ++(0.:0.04 and 0.04) arc(0.:360.:0.04 and 0.04);
\draw [fill,rotate around={0.:(10.3,11.2)}] (10.3,11.2) ++(0.:0.04 and 0.04) arc(0.:360.:0.04 and 0.04);
\draw [fill,rotate around={0.:(10.3,10.5)}] (10.3,10.5) ++(0.:0.04 and 0.04) arc(0.:360.:0.04 and 0.04);
\node [rotate=0.] at (10.6,9.6) {\scriptsize $\mu$};
\node [rotate=0.] at (9.9,10.9) {\scriptsize $\nu$};
\node [rotate=0.] at (10.6,10.9) {\scriptsize $\gamma$};
\node [rotate=0.] at (10.7,12.) {\scriptsize $\lambda$};
\node [rotate=0.] at (10.7,11.4) {\scriptsize $s$};
\node [circle, inner sep=1pt, fill=white,draw, inner sep=1pt,rotate=0.] at (10.3,11.6) {\tiny $l$};
\node [circle, inner sep=1pt, fill=white,draw, inner sep=1pt,rotate=0.] at (9.9,10.5) {\tiny $h$};
\node [circle, inner sep=1pt, fill=white,draw, inner sep=1pt,rotate=0.] at (10.3,9.2) {\tiny $g$};
\node [circle, inner sep=1pt, fill=white,draw, inner sep=1pt,rotate=0.] at (7.7,10.5) {\tiny $h$};
\node [circle, inner sep=1pt, fill=white,draw, inner sep=1pt,rotate=0.] at (7.,11.6) {\tiny $l$};
\node [circle, inner sep=1pt, fill=white,draw, inner sep=1pt,rotate=0.] at (7.,9.2) {\tiny $g$};
\node [rotate=0.] at (7.4,12.) {\scriptsize $\lambda'$};
\node [rotate=0.] at (6.6,11.5) {\scriptsize $s'$};
\node [rotate=0.] at (6.7,10.9) {\scriptsize $\gamma'$};
\node [rotate=0.] at (7.4,10.8) {\scriptsize $\nu'$};
\node [rotate=0.] at (7.4,9.5) {\scriptsize $\mu'$};
\node [rotate=0.] at (11.5,11.4) {\tiny $m_s$};
\node [rotate=0.] at (5.9,11.4) {\tiny $m_{s'}$};
\node [rotate=0.] at (10.6,12.7) {\tiny $n_{\lambda}$};
\node [rotate=0.] at (7.4,12.7) {\tiny $n_{\lambda'}$};
\node [rotate=0.] at (7.4,8.7) {\tiny $n_{\mu'}$};
\node [rotate=0.] at (10.6,8.7) {\tiny $n_{\mu}$};
\node [rotate=0.] at (9.1,10.8) {\tiny $n_{\nu}$};
\node [rotate=0.] at (8.3,10.8) {\tiny $n_{\nu'}$};
\end{tikzpicture}
}

\newcommand{\repBasisD}{
\begin{tikzpicture}[baseline={([yshift=-.8ex]current bounding box.center)},scale=0.8]
\draw [->-] (3.8,9.4) ..controls (3.8,9.56) and (3.8,9.72) .. (3.7,9.8)
			 ..controls (3.6,9.88) and (3.4,9.88) .. (3.3,9.8)
			 ..controls (3.2,9.72) and (3.2,9.56) .. (3.2,9.4);
\draw [->-] (3.4,8.3) -- (3.5,8.3);
\draw [->-] (3.5,8.) -- (3.4,8.);
\draw [->-] (3.5,12.3) -- (3.6,12.3);
\draw [->-] (3.5,12.) -- (3.4,12.);
\draw [->-] (1.8,10.7) -- (3.,10.7);
\draw [->-] (4.,10.7) -- (5.2,10.7);
\draw [->-] (3.,10.) -- (4.,10.);
\draw [->-] (4.,10.7) -- (4.,11.5);
\draw [->-] (3.,11.5) -- (3.,10.7);
\draw [->-] (3.,10.7) -- (3.,10.);
\draw [->-] (3.,10.) -- (3.,8.8);
\draw [->-] (4.,8.8) -- (4.,10.);
\draw [->-] (4.,10.) -- (4.,10.7);
\draw [rotate around={0.:(3.5,11.5)}] (3.5,11.5) ++(0.:0.5 and 0.5) arc(0.:180.:0.5 and 0.5);
\draw [rotate around={0.:(3.5,8.8)}] (3.5,8.8) ++(180.:0.5 and 0.5) arc(180.:360.:0.5 and 0.5);
\draw [rotate around={0.:(3.5,12.8)}] (3.5,12.8) ++(180.:0.5 and 0.5) arc(180.:360.:0.5 and 0.5);
\draw [rotate around={0.:(3.5,7.5)}] (3.5,7.5) ++(0.:0.5 and 0.5) arc(0.:180.:0.5 and 0.5);
\draw [fill,rotate around={0.:(4.,10.7)}] (4.,10.7) ++(0.:0.04 and 0.04) arc(0.:360.:0.04 and 0.04);
\draw [fill,rotate around={0.:(4.,10.)}] (4.,10.) ++(0.:0.04 and 0.04) arc(0.:360.:0.04 and 0.04);
\draw [fill,rotate around={0.:(3.,10.7)}] (3.,10.7) ++(0.:0.04 and 0.04) arc(0.:360.:0.04 and 0.04);
\draw [fill,rotate around={0.:(3.,10.)}] (3.,10.) ++(0.:0.04 and 0.04) arc(0.:360.:0.04 and 0.04);
\node [rotate=0.] at (3.6,10.2) {\scriptsize $\nu$};
\node [rotate=0.] at (4.3,7.9) {\scriptsize $\mu$};
\node [rotate=0.] at (4.3,8.5) {\scriptsize $\mu$};
\node [rotate=0.] at (4.3,11.7) {\scriptsize $\lambda$};
\node [rotate=0.] at (4.3,12.4) {\scriptsize $\lambda$};
\node [rotate=0.] at (4.3,10.4) {\scriptsize $\gamma$};
\node [rotate=0.] at (4.3,10.9) {\scriptsize $s$};
\node [rotate=0.] at (2.7,10.4) {\scriptsize $\gamma'$};
\node [rotate=0.] at (2.7,10.9) {\scriptsize $s'$};
\node [rotate=0.] at (2.7,11.7) {\scriptsize $\lambda'$};
\node [rotate=0.] at (2.7,12.4) {\scriptsize $\lambda'$};
\node [rotate=0.] at (2.7,8.5) {\scriptsize $\mu'$};
\node [rotate=0.] at (2.7,7.9) {\scriptsize $\mu'$};
\node [rotate=0.] at (2.7,7.5) {\tiny $n_{\mu'}$};
\node [rotate=0.] at (4.3,7.5) {\tiny $n_{\mu}$};
\node [rotate=0.] at (3.8,9.1) {\tiny $n_{\nu}$};
\node [rotate=0.] at (3.3,9.1) {\tiny $n_{\nu'}$};
\node [rotate=0.] at (4.3,12.8) {\tiny $n_{\lambda}$};
\node [rotate=0.] at (2.7,12.8) {\tiny $n_{\lambda'}$};
\node [rotate=0.] at (5.1,10.9) {\tiny $m_s$};
\node [rotate=0.] at (1.9,10.9) {\tiny $m_{s'}$};
\node [rotate=0.] at (3.4,9.4) {\tiny $\nu'$};
\node [rotate=0.] at (3.7,9.4) {\tiny $\nu$};
\end{tikzpicture}
}

\newcommand{\repBasisE}{
\begin{tikzpicture}[baseline={([yshift=-.8ex]current bounding box.center)},scale=0.7]
\draw [->-, fill] (5.6,10.3) -- (6.9,10.3);
\draw [->-] (2.3,12.) -- (2.3,10.3);
\draw [->-] (5.6,8.) -- (5.6,9.6);
\draw [->-, fill] (5.6,11.3) -- (5.6,12.);
\draw [->-, fill] (2.3,9.6) -- (2.3,8.);
\draw [->-, fill] (2.3,10.3) -- (2.3,9.6);
\draw [->-, fill] (1.1,10.3) -- (2.3,10.3);
\draw [->-, fill] (5.6,10.3) -- (5.6,11.3);
\draw [->-, fill] (3.2,9.6) -- (3.6,9.6);
\draw [->-, fill] (4.3,9.6) -- (5.6,9.6);
\draw [->-, fill] (2.3,9.6) -- (3.6,9.6);
\draw [->-, fill] (5.6,9.6) -- (5.6,10.3);
\draw [fill,rotate around={0.:(2.3,9.6)}] (2.3,9.6) ++(0.:0.04 and 0.04) arc(0.:360.:0.04 and 0.04);
\draw [fill,rotate around={0.:(2.3,10.3)}] (2.3,10.3) ++(0.:0.04 and 0.04) arc(0.:360.:0.04 and 0.04);
\draw [fill,rotate around={0.:(5.6,10.3)}] (5.6,10.3) ++(0.:0.04 and 0.04) arc(0.:360.:0.04 and 0.04);
\draw [fill,rotate around={0.:(5.6,9.6)}] (5.6,9.6) ++(0.:0.04 and 0.04) arc(0.:360.:0.04 and 0.04);
\node [rotate=0.] at (5.9,8.8) {\scriptsize $\mu$};
\node [rotate=0.] at (4.7,9.3) {\scriptsize $\nu$};
\node [rotate=0.] at (5.3,10.) {\scriptsize $\gamma$};
\node [rotate=0.] at (5.3,11.2) {\scriptsize $\lambda$};
\node [rotate=0.] at (6.,10.5) {\scriptsize $s$};
\node [circle, inner sep=1pt, fill=white,draw, inner sep=1pt,rotate=0.] at (5.6,11.) {\tiny $l$};
\node [circle, inner sep=1pt, fill=white,draw, inner sep=1pt,rotate=0.] at (5.1,9.6) {\tiny $h$};
\node [circle, inner sep=1pt, fill=white,draw, inner sep=1pt,rotate=0.] at (5.6,8.4) {\tiny $g$};
\node [rotate=0.] at (2.8,9.3) {\scriptsize $\nu$};
\node [rotate=0.] at (2.,10.5) {\scriptsize $s$};
\node [rotate=0.] at (2.7,11.2) {\scriptsize $\lambda$};
\node [rotate=0.] at (2.5,9.8) {\scriptsize $\gamma$};
\node [rotate=0.] at (2.6,8.8) {\scriptsize $\mu$};
\node [fill=white,draw, circle, inner sep=1pt, inner sep=1pt,rotate=0.] at (2.3,8.6) {\tiny $x$};
\node [fill=white,draw, circle, inner sep=1pt, inner sep=1pt,rotate=0.] at (3.2,9.6) {\tiny $y$};
\node [fill=white,draw, circle, inner sep=1pt, inner sep=1pt,rotate=0.] at (2.3,10.9) {\tiny $z$};
\node [rotate=0.] at (6.8,10.5) {\tiny $m_s$};
\node [rotate=0.] at (1.4,10.5) {\tiny $m_s$};
\node [rotate=0.] at (2.6,11.9) {\tiny $n_\lambda$};
\node [rotate=0.] at (5.9,11.9) {\tiny $n_\lambda$};
\node [rotate=0.] at (2.6,8.1) {\tiny $n_\mu$};
\node [rotate=0.] at (5.9,8.1) {\tiny $n_\mu$};
\node [rotate=0.] at (4.4,9.8) {\tiny $n_\nu$};
\node [rotate=0.] at (3.6,9.8) {\tiny $n_\nu$};
\end{tikzpicture}
}

\newcommand{\repBasisF}{
\begin{tikzpicture}[baseline={([yshift=-.8ex]current bounding box.center)},scale=0.7]
\draw [->-, fill] (5.1,11.8) -- (6.4,11.8);
\draw [->-] (2.9,12.8) -- (2.9,11.8);
\draw [->-] (5.1,9.8) -- (5.1,11.1);
\draw [->-, fill] (2.9,11.1) -- (2.9,9.8);
\draw [->-, fill] (2.9,11.8) -- (2.9,11.1);
\draw [->-, fill] (1.6,11.8) -- (2.9,11.8);
\draw [->-, fill] (5.1,11.8) -- (5.1,12.8);
\draw [->-, fill] (2.9,11.1) -- (3.5,11.1);
\draw [->-, fill] (5.1,11.1) -- (5.1,11.8);
\draw [->-] (3.5,11.1) -- (3.7,11.1);
\draw [->-] (4.3,11.1) -- (5.1,11.1);
\draw [fill,rotate around={0.:(2.9,11.1)}] (2.9,11.1) ++(0.:0.04 and 0.04) arc(0.:360.:0.04 and 0.04);
\draw [fill,rotate around={0.:(2.9,11.8)}] (2.9,11.8) ++(0.:0.04 and 0.04) arc(0.:360.:0.04 and 0.04);
\draw [fill,rotate around={0.:(5.1,11.8)}] (5.1,11.8) ++(0.:0.04 and 0.04) arc(0.:360.:0.04 and 0.04);
\draw [fill,rotate around={0.:(5.1,11.1)}] (5.1,11.1) ++(0.:0.04 and 0.04) arc(0.:360.:0.04 and 0.04);
\draw [->-,rotate around={0.:(4.,9.8)}] (4.,9.8) ++(180.:1.1 and 1.1) arc(180.:360.:1.1 and 1.1);
\draw [->-,rotate around={0.:(4.,12.8)}] (4.,12.8) ++(0.:1.1 and 1.1) arc(0.:180.:1.1 and 1.1);
\draw [->-,rotate around={0.:(4.,11.1)}] (4.,11.1) ++(180.:0.3 and 0.3) arc(180.:360.:0.3 and 0.3);
\draw [->-,rotate around={0.:(4.,11.8)}] (4.,11.8) ++(0.:2.4 and 2.4) arc(0.:180.:2.4 and 2.4);
\node [rotate=0.] at (5.4,10.3) {\scriptsize $\mu$};
\node [rotate=0.] at (4.8,10.7) {\scriptsize $\nu$};
\node [rotate=0.] at (5.4,11.4) {\scriptsize $\gamma$};
\node [rotate=0.] at (4.8,12.7) {\scriptsize $\lambda$};
\node [rotate=0.] at (5.5,12.) {\scriptsize $s$};
\node [circle, inner sep=1pt, fill=white,draw, inner sep=1pt,rotate=0.] at (5.1,12.5) {\tiny $l$};
\node [circle, inner sep=1pt, fill=white,draw, inner sep=1pt,rotate=0.] at (4.6,11.1) {\tiny $h$};
\node [circle, inner sep=1pt, fill=white,draw, inner sep=1pt,rotate=0.] at (5.1,10.1) {\tiny $g$};
\node [rotate=0.] at (2.6,12.) {\scriptsize $s$};
\node [rotate=0.] at (3.3,12.7) {\scriptsize $\lambda$};
\node [rotate=0.] at (2.6,11.4) {\scriptsize $\gamma$};
\node [rotate=0.] at (3.2,10.3) {\scriptsize $\mu$};
\node [fill=white,draw, circle, inner sep=1pt, inner sep=1pt,rotate=0.] at (2.9,10.1) {\tiny $x$};
\node [fill=white,draw, circle, inner sep=1pt, inner sep=1pt,rotate=0.] at (3.5,11.1) {\tiny $y$};
\node [fill=white,draw, circle, inner sep=1pt, inner sep=1pt,rotate=0.] at (2.9,12.5) {\tiny $z$};
\end{tikzpicture}
}

\newcommand{\repBasisG}{
\begin{tikzpicture}[baseline={([yshift=-.8ex]current bounding box.center)},scale=0.7]
\draw [->-, fill] (5.1,11.8) -- (6.4,11.8);
\draw [->-] (2.9,12.8) -- (2.9,11.8);
\draw [->-, fill] (1.6,11.8) -- (2.9,11.8);
\draw [->-, fill] (5.1,11.8) -- (5.1,12.8);
\draw [->-] (3.9,9.1) -- (4.1,9.1);
\draw [->-] (3.9,7.3) -- (4.1,7.3);
\draw [->-] (4.1,10.7) -- (3.9,10.7);
\draw [->-] (4.1,8.9) -- (3.9,8.9);
\draw [fill,rotate around={0.:(2.9,11.8)}] (2.9,11.8) ++(0.:0.04 and 0.04) arc(0.:360.:0.04 and 0.04);
\draw [fill,rotate around={0.:(5.1,11.8)}] (5.1,11.8) ++(0.:0.04 and 0.04) arc(0.:360.:0.04 and 0.04);
\draw [->-,rotate around={0.:(4.,12.8)}] (4.,12.8) ++(0.:1.1 and 1.1) arc(0.:180.:1.1 and 1.1);
\draw [->-,rotate around={0.:(4.,11.8)}] (4.,11.8) ++(0.:2.4 and 2.4) arc(0.:180.:2.4 and 2.4);
\draw [->-,rotate around={0.:(4.,8.1)}] (4.,8.1) ++(0.:1.1 and 0.8) arc(0.:360.:1.1 and 0.8);
\draw [->-,rotate around={0.:(4.,9.9)}] (4.,9.9) ++(0.:1.1 and 0.8) arc(0.:360.:1.1 and 0.8);
\draw [->-,rotate around={0.:(4.,11.8)}] (4.,11.8) ++(180.:1.1 and 0.5) arc(180.:360.:1.1 and 0.5);
\node [rotate=0.] at (5.4,8.) {\scriptsize $\mu$};
\node [rotate=0.] at (5.3,10.3) {\scriptsize $\nu$};
\node [rotate=0.] at (5.4,11.4) {\scriptsize $\gamma$};
\node [rotate=0.] at (4.8,12.7) {\scriptsize $\lambda$};
\node [rotate=0.] at (5.5,12.) {\scriptsize $s$};
\node [circle, inner sep=1pt, fill=white,draw, inner sep=1pt,rotate=0.] at (5.1,12.5) {\tiny $l$};
\node [circle, inner sep=1pt, fill=white,draw, inner sep=1pt,rotate=0.] at (5.1,9.9) {\tiny $h$};
\node [circle, inner sep=1pt, fill=white,draw, inner sep=1pt,rotate=0.] at (5.1,8.2) {\tiny $g$};
\node [rotate=0.] at (2.6,12.) {\scriptsize $s$};
\node [rotate=0.] at (3.3,12.7) {\scriptsize $\lambda$};
\node [rotate=0.] at (2.6,11.4) {\scriptsize $\gamma$};
\node [fill=white,draw, circle, inner sep=1pt, inner sep=1pt,rotate=0.] at (2.9,8.) {\tiny $x$};
\node [fill=white,draw, circle, inner sep=1pt, inner sep=1pt,rotate=0.] at (2.9,9.9) {\tiny $y$};
\node [fill=white,draw, circle, inner sep=1pt, inner sep=1pt,rotate=0.] at (2.9,12.5) {\tiny $z$};
\node [fill=white,draw, circle, inner sep=1pt, inner sep=1pt,rotate=0.] at (4.6,8.8) {\tiny $g'$};
\node [fill=white,draw, circle, inner sep=1pt, inner sep=1pt,rotate=0.] at (4.6,10.6) {\tiny $g'$};
\node [fill=white,draw, circle, inner sep=1pt, inner sep=1pt,rotate=0.] at (4.6,11.4) {\tiny $g'$};
\end{tikzpicture}
}

\newcommand{\repDelta}{
\begin{tikzpicture}[baseline={([yshift=-.8ex]current bounding box.center)},scale=0.7]
\draw [->-] (3.6,8.9) ..controls (3.565,9.011) and (3.53,9.123) .. (3.5,9.2)
			 ..controls (3.47,9.277) and (3.446,9.321) .. (3.4,9.4)
			 ..controls (3.354,9.479) and (3.287,9.595) .. (3.2,9.7)
			 ..controls (3.113,9.805) and (3.007,9.9) .. (2.9,9.9)
			 ..controls (2.793,9.9) and (2.687,9.805) .. (2.6,9.7)
			 ..controls (2.513,9.595) and (2.446,9.479) .. (2.4,9.4)
			 ..controls (2.354,9.321) and (2.33,9.277) .. (2.3,9.2)
			 ..controls (2.27,9.123) and (2.235,9.011) .. (2.2,8.9);
\draw [->-, fill] (3.7,8.6) -- (3.6,8.9);
\draw [->-, fill] (2.2,8.9) -- (2.1,8.6);
\node [rotate=0.] at (2.4,8.7) {\scriptsize $m$};
\node [rotate=0.] at (3.4,8.7) {\scriptsize $n$};
\node [rotate=0.] at (2.1,9.9) {\scriptsize $\mu$};
\node [rotate=0.] at (3.9,9.9) {\scriptsize $\nu$};
\end{tikzpicture}
}

\newcommand{\representation}{
\begin{tikzpicture}[baseline={([yshift=-.8ex]current bounding box.center)},scale=1]
\draw [->-] (1.6,11.2) -- (1.6,12.);
\draw [->-] (1.6,12.) -- (1.6,12.8);
\node [circle, inner sep=1pt, fill=white,draw, inner sep=1pt,rotate=0.] at (1.6,12.) {\scriptsize $g$};
\node [rotate=0.] at (1.9,11.3) {\scriptsize $n_{\mu}$};
\node [rotate=0.] at (1.9,12.7) {\scriptsize $m_{\mu}$};
\node [rotate=0.] at (1.8,11.7) {\scriptsize $\mu$};
\end{tikzpicture}
}

\newcommand{\representationDualA}{
\begin{tikzpicture}[baseline={([yshift=-.8ex]current bounding box.center)},scale=1]
\draw [->-] (1.6,12.8) -- (1.6,12.);
\draw [->-] (1.6,12.) -- (1.6,11.2);
\node [circle, inner sep=1pt, fill=white,draw, inner sep=1pt,rotate=0.] at (1.6,12.) {\scriptsize $g$};
\node [rotate=0.] at (1.8,11.7) {\scriptsize $\mu^*$};
\node [rotate=0.] at (1.9,12.7) {\scriptsize $m_{\mu^*}$};
\node [rotate=0.] at (1.9,11.3) {\scriptsize $n_{\mu^*}$};
\end{tikzpicture}
}

\newcommand{\representationDualB}{
\begin{tikzpicture}[baseline={([yshift=-.8ex]current bounding box.center)},scale=1]
\draw [->-] (1.6,12.3) -- (2.,12.8);
\draw [->-] (2.,11.2) -- (2.4,11.6);
\draw [->-] (2.,11.2) -- (2.,12.8);
\draw [fill,rotate around={0.:(2.,12.8)}] (2.,12.8) ++(0.:0.04 and 0.04) arc(0.:360.:0.04 and 0.04);
\draw [fill,rotate around={0.:(2.,11.2)}] (2.,11.2) ++(0.:0.04 and 0.04) arc(0.:360.:0.04 and 0.04);
\node [circle, inner sep=1pt, fill=white,draw, inner sep=1pt,rotate=0.] at (2.,11.7) {\tiny $g$};
\node [rotate=0.] at (1.7,12.7) {\tiny $\mu^*$};
\node [rotate=0.] at (2.2,12.2) {\tiny $\mu$};
\node [rotate=0.] at (1.4,12.2) {\tiny $m_{\mu^*}$};
\node [rotate=0.] at (2.6,11.7) {\tiny $n_{\mu^*}$};
\node [rotate=0.] at (2.4,11.3) {\tiny $\mu^*$};
\end{tikzpicture}
}

\newcommand{\representationprodct}{
\begin{tikzpicture}[baseline={([yshift=-.8ex]current bounding box.center)},scale=1]
\draw [->-] (1.6,8.) -- (1.6,8.7);
\draw [->-] (1.6,8.7) -- (1.6,9.7);
\draw [->-] (1.6,9.7) -- (1.6,10.4);
\node [fill=white,draw, circle, inner sep=1pt, inner sep=1pt,rotate=0.] at (1.6,8.7) {\scriptsize $g$};
\node [fill=white,draw, circle, inner sep=1pt, inner sep=1pt,rotate=0.] at (1.6,9.7) {\scriptsize $h$};
\node [rotate=0.] at (1.8,9.1) {\scriptsize $\mu$};
\node [rotate=0.] at (2.,8.1) {\scriptsize $m_{\mu}$};
\node [rotate=0.] at (1.9,10.3) {\scriptsize $n_{\mu}$};
\end{tikzpicture}
}

\newcommand{\schurLemma}{
\begin{tikzpicture}[baseline={([yshift=-.8ex]current bounding box.center)},scale=0.7]
\draw [->-, fill] (2.5,7.2) -- (2.5,8.3);
\draw [->-, fill] (2.5,9.5) -- (2.5,10.4);
\draw [->-] (3.1,8.8) -- (3.1,9.);
\draw [->-] (1.9,8.8) -- (1.9,9.);
\draw [fill,rotate around={0.:(2.5,9.5)}] (2.5,9.5) ++(0.:0.08 and 0.08) arc(0.:360.:0.08 and 0.08);
\draw [fill,rotate around={0.:(2.5,8.3)}] (2.5,8.3) ++(0.:0.08 and 0.08) arc(0.:360.:0.08 and 0.08);
\draw [rotate around={0.:(2.5,8.9)}] (2.5,8.9) ++(0.:0.6 and 0.6) arc(0.:360.:0.6 and 0.6);
\node [rotate=0.] at (1.6,9.) {\scriptsize $\mu$};
\node [rotate=0.] at (3.4,9.) {\scriptsize $\nu$};
\node [rotate=0.] at (2.8,10.3) {\scriptsize $m_{\gamma}$};
\node [rotate=0.] at (2.8,7.3) {\scriptsize $n_{\gamma}$};
\node [rotate=0.] at (2.7,9.9) {\scriptsize $\gamma$};
\node [rotate=0.] at (2.7,7.8) {\scriptsize $\gamma'$};
\end{tikzpicture}
}

\newcommand{\singleLineAA}[3]{
\begin{tikzpicture}[baseline={([yshift=-.8ex]current bounding box.center)},scale=1]
\draw [->-] (1.6,8.8) -- (1.6,10.4);
\node [rotate=0.] at (1.8,9.6) {\scriptsize $#1$};
\node [rotate=0.] at (2.,8.9) {\scriptsize $#2$};
\node [rotate=0.] at (1.9,10.3) {\scriptsize $#3$};
\end{tikzpicture}
}

\newcommand{\sixJGraph}[6]{
\begin{tikzpicture}[baseline={([yshift=-.8ex]current bounding box.center)},scale=0.6]
\draw [->-] (2.9,12.8) -- (2.8,12.8);
\draw [->-] (1.7,10.8) -- (1.7,10.6);
\draw [->-] (2.9,8.8) -- (3.1,8.8);
\draw [->-] (3.5,10.1) -- (3.4,10.2);
\draw [->-] (4.5,10.8) -- (4.5,10.6);
\draw [->-] (4.3,11.5) -- (4.2,11.7);
\draw [->-] (3.3,11.) -- (3.3,10.8);
\draw [->-, fill] (4.4,11.3) -- (3.3,10.6);
\draw [rotate around={0.:(3.9,10.8)}] (3.9,10.8) ++(0.:0.6 and 1.) arc(0.:360.:0.6 and 1.);
\draw [rotate around={0.:(2.9,10.8)}] (2.9,10.8) ++(31.:1.2 and 2.) arc(31.:330.:1.2 and 2.);
\draw [fill,rotate around={0.:(3.9,11.8)}] (3.9,11.8) ++(0.:0.04 and 0.04) arc(0.:360.:0.04 and 0.04);
\draw [fill,rotate around={0.:(4.4,11.3)}] (4.4,11.3) ++(0.:0.04 and 0.04) arc(0.:360.:0.04 and 0.04);
\draw [fill,rotate around={0.:(3.3,10.6)}] (3.3,10.6) ++(0.:0.04 and 0.04) arc(0.:360.:0.04 and 0.04);
\draw [fill,rotate around={0.:(3.9,9.8)}] (3.9,9.8) ++(0.:0.04 and 0.04) arc(0.:360.:0.04 and 0.04);
\node [rotate=0.] at (3.8,11.2) {\tiny $#1$};
\node [rotate=0.] at (3.4,11.7) {\tiny $#2$};
\node [rotate=0.] at (3.8,10.2) {\tiny $#3$};
\node [rotate=0.] at (1.6,9.8) {\tiny $#4$};
\node [rotate=0.] at (4.7,10.2) {\tiny $#5$};
\node [rotate=0.] at (4.4,11.8) {\tiny $#6$};
\end{tikzpicture}
}

\newcommand{\tetrahedron}{
\begin{tikzpicture}[baseline={([yshift=-.8ex]current bounding box.center)},scale=0.7]
\draw [->-] (4.,11.2) -- (2.8,12.8);
\draw [->-] (2.8,12.8) -- (1.6,11.2);
\draw [->-] (2.8,9.6) -- (1.6,11.2);
\draw [->-] (4.,11.2) -- (2.8,9.6);
\draw [->-] (4.,11.2) -- (1.6,11.2);
\draw [->-] (2.8,12.8) -- (2.8,11.4);
\draw [->-] (2.8,11.) -- (2.8,9.6);
\draw [fill,rotate around={0.:(2.8,12.8)}] (2.8,12.8) ++(0.:0.04 and 0.04) arc(0.:360.:0.04 and 0.04);
\draw [fill,rotate around={0.:(1.6,11.2)}] (1.6,11.2) ++(0.:0.04 and 0.04) arc(0.:360.:0.04 and 0.04);
\draw [fill,rotate around={0.:(4.,11.2)}] (4.,11.2) ++(0.:0.04 and 0.04) arc(0.:360.:0.04 and 0.04);
\draw [fill,rotate around={0.:(2.8,9.6)}] (2.8,9.6) ++(0.:0.04 and 0.04) arc(0.:360.:0.04 and 0.04);
\node [rotate=0.] at (3.7,12.2) {\tiny $\rho$};
\node [rotate=0.] at (3.7,10.2) {\tiny $\kappa$};
\node [rotate=0.] at (1.9,10.2) {\tiny $\lambda$};
\node [rotate=0.] at (3.,10.6) {\tiny $\eta$};
\node [rotate=0.] at (2.3,11.4) {\tiny $\mu$};
\node [rotate=0.] at (2.,12.2) {\tiny $\nu$};
\end{tikzpicture}
}

\newcommand{\threejSymbol}{
\begin{tikzpicture}[baseline={([yshift=-.8ex]current bounding box.center)},scale=0.6]
\draw [->-] (4.8,10.4) -- (3.2,12.8);
\draw [->-] (3.2,10.4) -- (3.2,12.8);
\draw [->-] (1.6,10.4) -- (3.2,12.8);
\draw [fill,rotate around={0.:(3.2,12.8)}] (3.2,12.8) ++(0.:0.08 and 0.08) arc(0.:360.:0.08 and 0.08);
\node [rotate=0.] at (3.,11.6) {\scriptsize $b$};
\node [rotate=0.] at (2.,11.6) {\scriptsize $a$};
\node [rotate=0.] at (4.4,11.6) {\scriptsize $c$};
\node [rotate=0.] at (5.2,10.5) {\scriptsize $m_c$};
\node [rotate=0.] at (3.6,10.5) {\scriptsize $m_b$};
\node [rotate=0.] at (2.1,10.5) {\scriptsize $m_a$};
\end{tikzpicture}
}

\newcommand{\threejSymbolConj}{
\begin{tikzpicture}[baseline={([yshift=-.8ex]current bounding box.center)},scale=0.6]
\draw [->-] (3.2,7.8) -- (1.6,10.);
\draw [->-] (3.2,7.8) -- (3.2,10.);
\draw [->-] (3.2,7.8) -- (4.8,10.);
\draw [fill,rotate around={0.:(3.2,7.8)}] (3.2,7.8) ++(0.:0.08 and 0.08) arc(0.:360.:0.08 and 0.08);
\node [rotate=0.] at (3.4,9.2) {\scriptsize $b$};
\node [rotate=0.] at (2.4,9.2) {\scriptsize $a$};
\node [rotate=0.] at (4.,9.2) {\scriptsize $c$};
\node [rotate=0.] at (5.2,9.9) {\scriptsize $m_c$};
\node [rotate=0.] at (3.6,9.9) {\scriptsize $m_b$};
\node [rotate=0.] at (2.1,9.9) {\scriptsize $m_a$};
\end{tikzpicture}
}

\newcommand{\traceA}{
\begin{tikzpicture}[baseline={([yshift=-.8ex]current bounding box.center)},scale=1]
\draw [->-] (2.1,11.4) -- (2.2,11.4);
\draw [->-] (2.3,12.6) -- (2.2,12.6);
\draw [rotate around={0.:(2.2,12.)}] (2.2,12.) ++(0.:0.6 and 0.6) arc(0.:360.:0.6 and 0.6);
\node [circle, inner sep=1pt, fill=white,draw, inner sep=1pt,rotate=0.] at (1.6,12.) {\tiny $h$};
\node [circle, inner sep=1pt, fill=white,draw, inner sep=1pt,rotate=0.] at (2.8,12.2) {\tiny $g'$};
\node [rotate=0.] at (2.2,12.8) {\tiny $\nu$};
\node [circle, inner sep=1pt, fill=white,draw, inner sep=1pt,rotate=0.] at (2.7,11.7) {\tiny $y$};
\end{tikzpicture}
}

\newcommand{\traceB}{
\begin{tikzpicture}[baseline={([yshift=-.8ex]current bounding box.center)},scale=1]
\draw [->-] (2.1,11.4) -- (2.2,11.4);
\draw [->-] (2.2,12.6) -- (2.1,12.6);
\draw [rotate around={0.:(2.2,12.)}] (2.2,12.) ++(0.:0.6 and 0.6) arc(0.:360.:0.6 and 0.6);
\node [circle, inner sep=1pt, fill=white,draw, inner sep=1pt,rotate=0.] at (2.8,12.) {\tiny $g$};
\node [circle, inner sep=1pt, fill=white,draw, inner sep=1pt,rotate=0.] at (2.6,12.4) {\tiny $g'$};
\node [circle, inner sep=1pt, fill=white,draw, inner sep=1pt,rotate=0.] at (1.6,12.) {\tiny $x$};
\node [rotate=0.] at (2.2,12.8) {\tiny $\mu$};
\end{tikzpicture}
}

\newcommand{\traceC}{
\begin{tikzpicture}[baseline={([yshift=-.8ex]current bounding box.center)},scale=1]
\draw [->-] (2.5,11.) -- (2.5,12.2);
\draw [->-] (1.7,12.2) -- (1.7,11.);
\draw [->-] (2.2,13.5) -- (2.,13.5);
\draw [->-,rotate around={0.:(2.1,11.)}] (2.1,11.) ++(180.:0.4 and 0.4) arc(180.:360.:0.4 and 0.4);
\draw [fill,rotate around={0.:(1.7,11.)}] (1.7,11.) ++(0.:0.04 and 0.04) arc(0.:360.:0.04 and 0.04);
\draw [fill,rotate around={0.:(2.5,11.)}] (2.5,11.) ++(0.:0.04 and 0.04) arc(0.:360.:0.04 and 0.04);
\draw [->-,rotate around={0.:(2.1,12.2)}] (2.1,12.2) ++(0.:0.4 and 0.4) arc(0.:180.:0.4 and 0.4);
\draw [rotate around={0.:(2.1,12.2)}] (2.1,12.2) ++(295.:1. and 1.3) arc(295.:605.:1. and 1.3);
\node [circle, inner sep=1pt, fill=white,draw, inner sep=1pt,rotate=0.] at (1.7,11.9) {\tiny $l$};
\node [circle, inner sep=1pt, fill=white,draw, inner sep=1pt,rotate=0.] at (2.5,11.9) {\tiny $z$};
\node [circle, inner sep=1pt, fill=white,draw, inner sep=1pt,rotate=0.] at (2.3,10.7) {\tiny $g'$};
\node [rotate=0.] at (2.1,12.8) {\tiny $\lambda$};
\node [rotate=0.] at (1.6,10.7) {\tiny $\gamma$};
\node [rotate=0.] at (2.1,13.7) {\tiny $s$};
\end{tikzpicture}
}

\newcommand{\traceD}{
\begin{tikzpicture}[baseline={([yshift=-.8ex]current bounding box.center)},scale=1]
\draw [->-] (2.2,11.4) -- (2.3,11.4);
\draw [->-] (2.2,12.6) -- (2.1,12.6);
\draw [rotate around={0.:(2.2,12.)}] (2.2,12.) ++(0.:0.6 and 0.6) arc(0.:360.:0.6 and 0.6);
\node [circle, inner sep=1pt, fill=white,draw, inner sep=1pt,rotate=0.] at (1.6,12.) {\tiny $l$};
\node [circle, inner sep=1pt, fill=white,draw, inner sep=1pt,rotate=0.] at (2.8,12.) {\tiny $z$};
\node [circle, inner sep=1pt, fill=white,draw, inner sep=1pt,rotate=0.] at (2.6,11.6) {\tiny $g''$};
\node [rotate=0.] at (2.2,12.8) {\tiny $\lambda$};
\end{tikzpicture}
}

\newcommand{\traceE}{
\begin{tikzpicture}[baseline={([yshift=-.8ex]current bounding box.center)},scale=1]
\draw [->-] (1.6,11.6) -- (1.7,11.6);
\draw [rotate around={0.:(1.6,12.2)}] (1.6,12.2) ++(0.:0.6 and 0.6) arc(0.:360.:0.6 and 0.6);
\node [circle, inner sep=1pt, fill=white,draw, inner sep=1pt,rotate=0.] at (1.6,12.8) {\tiny $g''$};
\node [rotate=0.] at (2.1,12.8) {\tiny $s$};
\end{tikzpicture}
}

\newcommand{\traceF}{
\begin{tikzpicture}[baseline={([yshift=-.8ex]current bounding box.center)},scale=1]
\draw [->-] (1.8,11.3) -- (1.9,11.3);
\draw [->-] (1.9,12.5) -- (1.8,12.5);
\draw [rotate around={0.:(1.9,11.9)}] (1.9,11.9) ++(0.:0.6 and 0.6) arc(0.:360.:0.6 and 0.6);
\node [circle, inner sep=1pt, fill=white,draw, inner sep=1pt,rotate=0.] at (2.2,11.4) {\tiny $g'$};
\node [circle, inner sep=1pt, fill=white,draw, inner sep=1pt,rotate=0.] at (2.3,12.3) {\tiny $g''$};
\node [rotate=0.] at (1.9,12.8) {\tiny $\gamma$};
\end{tikzpicture}
}

\newcommand*\circled[1]{\tikz[baseline=(char.base)]{
   \node[shape=circle,draw,inner sep=0.5pt] (char) {$#1$};}}
\newcolumntype{C}{>{\centering\arraybackslash} m{1.5em} }

\makeatletter
\newcommand*{\Relbarfill@}{\arrowfill@\Relbar\Relbar\Relbar}
\newcommand*{\xeq}[2][]{\ext@arrow 0055\Relbarfill@{#1}{#2}}
\makeatother

\allowdisplaybreaks
\title{Electric-magnetic duality in the quantum double models of topological orders with gapped boundaries}
\date{\today}
\author[a,b]{Hongyu Wang}
\author[a,b]{Yingcheng Li}
\author[c,d]{Yuting Hu}
\author[a,b,c,1]{Yidun Wan,\note{Corresponding author}}
\affiliation[a]{State Key Laboratory of Surface Physics, Fudan University, Shanghai 200433, China}
\affiliation[b]{Department of Physics, Center for Field Theory and Particle Physics, and Institute for Nanoelectronic devices and Quantum computing, Fudan University, Shanghai 200433, China}
\affiliation[c]{Department of Physics and Institute for Quantum Science and Engineering, South University of Science and Technology, Shenzhen 518055, China}
\affiliation[d]{CAS Key Laboratory of Microscale Magnetic Resonance and Department of Modern Physics, University of Science and Technology of China, Hefei,
Anhui 230026, China}
\emailAdd{wanghy17@fudan.edu.cn,ycli17@fudan.edu.cn,yuting.phys@gmail.com,\\ ydwan@fudan.edu.cn}

\abstract{
        We generalize the Electric-magnetic (EM) duality in the quantum double (QD) models to the case of topological orders with gapped boundaries. We also map the QD models with boundaries to the Levin-Wen (LW) models with boundaries. To this end, we Fourier transform and rewrite the extended QD model with a finite gauge group $G$ on a trivalent lattice with a boundary. Gapped boundary conditions of the model before the transformation are known to be characterized by the subgroups $K \subseteq G$. We find that after the transformation, the boundary conditions are then characterized by the Frobenius algebras $A_{G,K}$ in $\rep_G$. An $A_{G,K}$ is the dual space of the quotient of the group algebra of $G$ over that of $K$, and $\rep_G$ is the category of the representations of $G$. The EM duality on the boundary is revealed by mapping the $K$'s to $A_{G,K}$'s. We also show that our transformed extended QD model can be mapped to an extended LW model on the same lattice via enlarging the Hilbert space of the extended LW model. Moreover, our transformed extended QD model elucidates the phenomenon of anyon splitting in anyon condensation.
}

\begin{document}

\maketitle 
\flushbottom
\section{Introduction}\label{sec:intro}
Two-dimensional phases of matter with intrinsic topological orders, or topological phases for short, can be well studied by effective topological field theories, whose Hamiltonian extensions are exactly solvable lattice models.  Two major families of such models are the quantum double (QD) models \cite{Kitaev2003a} and the string-net models or Levin-Wen (LW) models\cite{Levin2004}.  The QD models have been generalized to be the twisted QD models\cite{Hu2012a,Mesaros2011}, and the LW models have also been generalized similarly\cite{Lin2014}. In this paper, we shall not deal with such generalizations. 

A QD model is a lattice gauge theory with a finite gauge group $G$ as its input data. The Hamiltonian of the model converts the input data to an output data---the quantum double $D(G)$ of good quantum numbers. The quasiparticle excitations of the topological phase described by the model, namely the anyons, carry representations $([c],\mu)$ of $D(G)$, where $[c]$ labels the conjugacy classes of $G$, and $\mu$ labels the irreducible representations of the centralizer of $[c]$ in $G$. Anyons of the type $([e],\mu)$ in which $e\in G$ is the identity element are the pure charges, which live at the vertices of $\Gamma$; the anyons of the type $([c],0)$ with $0$ being the trivial representation of the centralizer of $[c]$ are the pure fluxes, which live in the plaquettes of $\Gamma$; the anyons of the mixed type are the dyons. Hence, intuitively, the QD models would exhibit an explicit EM duality. Indeed, to every QD model on a lattice $\Gamma$, there corresponds a QD model on the dual lattice $\Gamma^*$, such that the charges (fluxes) on $\Gamma$ are the fluxes (charges) on $\Gamma^*$, and vice versa\cite{Buerschaper2013}. This EM duality is immediately understood in the cases of Abelian groups. In such a case, the irreducible representations of $G$ are all $1$-dimensional and form a group isomorphic to $G$ itself. Denote the set of all irreducible representations of $G$ by $\rep_G$. While the QD model on $\Gamma$ takes $G$ as its input data, the dual model on $\Gamma^*$ takes $\rep_G\cong G$ as the input data. That is, the dual model is truly a QD model, which by definition has a finite group as its input data. If $G$ is non-Abelian, however, $\rep_G$ cannot be a group and thus cannot serve as the input data of a QD model. In such cases, the QD models must be generalized to allow Hopf algebras as their input data to exhibit the EM duality\cite{Buerschaper2013}.

On the other hand, a QD model with a finite group $G$ can also be mapped via Fourier transform to an LW model with $\rep_G$ as its input data on the same lattice\cite{Buerschaper2009}. A subtlety is that when $G$ is non-Abelian, a truncation of the Hilbert space of the QD model must be done to complete the mapping\cite{Buerschaper2009} unless one enlarges the Hilbert spaces of the LW model\cite{Buerschaper2013,Hu2018}. Via this mapping, the LW models with input data $\rep_G$ bear an EM duality as well. 

The EM duality and the aforementioned mapping to the LW models of the QD models are restricted to topological phases on closed surfaces only. Nevertheless, topological phases on surfaces with boundaries are of more practical and theoretical importance because 1) materials with boundaries are much more available than the closed ones, 2)boundary modes are easier to measure experimentally, and 3) a dynamical theory of topological phases is incomplete if unable to encompass different boundary conditions. Recently the QD models and LW models have been extended to be defined on lattices with boundaries by adding appropriate boundary Hamiltonians and are called the extended QD models\cite{Beigi2011,Bullivant2017} and the extended LW models\cite{Hu2017,Hu2017a}. This motivates us to examine in this paper whether and how the extended QD models still possess an EM duality, in particular along the boundaries, and can be mapped to the extended LW models. 

Because of topological invariance and for convenience, we consider a trivalent lattice $\Gamma$ with a single boundary (see Fig. \ref{fig:trivalentLattice}). The input data of an extended QD model on $\Gamma$ is still a finite group $G$; however, the boundary Hamiltonian projects the boundary degrees of freedom into a subgroup $K\subseteq G$.  Each $K$ characterizes a gapped boundary condition. We first Fourier transform the Hilbert space on $\Gamma$. The Fourier-transformed Hilbert space basis begs us to rewrite the Fourier-transformed model on a slightly different lattice $\tilde\Gamma$, which modifies $\Gamma$ by adding near to each vertex of $\Gamma$ a dangling edge (see Fig. \ref{fig:basisTrans}(d)). While the bulk degrees of freedom after the Fourier transform become $\rep_G$, the boundary degrees of freedom would be projected by the Fourier-transformed boundary Hamiltonian into a Frobenius algebra $A_{G,K} = \left(\mathds{C}[G]/\mathds{C}[K]\right)^*$, the dual space of the quotient of the group algebra of $G$ over that of a given $K$.  When $G$ is Abelian,  the emergent Frobenius algebra $A_{G,K}$ happens to be an Abelian group too; hence, the boundary EM duality can be understood as one between the Fourier-transformed extended QD model on $\tilde\Gamma$ with boundary condition specified by $A_{G,K}$ and the extended QD model with boundary condition specified by $K$. When $G$ is non-Abelian, $A_{G,K}$ cease to being a group but truly an algebra. 

We also show that our Fourier-transformed extended QD model on $\tilde\Gamma$ can be mapped to an extended LW model on the same lattice. In doing so, instead of truncating the Hilbert space of the extended QD model, we enlarge the extended LW model. This enlargement is necessary because the Hilbert space of the original extended LW model is too small to contain the full spectrum of excited states with charges; it has already been done for the original LW model on a closed surface\cite{Hu2018}.
Since the extended QD model and the extended LW model are Hamiltonian extension of the extended Dijkgraaf-Witten and extended Turaev-Viro types of topological field theories, our results also offer a correspondence between the two types of topological field theories. 

The EM duality on the boundary and the mapping to the extended LW model can be revealed by mapping a $K\subseteq G$ to $(\mathds{C}[G]/\mathds{C}[K])^*$. Three cases of such mappings are listed in Table \ref{tab:EMdualityOnBdry}. We explain them in order.
\begin{enumerate}
        \item[(a)] The extended QD model for any $G$ has a rough boundary condition specified by $K=\{e\}$, indicating charge condensation at the boundary. By a Fourier transform, the extended QD model is mapped to the extended LW model with a boundary Hamiltonian given by the Frobenius algebra $A_{G,\{e\}} =\mathds{C}[G]^*$. As the function space over $G$, $\mathds{C}[G]^*$ is the regular representation in $\rep_G$, which has a canonical Frobenius algebra structure that has a decomposition $A_{G,\{e\}}=\displaystyle\bigoplus_{j\in \mathrm{Irrep}_G}V_j^{\oplus \dim V_j}$.
        \item[(b)] For $K=G$, the boundary condition is a smooth one and due to flux condensation at the boundary. The transformed boundary Hamiltonian has the trivial input Frobenius algebra $A_{G,G}=0$, the trivial representation in $\rep_G$.
        \item[(c)] For a nontrivial subgroup $K$, the corresponding Frobenius algebra is $(\mathds{C}[G]/\mathds{C}[K])^*$, which is defined by the function space $\{f| f(kgk')=f(g) \forall g\in G,\forall k,k'\in K\}$. It is the largest sub-representation space of $\mathds{C}[G]^*$, such that $\rho(k)=\id,\forall k\in K$.
\end{enumerate}

\begin{table}[!h]
        \small
        \centering
        %       \resizebox{\textwidth}{31mm}{
        \begin{tabular}{|c|c|c|}
                \hline
                boundary condition & extended QD model & extended LW model              
                \\
                \hline
                charge condensation & $K=\{e\}$  & $A_{G,\{e\}}=\mathds{C}[G]^*$ (regular rep)
                \\
                \hline 
                flux condensation & $K=G$  & $A_{G,G}=0$ (trivial rep)
                \\
                \hline
                generic dyon condensation & $K$ & $A_{G,K}=(\mathds{C}[G]/\mathds{C}[K])^*$
                \\
                \hline
        \end{tabular}
        %}
        \caption{EM duality on the boundary.}
        \label{tab:EMdualityOnBdry}
\end{table} 

Another understanding of the boundary EM duality would require generalizing the entire extended model to one with input data being a Hopf algebra, similar to what is done only to the bulk as in Ref.\cite{Buerschaper2013}. We shall not discuss such generalizations in this paper but offer a justification of the boundary EM duality when $G$ is non-Abelian.

According to the mechanism of anyon condensation\cite{Bais2009}, a gapped boundary condition of a topological phase can be accounted for by a condensate at the boundary formed by certain types of anyons from the bulk\cite{Kitaev2012,HungWan2015a}. To condense at the boundary, certain types of bulk anyons would have to first split into a few pieces, several or all of which are allowed to condense at the boundary, depending on the structure of the condensate. Current understandings of this phenomenon are categorical and abstract. It would be interesting to understand the phenomenon of splitting and partial condensation based on concrete lattice models of topological phases. As we will show, our Fourier transform of the extended QD model explains this phenomenon in terms of solely the input data of the model.    

To facilitate our studies in the paper, we introduce also a graphical tool of group representation theory. We provide concrete examples, one for Abelian groups $G$ and one for the non-Abelian group $S_3$, to illustrate our results.

Our paper is organized as follows. Section \ref{sec:rev} reviews the extended QD model. Section \ref{sec:FTeQD} Fourier-transforms and rewrites the extended QD model. Section \ref{sec:FA} verifies the emergent Frobenius algebra structure on the boundary and elucidates the phenomenon of anyon splitting in boundary anyon condensation. Section \ref{sec:duality} illuminates the boundary EM duality. Section \ref{sec:mapToLW} maps the Fourier-transformed extended QD model to the extended LW model on the same lattice. Section \ref{sec:examples} provides two concrete examples of our results. Finally, the appendices collect a review of the extended LW model and certain details to avoid clutter in the main text.

\section{Extended quantum double model}\label{sec:rev}
\begin{figure}[h!]
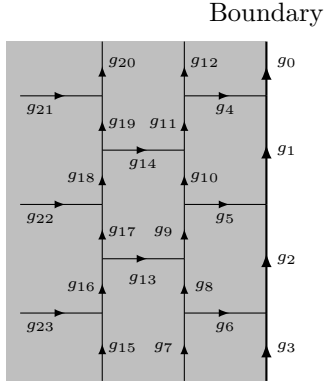

\centering
\qdReview
\caption{A portion of an oriented trivalent lattice, on which the extended QD model is defined. Each edge of the lattice is graced with a group element of a finite gauge group $G$. Grey region is the bulk, to the left of the boundary (thick line).}
\label{fig:trivalentLattice}
\end{figure}

An extended QD model\cite{Beigi2011,Bullivant2017} is an extension of the QD model to the case with boundaries by adding boundary Hamiltonians to the QD Hamiltonian. The model is a Hamiltonian extension of the Dijkgraaf-Witten topological gauge theory with a finite gauge group. The model can be defined on an arbitrary lattice with one or multiple boundaries. Topological invariance allows the model to be defined on a fixed lattice for computational convenience. In this paper, we consider an oriented trivalent lattice
$\Gamma$, part of which is shown in Fig. \ref{fig:trivalentLattice}.

The input data of the model is a finite gauge group $G$, whose elements are assigned to the edges of $\Gamma$. The total Hilbert space is spanned by all possible configurations of the group elements of $G$ on the edges of $\Gamma$ and is the tensor product of the local Hilbert spaces respectively on the edges. Namely,
\be\label{eq:eQDHil}
\Hil^{\QD}_G = \bigotimes_{ \e \in\Gamma} \Hil_\e=\bigotimes_{\e\in \Gamma} \mathrm{span} \{\ket{g_\e}| g_\e\in G \},
\ee
where $\e$ is an edge in $\Gamma$. Note that reversing the orientation of an edge graced with a group element $g$ changes the group element to $\bar g := g^{-1}$; however, we work with a fixed lattice with a fixed orientation. The Hamiltonian of the model is the sum of a bulk Hamiltonian and a boundary Hamiltonian:
\be\label{eq:eQDHam}
H_{G,K}^{\QD} = H^{\rm{QD}}_G + H^{\rm{QD}}_K .
\ee
The bulk Hamiltonian consists of two sums of local operators:
\be\label{eq:eQDbulkHam}
H^{\rm{QD}}_G =  -\sum_{v\in \Gamma\backslash \partial{\Gamma}} A_{v}^{\QD} - \sum_{p\in \Gamma\backslash \partial{\Gamma}}B_{p}^{\QD},
\ee
where the two sums are respectively over all vertices and all plaquettes in the bulk of $\Gamma$. A local vertex operator $A_v^\QD$ acts locally on the three edges incident at the vertex $v$ as follows. 
\be\label{eq:qdbuaOperators}
A_v^{\QD}\bket{\bulkHamiltonian{g}{h}{l}{v}}=\frac{1}{|G|}\sum_{x\in G}T(x)\bket{\bulkHamiltonian{g}{h}{l}{v}}=\frac{1}{|G|}\sum_{x\in G}\bket{\bulkHamiltonian{xg}{xh}{l\bar x}{v}},
\ee
which is understood as a discrete gauge transformation averaged over $G$. Clearly, $A_v^\QD$ is a projector because $(A_v^\QD)^2 = A_v^\QD$, and it projects out any such local states that are not invariant under the gauge transformation. In gauge theory terminologies, $A_v^\QD$ imposes a local Gauss constraint. A local plaquette operator acts locally on the six edges outlining the plaquette $p$ as follows.
\be\label{eq:qdbubOperators}
B_p^{\QD}\bket{\bulkBpOperator{g_1}{g_2}{g_3}{g_4}{g_5}{g_6}}=\delta_{g_1\cdot g_2\cdot g_3...g_6,e}\bket{\bulkBpOperator{g_1}{g_2}{g_3}{g_4}{g_5}{g_6}},
\ee
which is also a projection. In gauge theory terminologies, $B_p^\QD$ imposes a local flatness condition in $p$. All plaquette operators and vertex operators commute.
A gapped boundary condition is specified by a subgroup $K\subseteq G$. The boundary Hamiltonian comprises boundary local operators: 
\be\label{eq:eQDbdryHam}
H_{K}^{QD}=-\sum_{v\in \partial\Gamma}\Ab_{v}^{QD} - \sum_{p\in \partial \Gamma} \Bb_{p}^{\QD},
\ee
where the two sums are respectively over all vertices and all virtual plaquettes (to be defined shortly) on the boundary of $\Gamma$.
We define\be\label{eq:qdaOperators}
\Ab_v^{\QD}\bket{\qdLocalBasisA{g}{h}{l}{v}}=\frac{1}{|K|}\sum_{k\in K}T(k)\bket{\qdLocalBasisA{g}{h}{l}{v}}=\frac{1}{|K|}\sum_{k\in K}\bket{\qdLocalBasisA{kg}{kh}{l\bar k}{v}},
\ee  
which is again a gauge transformation averaged instead in a subgroup $K\subseteq G$. An $\Ab_v^{\QD}$ is clearly also a projector that projects out any non-invariant states under its action. All boundary vertex operators commute with each other and with all other operators in the total Hamiltonian. An operator $\Bb_v^\QD$ simply does the following projection.
\be\label{eq:qdbOperators}
\Bb_p^{\QD}\bket{\qdLocalBasisB{h}{l}{x}{y}{z}}=\delta_{x\in K}\bket{\qdLocalBasisB{h}{l}{x}{y}{z}},
\ee
where the virtual plaquette is evidently defined. 
The boundary plaquette operators all commute. Therefore, the total Hamiltonian \eqref{eq:eQDHam} is exactly solvable. The ground states are the common $+1$ eigenstates of all operators $A_v^{\QD}$, $B_p^{\QD}$, $\Ab_v^{\QD}$ and $\Bb_p^{\QD}$. The ground state degeneracy (GSD) can be computed by
\be\label{eq:qdGSD}
\mathrm{GSD} = \mathrm{Tr} \prod_{v\in \Gamma\backslash \partial{\Gamma}}A_v^{\QD}\prod_{v'\in \partial\Gamma}\Ab_{v'}^{QD}\prod_{p\in \Gamma\backslash \partial{\Gamma}}B_p^{\QD}\prod_{p'\in \partial\Gamma}\Bb^{QD}_{p'},
\ee
where the trace is taken over the total Hilbert space \eqref{eq:eQDHil}. Quasiparticle excitations of the model are charges on the vertices, fluxes through plaquettes, and dyons as bound-states of charges and fluxes. A charge at a vertex $v$ arises when the local Gauss constraint is violated; a flux through a plaquette $p$ occurs when the local flatness condition is violated; when both constraints are violated in $p$, a dyon shows up in $p$. Other properties of the model and topological phases that are classified by this model can be found in Ref.\cite{Bullivant2017} and references therein.

\section{Fourier transforming and rewriting the extended QD model}\label{sec:FTeQD}
In this section, we Fourier-transform the basis of the Hilbert space of the extended QD model from the group space to the representation space, and as urged by this transformation, rewrite the extended QD model on a slightly different lattice.  

\subsection{A graphical tool for group representation theory}
To facilitate the derivations in this paper, we introduce the following  graphical tool for group representation theory\cite{Hu2013a}. Let $L_G$ be the set of all (representatives of equivalence classes of) irreducible representations of a finite group $G$. For simplicity, we shall define $d_\mu=\dim V_\mu$ for $\mu\in L_G$. A representation matrix $D^{\mu}_{m_{\mu}n_{\mu}}(g)$ acting on $V$ is depicted as 
\be\label{eq:repMatrix}
\representation .
\ee
Here, the line is oriented from the right index $n_{\mu}$ to the left index $m_{\mu}$ of the representation matrix. The $\circled{g}$ insertion on the line indicates the group action by $g$. In this graphical presentation, multiplying two representation matrices is done by simply concatenating two such lines: 
\be\label{eq:repProduct}
\sum_{p_{\mu}}D^{\mu}_{n_{\mu}p_{\mu}}(h)D^{\mu}_{p_{\mu}m_{\mu}}(g)=\representationprodct .
\ee
A line with group action $g=e$ reads as an identity matrix, as in Eq. \eqref{eq:singleLine}. 
\be\label{eq:singleLine}
\singleLineAA{\mu}{m_{\mu}}{n_{\mu}}=\delta_{n_{\mu},m_{\mu}},
\ee
which serves as presenting the $\dim V_\mu$ basis vectors of the representation space $V_\mu$. In the representation space, we can define the inner product $\delta_{m,n}=\langle e_m,e_n\rangle$, which has a graphical expression shown in Eq. \ref{eq:delta}.
\be\label{eq:delta}
\delta_{m,n}\delta_{\mu,\nu} =\repDelta,
\ee
where we invoke that two bases belong to different representation spaces are orthogonal  to each other. Using the above inner product, we can define the complex conjugate of  an irreducible representation diagrammatically as
\be\label{eq:ccRep}
\representation \xrightarrow{\text{c. conj}}\quad \conjugateRep \equiv \conjugate,
\ee
where the arrow is reversed after the complex conjugation. This way, the complex-conjugated group action should be read upward as $D^{\mu}_{m_{\mu}n_{\mu}}(g)^*=D^{\mu}_{n_{\mu}m_{\mu}}(\bar g)$, with $\bar g :=g^{-1}$, because of the unitarity of $\mu$.
As such, the great orthogonality theorem is presented by
\be\label{eq:greatOrTheom}
\sum_{g}\greatOrTheomA=\frac{|G|}{d_{\mu}}\greatOrTheomB.
\ee
Every irreducible representation $\mu\in L_G$ has a dual $\mu^*\in L_G$, such that $\mu^*$ is equivalent to (but not necessarily identical to) the complex conjugate representation of $\mu$.
% defined on the dual vector space $V^*_{\mu^*}$. The basis vectors of $V^*_{\mu^*}$ are denoted by $e_{m_{\mu^*}}=\langle e_m,\cdot\rangle $.  
There is an invertible duality map
\be\label{eq:dualMap}
\omega_\mu : \mathbb{C}\mapsto V_\mu\otimes V_{\mu^*}; 1 \mapsto \sum_{m_{\mu},n_{\mu^*}}\Omega^{\mu}_{{m_\mu}n_{\mu^*}}e_{m_\mu}\otimes e_{n_{\mu^*}}, 
\ee
where the $\Omega^{\mu}_{{m_\mu}n_{\mu^*}}$ is a complex matrix satisfies
normalization $\Omega^\dagger \Omega=\mathbb{I}$ and maps $\mu$ to $\mu^*$ by similarity transformation
\be\label{eq:dualMatrix}
(\Omega^\mu)^{-1}D^\mu(g)\Omega^\mu=(D^{\mu^*}(g))^*.
\ee
Graphically, the duality map and its inverse has  presentations
\be\label{eq:dualMapGraph}
\Omega^\mu_{m_\mu n_{\mu^*}}=\dualMap, \quad (\Omega^\mu)^{-1}_{n_{\mu^*}m_\mu}=\dualMapInverse.
\ee
  And the similarity transformation in Eq. \eqref{eq:dualMatrix} is presented
by
\be\label{eq:dualConjMap}
\representationDualB=\representationDualA.
\ee
For the uniqueness of duality map, the matrix $\Omega^\mu$ is either symmetric or antisymmetric depends on whether $\mu$ is pseudo real or not.
This is an intrinsic property specified by a number $\beta_\mu$ called the Frobenius-Schur (FS) indicator. We have $(\Omega^{\mu^*})^T=\beta_\mu\Omega^\mu$ with $\beta_\mu=\pm 1$ if $\mu$ is real or pseudo real.

Frequently in later derivations, we will need $3j$-symbols to deal with the coupling of three representations of $G$. A $3j$-symbol is a tensor $w^{abc}_{m_am_bm_c}$ that is defined as an intertwiner:
\be\begin{aligned}\label{eq:threejSymbol}
&V_a\otimes V_b\otimes V_c \rightarrow \mathbb{C};\\
&\ket{am_a,bm_b,cm_c}\mapsto w^{abc}_{m_am_bm_c},
\end{aligned}\ee
where $(a,V_a), (b,V_b)$, and $(c,V_c)$ are three irreducible representations of $G$, and $m_j$  labels the basis $e_{m_j}$ in the vector space $V_j$. Graphically, a $3j$-symbol is presented by
\be\label{eq:threejSymbol}
w^{abc}_{m_am_bm_c}= \threejSymbol,\quad \left( w^{abc}_{m_am_bm_c} \right)^{*}= \threejSymbolConj.
\ee

Being an intertwiner, by definition a $3j$-symbol is invariant under group actions, namely,
\be\label{eq:intertwiner}
\intertwiner=\threejSymbol.
\ee 
In this paper, we will also use a lot of Clebsch-Gordan coefficients, which are values of intertwiners that can be defined using the $3j$-symbols \eqref{eq:threejSymbol} and duality map \eqref{eq:dualMapGraph} as
\be\label{eq:cGCoe}
\cGCoefficientA=\cGCoefficientB, \quad \cGCoefficientC=\cGCoefficientD.
\ee
For later convenience, we list a few properties of $3j$-symbols as follows. For a generic finite group $G$, we can always construct such $3j$-symbols satisfying the following properties\cite{Hu2013a}. 
\be\label{eq:colsedGraph}
\idnetityA=\frac{1}{|G|}\sum_g\identityB=\identityC,
\ee 
where the part in a dashed cap does not have any open edges. 

\be\label{eq:ortho}
\sum_c \beta_c \ d_c\identityD=\identityE,
\ee

\be\label{eq:schursLemma}
\schurLemma=\frac{1}{d_{\gamma}}\beta_\gamma\delta_{m_{\gamma},n_{\gamma}}\delta_{\gamma,\gamma'},
\ee

 \be\label{eq:normalizationThreej}
\sum_{m_a,m_b,m_c}w^{abc}_{m_am_bm_c}(w^{abc}_{m_am_bm_c})^*=\normalizeThreej=1.
\ee 
Equations. \eqref{eq:ortho} and \eqref{eq:schursLemma} are the orthogonality conditions, while Eq. \eqref{eq:normalizationThreej} is the normalization condition.
The symmetrized $6j$-symbols in terms of $3j$-symbols and duality map are
\be\begin{aligned}\label{eq:SixjSymbol}
G^{\mu\nu\lambda}_{\eta\kappa\rho}=&\sum_{m_\mu,n_{\mu^*},m_{\nu},n_{\nu^*},...,m_{\rho},n_{\rho^*}}\Omega^\mu_{m_\mu n_{\mu^*}}\Omega^\nu_{m_\nu n_{\nu^*}}\Omega^\lambda_{m_\lambda n_{\lambda^*}}\Omega^\eta_{m_\eta n_{\eta^*}}\Omega^\kappa_{m_\kappa n_{\kappa^*}}\Omega^\rho_{m_\rho n_{\rho^*}}
\\
&\times w^{\kappa\lambda^*\eta}_{m_\kappa n_{\lambda^*} m_\eta}w^{\eta^*\nu^*\rho}_{n_{\eta^*}n_{\nu^*}m_{\rho}}w^{\rho^*\mu^*\kappa^*}_{n_{\rho^*}n_{\mu^*}n_{\kappa^*}}w^{\mu \nu \lambda}_{m_\mu m_\nu m_\lambda},
\end{aligned}\ee
which can be presented either by a planar graph or a tetrahedron 
\be\label{sixJGraph}
G^{\mu\nu\lambda}_{\eta\kappa\rho}=\beta_\eta \sixJGraph{\mu}{\nu}{\lambda}{\eta}{\kappa}{\rho}=\tetrahedron.
\ee

\subsection{Fourier transform on the Hilbert space}\label{subsec:HilbertFT}

\begin{figure}[h!]
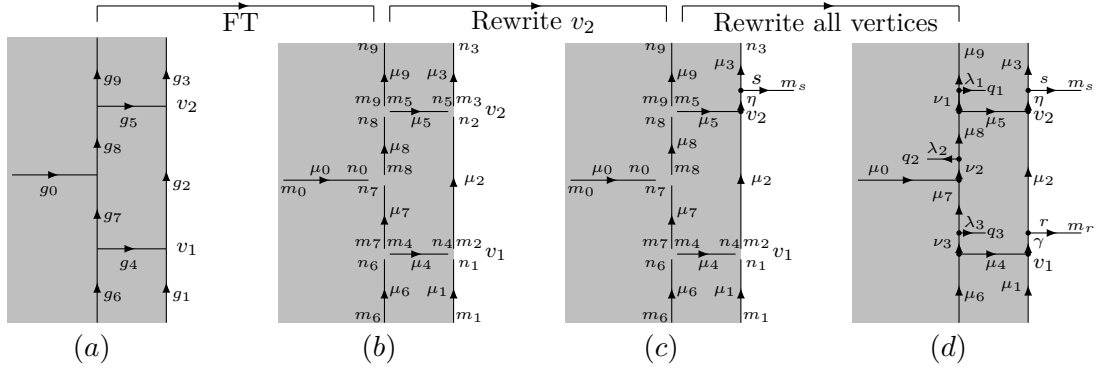

\centering
\hilbertBasisA
\caption{Basis transformations of the Hilbert space of the extended QD model. (a) A group-basis state on a part of the original lattice $\Gamma$. (b) Basis states in the Fourier transformation of (a). (c) Basis states obtained by rewriting vertex $v_2$ in (b). (d) Basis states obtained by rewriting all vertices in (b), obtaining a new lattice $\tilde\Gamma$.}
\label{fig:basisTrans}
\end{figure}

Let us first Fourier-transform the Hilbert space of an extended QD model with a finite gauge group $G$. The total Hilbert space of the model is defined in Eq. \eqref{eq:eQDHil}. Since the degrees of freedom of the extended QD model live on the edges of the lattice, the total Hilbert space $\Hil$ is the tensor product of all local Hilbert spaces $\Hil_\e$ on the edges.  The local basis state on an edge is $\ket{g}$, which can be Fourier transformed to the basis states in terms of representations $\ket{\mu,m_{\mu},n_{\nu}}$, called a rep-basis,
by performing the following Fourier transform (FT): 
\be\label{eq:fouriertrans}
\ket{\mu,m_{\mu},n_{\mu}}=\frac{v_{\mu}}{\sqrt{|G|}}\sum_{g\in G}D^{\mu}_{m_{\mu}n_{\mu}}(g)\ket{g},
\ee
where $(\mu,V_\mu)$ pairs a unitary irreducible representation of $G$ with representation space $V_\mu$, $v_{\mu}=\sqrt{d_\mu}$, $|G|$ is the order of $G$, and $D^{\mu}_{m_{\mu}n_{\mu}}$ is the representation matrix in $V_{\mu}$. The local rep-basis and the group-basis have the same dimension because $\sum_\mu d_\mu^2 = |G|$.  
Note again that the total Hilbert space of the model is defined regardless of the Hamiltonian, which when imposed would separate the Hilbert space into ground and excited states. Hence, the Fourier transformation of the group-basis of total Hilbert space on the trivalent lattice can be done by transforming the local basis of the $\Hil_\e$ on each individual edge independently. Figure \ref{fig:basisTrans}(a) shows a small part of the lattice $\Gamma$, including both bulk and boundary, and a corresponding group-basis state on the part of $\Gamma$. Fourier-transforming the group-basis state in Figure \ref{fig:basisTrans}(a) results in a linear superposition of the rep-basis states, shown in Figure \ref{fig:basisTrans}(b). The explicit transformation will be dealt with a little later. Here we focus on the logic and physics of the basis transformations.

While the basis state in Figure \ref{fig:basisTrans}(a) is directly defined on the original trivalent lattice of the model, the basis states in Figure \ref{fig:basisTrans}(b) lack the attachment to an actual trivalent lattice. Here is the reason. The Fourier transform turns the three group elements on the three lattice edges meeting at a vertex independently into three linear combinations of all irreducible representations of $G$. Taking the vertex $v_2$ in Figure \ref{fig:basisTrans}(a) as an example, the three group elements $g_2,g_3$ and $g_5$ are respectively transformed into three independent representations $\mu_2,\mu_3$, and $\mu_5$ in Figure \ref{fig:basisTrans}(b); however, there is no \textit{a priori} a reason the three lines graced with $\mu_2,\mu_3$, and $\mu_5$ should simply just meet at $v_2$. If we did so and removed the matrix indices $n_2, m_3$, and $n_5$, one would misinterpret the vertex $v_2$ as an intertwiner of the three representations and would certainly be wrong because the dimension of the basis would be lower than that of the group-basis. Hence, we leave the vertex $v_2$ and all other vertices open in Fig. \ref{fig:basisTrans}(b), and the basis becomes detached from the original lattice. 

Now the question is: Can we rewrite the basis in Fig. \ref{fig:basisTrans}(b) to one that has the same dimension and still attached to an actual trivalent lattice? The answer is ``Yes''. Let us again stare at the vertex $v_2$ in Fig. \ref{fig:basisTrans}(b). The strategy is to fuse (i.e., couple) the three representations in an order at $v_2$. Since coupling representations is an associative linear transformation, we can choose our convention. Let us first fuse  $\mu_2$ and $\mu_5$ by contracting their indices $n_2$ and $n_5$, resulting in a linear combination of irreducible representations $\{\eta\}$, each member of which is graphically a line with a free end labeled by $m_\eta$. Then, we can fuse an $\eta$ with $\mu_3$ by contracting $m_\eta$ and $m_3$, resulting in a linear combination of representations $\{s\}$, each member of which is a line with a free end labeled by $m_s$. For each $\eta$ and each $s$, we obtain a new basis state, as in Fig. \ref{fig:basisTrans}(c). A pivotal point is that the degrees of freedom $\eta$ and $s$ are both local with respect to the vertex $v_2$ because they arise from fusing the three representations $\mu_2,\mu_5$, and $\mu_3$ at $v_2$. Hence, the line graced with $s$ can be considered an induced degree of freedom at $v_{2}$, and we place the line very close to $v_2$, as a dangling edge. In this procedure of rewriting the basis, there is no loss of dimension because the degrees of freedom associated with $v_2$ in Fig. \ref{fig:basisTrans}(c) contribute dimension $\dim= \sum_{\eta,\mu_5,\mu_2,\mu_3,s}N_{\mu_5\mu_2}^{\eta}d_{\mu_5}d_{\mu_2}N^{\eta}_{\mu_3 s}d_{\mu_3}d_{s}=\sum_{\mu_2}d_{\mu_2}^2\sum_{s}d_{s}^2\sum_{\eta}d_{\eta}^2=|G|^3$, which agrees with that contributed by $g_2,g_3$, and $g_5$  in Fig. \ref{fig:basisTrans}(a) and that contributed by $\mu_2,\mu_3$, and $\mu_5$ in Fig. \ref{fig:basisTrans}(b).
Here we used the identities \eqref{eq:quantumdim} and $\sum_a d_a^2 = |G|$.

We can go through the procedure above on all the other open vertices in Fig. \ref{fig:basisTrans}(c) and obtain the basis states in Fig. \ref{fig:basisTrans}(d), which we shall call a rep-basis state. Now the basis states are again defined on an actual trivalent lattice $\tilde\Gamma$, which differs from the original lattice $\Gamma$ by having a tail attached to each of the original vertices. The tails are necessary to maintain the correct number of local degrees of freedom. On the new lattice, each vertex can indeed be interpreted as where three representations fuse. The new lattice $\tilde\Gamma$ is in fact the right lattice for defining a LW model, such that the Hilbert space contains both ground states and charge excitations\cite{Hu2018}. We will come back to this point later when we map our Fourier-transformed model to the extended LW model in Section \ref{sec:mapToLW}.

Since the vertex and plaquette operators comprising the Hamiltonian of the extend QD model are local operators, the action of such an operator only affects a few local degrees of freedom, without affecting others. Consider a boundary vertex operator \eqref{eq:qdaOperators} acting on the vertex $v_2$ in Fig. \ref{fig:basisTrans}(a), it affects only the three group elements meeting at $v_2$. When transformed into the basis in Fig. \ref{fig:basisTrans}(d), the vertex operator acting on $v_2$ would possibly affect at most the degrees of freedom $\mu_2,\mu_3,\mu_5,\eta,s,m_s$, which are local at $v_2$. In other words, the other degrees of freedom other than these six are diagonal indices when the vertex operator is represented in the Hilbert space. Therefore, when studying a vertex operator action at a vertex, we can simply single out a \textit{local basis state} consisting of only the degrees of freedom local at the vertex.  

Let us study the Fourier transform and basis rewriting depicted in Fig. \ref{fig:basisTrans} by focusing on a local state at a boundary vertex in detail to retrieve the linear transformations of basis with precise coefficients. Consider a local basis state $\ket{g,h,l}$ drawn on the right hand side of Eq. \eqref{eq:basisFTInverse}. In our graphical presentation, Fourier-transforming this basis to one in the representation space is as trivial as in Eq. \eqref{eq:basisFTInverse}, and the inverse transformation in Eq. \eqref{eq:basisFT}. 
\be\label{eq:basisFTInverse}
\bket{\ftBasisA}=\frac{v_{\mu}v_{\nu}v_{\lambda}}{\sqrt{|G|^3}}\sum_{g,h,l} \ftBasisB \bket{\qdLocalBasisPlus{g}{h}{l}{}{}{}{}{}{}},
\ee
\be\label{eq:basisFT}
\bket{\qdLocalBasisPlus{g}{h}{l}{}{}{}{}{}{}}=\sum_{\substack{\mu,\nu,\lambda\\m_{\mu},m_{\nu},m_{\lambda}}}\frac{v_{\mu}v_{\nu}v_{\lambda}}{\sqrt{|G|^3}}\left( \ftBasisB \right)^*\bket{\ftBasisA}.
\ee

In the basis state on the left hand side of Eq. \ref{eq:basisFTInverse}, for any fixed $\mu,\nu$, and $\lambda$, the degrees of freedom are only at the ends of the three lines. We then rewrite the local basis state on the left hand side of Eq. \eqref{eq:basisFTInverse} by first fusing the two representations $\mu$ and $\nu$ via contracting their indices $m_\mu$ and $m_\nu$, resulting in a set of representations $\{\gamma\}$, Then, we can fuse a $\gamma$ with $\lambda$ by contracting $m_\gamma$ and $m_\lambda$, resulting in a set of representations $\{s\}$. This procedure yields two $3j$-symbols with a pair of indices contracted, resulting in the coefficients of the expansion on the right hand side of \eqref{eq:basisLTInvers}. This procedure produces the linear combination of the new local basis states on the right hand side of Eq. \eqref{eq:basisLTInvers} (also recall Fig. \ref{fig:basisTrans}(d)). We can denote the new local basis states at the vertex  by $\ket{\Psi_{sm_s}}$ and write down the inverse transformation in
Eq. \eqref{eq:basisLT}.    
\be\begin{aligned}\label{eq:basisLTInvers}
\bket{\ftBasisA}=\sum_{\gamma,s,m_s}v_{\gamma}v_{s}\ftBasisD\bket{\ftBasisE{\mu}{\nu}{\gamma}{s}{\lambda}{n_\mu}{n_\nu}{n_\lambda}{m_{s}}}.
\end{aligned}\ee
\be\begin{aligned}\label{eq:basisLT}
\ket{\Psi_{sm_s}}:=\bket{\ftLocalBasis{\mu}{\nu}{\gamma}{s}{\lambda}{n_\mu}{n_\nu}{n_\lambda}{m_{s}}}
=\sum_{\substack{m_{\mu},m_{\nu},m_{\lambda}}}v_{\gamma}v_s\ftBasisC\bket{\ftBasisA}.
\end{aligned}\ee 

Each black dot in Eqs. \eqref{eq:basisLTInvers} and \eqref{eq:basisLT} and in derivations hereafter represents a $3j$-symbol assigned to each vertex to ensure that the three representations can couple to trivial representation. Explanations of the notation $\ket{\Psi_{sm_s}}$ must follow. The linear transformation in Eq. \eqref{eq:basisLTInvers} rewrites only a subspace of the local Hilbert space spanned by the basis on the left hand side of the equation. The subspace is the one comprised by the degrees of freedom $m_\mu, m_\nu$, and $m_\lambda$, which are transformed into $\gamma,s$, and $m_s$. Hence, the two bases before and after the rewriting both have the same labels $\mu,\nu,\lambda, n_\mu, n_\nu$, and $n_\lambda$. Moreover, staring at the right hand side of Eq. \eqref{eq:basisLTInvers}, one can see that the two degrees of freedom $\gamma$ and $s$ cannot be both independent. If we choose $s$ independent, then $\gamma$ is determined by $\mu,\nu,\lambda$, and $s$. Therefore, we can denote the new basis after the rewriting by $\ket{\Psi_{sm_s}}$ for simplicity, while keeping all the other labels in the graph inexplicit. This simplification causes no confusion because in actual calculations, e.g., in computing an inner product of two such local basis states, i.e., $\langle \Psi'_{s'm_{s'}} \ket{\Psi_{sm_s}}$, the prime in $\Psi'$ implies that the hidden labels in $\Psi'$ should all be the primed version of those in $\Psi$. In the next subsection, we will see another advantage of this simplified notation.

Equations \eqref{eq:basisLT} and \eqref{eq:basisFT} lead to a direct linear transformation between the local basis states in the rep-basis and those in the group-basis at a boundary vertex $v$:
\be\label{eq:repLocal}
\ket{\Psi_{sm_s}}=\sum_{g,h,l\in G} \frac{v_{\mu}v_{\nu}v_{\lambda} v_{\gamma}v_s}{\sqrt{|G|^{3}}}\repBasisA\bket{\qdLocalBasisPlus{g}{h}{l}{}{}{}{}{}{}}.
\ee 
We verify in Appendix \ref{appd:proof} that the local basis states $\ket{\Psi_{sm_s}}$ indeed form a well-defined local basis by showing that it is orthonormal and complete.

\subsection{Fourier transform of the vertex operators}\label{subsec:AvFT}
We are now ready to study how a boundary vertex operator $\Ab_v^{\QD}$ acts on a local basis state  $\ket{\Psi_{sm_s}}$ in the rep-basis. In other words, we  need to find the Fourier-transformed version $\widetilde{\Ab_v^\QD}$ of  $\Ab_v^{\QD}$. Two subtleties are in order.

First, a boundary vertex operator $\Ab_{v}^{QD}$ acts on a group-basis local state $\ket{g,h,l}\in\Hil_v$ by modifying the group elements $g,h$, and $l$ on the three edges incident at $v$ via gauge transformation. When the group-basis local state is expanded in terms of the local states $\ket{\Psi_{sm_s}}$, the action of $\widetilde{\Ab_v^\QD}$ are expected to spread over all the degrees of freedom $\mu,\nu,\lambda,\gamma, s$, and $m_s$ local to the vertex $v$. Nevertheless, as we will see, the indices $\mu,\nu,\lambda,\gamma$ are all readily diagonal 
with respect to the $\widetilde{\Ab_v^\QD}$ represented in this local Hilbert space. This further renders $\ket{\Psi_{sm_s}}$ a good notation of the local basis states for $\widetilde{\Ab_v^\QD}$. 

Second, recall that a bulk vertex operator $A_v^{\QD}$ \eqref{eq:qdbuaOperators} acts as a gauge transformation averaged over the entire group $G$; it is a projector that projects out non-intertwiner states of $\mu,\nu$, and $\lambda$ but keep the intertwiner states as its $+1$ eigenstates.
A boundary vertex operator $\Ab_{v}^{QD}$ \eqref{eq:qdaOperators} however performs a gauge transformation averaged over a subgroup $K\subseteq G$. Consequently, $\widetilde{\Ab_v^\QD}$ may not project out all non-intertwiner states of $\mu,\nu$, and $\lambda$. To see this point, let us consider two extreme cases. When $K=G$, $\widetilde{\Ab_v^\QD}$ acts exactly the same as a bulk vertex operator and will project out all non-intertwiner states of $\mu,\nu$, and $\lambda$, and leaves no degrees of freedom for $s$. That is, the $s$ on the dangling edge (tail) attached to $v$ must be the trivial representation $0$ of $G$. Note that the $\gamma$, as a degree of freedom on an internal edge, is not independent; hence, if $s$ is trivial, $\gamma=\nu$. When $K={\{e\}}$, the action of $\widetilde{\Ab_v^\QD}$ is trivial and thus does not project out any states in the fusion space of $\mu,\nu$, and $\lambda$. That is, the $s$ can be any irreducible representation of $G$. Now for a generic nontrivial subgroup $K\subset G$, the action of $\widetilde{\Ab_v^\QD}$ may render $s$ taking value in certain subset of the set $L_G$ of all irreducible representations of $G$. Moreover, certain states in $V_s$ may also be projected out. Therefore, the states that survive the action of $\widetilde{\Ab_v^\QD}$ may be labeled by pairs $(s,\alpha_s)$ and collected into a set $L_A = \{(s,\alpha_s)\}$, which is to be defined shortly. The intuitive argument above can be explicitly verified as follows, using Eqs. \eqref{eq:qdaOperators} and \eqref{eq:repLocal}, and the completeness of the local rep-basis.
\be\begin{aligned}\label{eq:avFourierBasis}
&\widetilde{\Ab_v^\QD}\ket{\Psi_{sm_s}} \\
=&\sum_{\substack{\mu',\nu',\gamma'\\ \lambda',s'm_{s'}\\n_{\mu'}n_{\nu'}n_{\lambda'}}} \ket{ \Psi'_{s'm_{s'}}} \bra{ \Psi'_{s'm_{s'}}} \sum_{g,h,l\in G}\Ab_v^{QD}\ket{g,h,l}\bra{g,h,l}\Psi_{sm_s}\rangle\\ =&\sum_{\substack{\mu',\nu',\gamma'\\ \lambda',s'm_{s'}\\n_{\mu'}n_{\nu'}n_{\lambda'}}} \ket{ \Psi'_{s'm_{s'}}} \bra{ \Psi'_{s'm_{s'}}} \sum_{g,h,l\in G}\sum_{k\in K}\frac{1}{|K|}\vert k{g,kh,l}\bar k\rangle\bra{g,h,l}\Psi_{sm_s}\rangle
\\ 
=& \sum_{\substack{\mu',\nu',\gamma'\\ \lambda',s'm_{s'}\\n_{\mu'}n_{\nu'}n_{\lambda'}}} \sum_{g,h,l\in G}\sum_{k\in K} \frac{1}{|K|}\bra{ \Psi'_{s'm_{s'}}}k{g,kh,l}\bar k\rangle\bra{g,h,l}\Psi_{sm_s}\rangle
\ket{\Psi'_{s'm'_{s'}}} \\
=& \sum_{m'_s} \frac{1}{|K|}\sum_{k\in K}D^{s}_{m_sm'_s}(k)\ket{\Psi_{sm'_s}}\\
=& \sum_{m'_s}(P^s_K )_{m_s m'_s}\ket{\Psi_{sm'_s}}=P^s_K \ket{\Psi_{sm_s}},
\end{aligned}\ee
where $P^s_K := \sum_{k\in K}D^{s}_{m_sm'_s}(k)/|K|$. The full version of the above derivation using the graphic tools can be found in Appendix \ref{appd:proof}. So, indeed, $\widetilde{\Ab_v^{QD}}$ is automatically diagonalized in almost the entire local Hilbert space spanned by $\ket{\Psi_{sm_s}}$ except in the small subspace spanned by $s$ and $m_s$. Besides, $\widetilde{\Ab_v^{QD}}$ is clearly block diagonalized by $s$ but within the representation space of $s$, $\widetilde{\Ab_v^{QD}}$ is yet not diagonalized. As we know that $\widetilde{\Ab_v^{QD}}$ is a projector, the operator $P^s_K$ is actually the matrix representation of $\widetilde{\Ab_v^{QD}}$ on the local Hilbert subspace $\Hil_v^s$ spanned by the states $\ket{\Psi_{sm_s}}$ for a given $s\in L_G$. A linear transformation can be applied to diagonalize the matrix $P^s_K$, which transforms the states $\ket{\Psi_{sm_s}}$ too. Let us rename the states that diagonalize $P^s_K$ by $\ket{\Psi_{s \tilde m_s}}$. Hence,
\be\label{eq:AbProjector}
\widetilde{\Ab_v^{QD}} \ket{\Psi_{s\tilde m_s}} = P^s_K \ket{\Psi_{s \tilde m_s}} =: \delta_{(s,\tilde m_s)\in L_A} \ket{\Psi_{s \tilde m_s}}.
\ee
Note that for certain $s\in L_G$, $P^s_K$ may project out the entire $\Hil_v^s$; therefore, the set $L_A$ collects all the +1 eigenstates of $\widetilde{\Ab_v^{QD}}$ or its representation $P^s_K$ for all possible $s$ that is not annihilated by $P^s_K$. That is,
\be\label{eq:LA}
L_A = \{(s,\alpha_s) |P^s_K \ket{\Psi_{s \alpha_s}}= \ket{\Psi_{s \alpha_s}},  s\in L_G\}.
\ee
The states $\ket{\Psi_{s \tilde m_s}}$ with $(s,\tilde m_s)\notin L_A$ are zero eigenstates of $\widetilde{\Ab_v^{QD}}$; they have higher energy than the $+1$ eigenstates according to the Hamiltonian \eqref{eq:eQDbdryHam} and thus are excited states. The excitations appear at the end of the dangling edge graced with representations $s$ and are the point-like charge excitations at the boundary.

\subsection{Fourier transform of the plaquette operators}\label{subsec:BpFT}

We can now proceed to check how a boundary plaquette operator $\Bb_p^{\QD}$ \eqref{eq:qdbOperators} of the extended QD model should be Fourier-transformed to act on a local basis state in the rep-basis. According to Eq. \eqref{eq:qdbOperators}, a $\Bb_p^{\QD}$ does not act on a vertex but on the edge connecting two vertices. Hence, in the rep-basis, a local basis state on which a $\widetilde{\Bb_p^{\QD}}$ acts would involve two neighbouring vertices and the local degrees of freedom associated with these two vertices. For example, if a $\Bb_p^{\QD}$ acts on the edge with $g_2$ in Fig. \ref{fig:basisTrans}(a), then in Fig. \ref{fig:basisTrans}(d), the operator $\widetilde{\Bb_p^{\QD}}$ would act on the degrees of freedom local to the vertices $v_1$ and $v_2$, and all other degrees of freedom correspond to diagonal indices when $\widetilde{\Bb_p^{\QD}}$ is represented in the Hilbert space. Hence, following the same logic as that of constructing the local basis states for the boundary vertex operators, we can denote the local basis states to be acted on by a boundary plaquette operator by $\ket{\Psi^{\eta\lambda}_{r\tilde m_r;s\tilde m_s}}$, defined as follows.
\be\label{eq:bpBasis}
\ket{\Psi^{\eta\lambda}_{r\tilde m_r; s\tilde m_s}} :=\bket{\bpTensorBasis{\mu}{\nu}{\gamma}{r}{\lambda}{\kappa}{\eta}{\rho}{s}},
\ee
where the indices $\tilde m_r$ and $\tilde m_s$ diagonalize the boundary vertex operators acting on $v_1$ and $v_2$ as in Eq. \eqref{eq:AbProjector}. In Eq. \eqref{eq:qdbOperators}, the boundary plaquette operator $\Bb_p^{\QD}$acts on a virtual boundary plaquette $p$, but now in the rep-basis local state \eqref{eq:bpBasis}, it acts on a real, open plaquette $p$ outlined by the edges with degrees of freedom $\tilde m_s, s, \eta, \lambda, r$, and $\tilde m_r$.

It is customary to directly compute the matrix elements of a $\widetilde{\Bb_p^{\QD}}$ in the local basis states:
\be\begin{aligned}\label{eq:bpOperator}
&\langle\Psi^{\eta'\lambda'}_{r'\tilde{m}_{r'};s' \tilde{m}_{s'}} \vert \widetilde{\Bb_p^{\QD}} \vert \ \Psi^{\eta\lambda}_{r\tilde{m}_{r};s \tilde{m}_{s}}\rangle
\\
=&\!\sum_{g,h,l,x,y}\! \langle\Psi^{\eta'\lambda'}_{r'\tilde{m}_{r'};s' \tilde{m}_{s'}} \vert\Bb^{QD}_p\vert g,h,l,x,y\rangle\langle g,h,l,x,y| \Psi^{\eta\lambda}_{r\tilde{m}_{r};s\tilde{m}_{s}}\rangle
\\
=&\sum_{g,h,l,x,y}
\sum_{(t,\alpha_t)\in L_A}\frac{|K|}{|G|}d_{t}D^{t}(l)_{\alpha_t\alpha_t}\langle\Psi^{\eta'\lambda'}_{r'\tilde{m}_{r'};s' \tilde{m}_{s'}} \vert g,h,l,x,y\rangle\langle g,h,l,x,y|\Psi^{\eta\lambda}_{r\tilde{m}_{r};s\tilde{m}_{s}}\rangle \\
=&\sum_{(t,\alpha_t)\in L_A}\sum_{\tilde n_{r'^*},\tilde n_{t^*},\tilde n_{s'^*}}\frac{1}{d_A}\tilde d_t \tilde v_\eta \tilde v_s \tilde v_r \tilde v_\lambda \tilde v_{\lambda'} \tilde v_{\eta'} \tilde v_{s'} \tilde v_{r'} (w^{rtr'^*}_{\tilde m_{r}\alpha_t \tilde n_{r'^*}})^*(\Omega^{r'})^{-1}_{\tilde n_{r'^*}\tilde m_{r'}}
\\
&\times(w^{t^*s'^*s}_{\tilde n_{t^*}\tilde n_{s'^*}\tilde m_s})^*(\Omega^{s'^*})^{-1}_{\tilde m_{s'}\tilde n_{s'^*}}(\Omega^{t^*})^{-1}_{\alpha_{t}\tilde n_{t^*}}G^{\rho^*\eta s^*}_{t^*s'^*\eta'}G^{\kappa\lambda\eta^*}_{t^*\eta'^*\lambda'}G^{\gamma r^*\lambda^*}_{t^*\lambda'^* r'^*}.
\end{aligned}\ee 
Here, $\tilde v_j = \sqrt{\tilde d_j}$ with $\tilde d_j = \beta_j d_j$, and the coefficients $G^{abc}_{def}$ are the symmetric $6j$-symbols of the irreducible representations of the group $G$; their properties can be found in Appendix \ref{appd:6j} or in Ref.\cite{Hu2013a}. The full derivation of the equation above is found in Appendix \ref{appd:proof}. The local basis states $\vert \ \Psi^{\eta\lambda}_{r\tilde{m}_{r};s \tilde{m}_{s}}\rangle $ may be $+1$ or zero eigenstate of the boundary vertex operators acting on the relevant vertices. It is useful to study the matrix elements of $\widetilde{\Bb_p^{\QD}}$ in the local basis states free of any charge excitations, such that boundary pure flux excitations can be identified. To this end, one can act the boundary vertex operators $\widetilde{\Ab_{v_1}^{QD}}$ and $\widetilde{\Ab_{v_2}^{QD}}$ on the states $\vert \ \Psi^{\eta\lambda}_{r\tilde{m}_{r};s \tilde{m}_{s}}\rangle $ to project out all the charge excitations. Equivalently, we can simply replace the indices $\tilde m_s$ and $\tilde m_r$ in Eq. \eqref{eq:bpOperator} by $\alpha_s$ and $\alpha_r$ in Eq. \eqref{eq:LA} and obtain
\be\label{eq:bdBpMatrixNoCharge}
\begin{aligned}
& \langle\Psi^{\eta'\lambda'}_{r'\alpha_{r'};s' \alpha_{s'}} \vert\widetilde{\Bb^{QD}_p}\vert \ \Psi^{\eta\lambda}_{r\alpha_{r};s \alpha_{s}}\rangle \\
=&\sum_{(t,\alpha_t)\in L_A}\sum_{\tilde n_{r'^*},\tilde n_{t^*},\tilde n_{s'^*}}\frac{1}{d_A}\tilde d_t \tilde v_\eta \tilde v_s \tilde v_r \tilde v_\lambda \tilde v_{\lambda'} \tilde v_{\eta'} \tilde v_{s'} \tilde v_{r'} (w^{rtr'^*}_{\alpha_{r}\alpha_t \tilde n_{r'^*}})^*(\Omega^{r'})^{-1}_{\tilde n_{r'^*}\alpha_{r'}}
\\
&\times(w^{t^*s'^*s}_{\tilde n_{t^*}\tilde n_{s'^*}\alpha_s})^*(\Omega^{s'^*})^{-1}_{\alpha _{s'}\tilde n_{s'^*}}(\Omega^{t^*})^{-1}_{\alpha_{t}\tilde n_{t^*}}G^{\rho^*\eta s^*}_{t^*s'^*\eta'}G^{\kappa\lambda\eta^*}_{t^*\eta'^*\lambda'}G^{\gamma r^*\lambda^*}_{t^*\lambda'^* r'^*} \\ 
=& \sum_{(t,\alpha_t)\in L_A}\frac{1}{d_A}\tilde v_{t}\tilde v_{\eta}\tilde u_{s}\tilde u_{r}\tilde v_{\lambda}\tilde v_{\lambda'}\tilde v_{\eta'}\tilde u_{s'}\tilde u_{r'}G^{\rho^* \eta s^*}_{t^*s'^*\eta'}G^{\kappa \lambda \eta^*}_{t^*\eta'^*\lambda'}G^{\gamma r^* \lambda^*}_{t^*\lambda'^*r'^*} \\
&\x f_{t^{*}\alpha_{t} s'^*\alpha_{s'} s\alpha_s}  f_{r\alpha_rt\alpha_t r'^{*}\alpha_{r'} },
\end{aligned}
\ee
where in the last equality we define
\be\label{eq:fCorrespondence}
f_{a\alpha_ab\alpha_bc^{*}\alpha_c}=\sum_{\tilde n_{c^*}}\tilde  u_a\tilde u_b\tilde u_c(w^{abc^{*}}_{\alpha_a\alpha_b\tilde n_{c^*}})^*(\Omega^{c})^{-1}_{\tilde n_{c^*}\alpha_c},
\ee
with $(a,\alpha_a), (b,\alpha_b)$, and  $(c,\alpha_c)$ being elements in the $L_A$ defined in Eq. \eqref{eq:LA}. 

We have successfully Fourier transformed and rewritten the extended QD model on a trivalent lattice $\tilde\Gamma$. In the following sections, we shall study the physical consequences.  

\section{Emergence of Frobenius algebras and anyon condensation} \label{sec:FA}
Interestingly, the Fourier transform and rewriting of the extended QD model leads to an emergent Frobenius algebra structure on the boundary of $\tilde\Gamma$. Namely, the set $L_A$ defined in Eq. \eqref{eq:LA} together with the symbols $f$ defined in Eq. \eqref{eq:fCorrespondence} form a Frobenius algebra, as an object in the UFC $\rep_G$---the category of linear representations of $G$, which is the tensor category generated by $L_G$---the set of irreducible representations of $G$.

Before we show the emergent Frobenius algebra, let us review what Frobenius algebras are. A Frobenius algebra $A$ is a pair $(L_A,f)$, where $L_A$ is the set of elements of $A$, and $f$ is the multiplication. An element of $L_A$ is a pair $(s, \alpha_s)$ (or just $s\alpha_s$ for short), where $s$ is a simple object of a UFC $\F$. We denote the number of different pairs $(s, \alpha_s)$ with the same $s$ by $|s|$ and call it the multiplicity of $s$ in $A$. The multiplication $f$ is a map : $L_A\times L_A\times L_A\rightarrow \mathbb{C}$ that satisfies
the following associativity and non-degeneracy. 
\begin{subequations}\label{eq:faCondi}
\be\label{eq:faAssoc}
\sum_{c\alpha_c}f_{a\alpha_a b\alpha_b c^*\alpha_{c}} f_{c\alpha_cr\alpha_rs^* \alpha_{s}} G^{abc^*}_{rs^*t} \tilde v_c \tilde v_t = \sum_{\alpha_t}f_{a\alpha_at\alpha_ts^*\alpha_{s}}f_{b\alpha_br\alpha_rt^*\alpha_{t}},
\ee
\be\label{eq:faNonDegen}
f_{b\alpha_bb'^*\alpha_{b'}\ 0} = \delta_{b,b'}\delta_{\alpha_b,\alpha_{b'}}\beta_b,
\ee
\end{subequations}
where $0$ is the unit element of $A$ and has multiplicity $1$, i.e. $0=(0,1)$. Here, $\tilde v_c = \sqrt{\tilde d_c}$, which is defined in Appendix \ref{appd:6j}.
 
That the Frobenius algebra $A$ defined above is an object of the corresponding UFC $\F$ is understood by writing $A$ as $A = \bigoplus_{s|_{(s,\alpha_s)\in L_A}} s^{\oplus |s|}$, which is in general a non-simple object in $\F$. For the sake of computation, one may also write $A = \bigoplus_{(s,\alpha_s) \in L_A} s_{\alpha_s}$, explicitly taking different appearances of $s$ as distinct elements of $A$. 

We can show that the symbols $f_{a\alpha_ab\alpha_bc\alpha_c}$ defined in Eq. \ref{eq:fCorrespondence} indeed satisfy the defining conditions \eqref{eq:faCondi} of a Frobenius algebra. We first prove the associativity \eqref{eq:faAssoc} as follows.  
\begin{align}
& \sum_{c\alpha_c}f_{a\alpha_a b\alpha_b c^*\alpha_{c}} f_{c\alpha_cr\alpha_rs^*\alpha_{s}} G^{abc^*}_{rs^*t}\tilde v_c \tilde v_t \nonumber\\
=& \sum_{c\alpha_c}\sum_{\tilde n_{c^*},\tilde n_{s^*}}\tilde d_c \tilde u_a \tilde u_b \tilde u_r \tilde u_s \tilde v_t \left( w^{abc^*}_{\alpha_a\alpha_b\tilde n_{c^*}} \right)^*(\Omega^{c})^{-1}_{\tilde n_{c^*}\alpha_c}\left( w^{crs^*}_{\alpha_c\alpha_r\tilde n_{s^*}} \right)^*(\Omega^{s})^{-1}_{\tilde n_{s^*}\alpha_s}G^{abc^*}_{rs^*t}
\nonumber\\
=&\sum_{c}\tilde d_c \tilde u_a \tilde u_b \tilde u_r \tilde u_s \tilde v_t \beta_s\assoCondA{a}{b}{c}{r}{s}{t}=\sum_{c}\sum_{t\in G}\frac{\tilde d_c \tilde u_a \tilde u_b \tilde u_r \tilde u_s \tilde v_t }{|G|}\beta_r\assoCondBB{a}{b}{c}{r}{s}{t}
\nonumber\\
=&\sum_{c}\tilde d_c \tilde u_a \tilde u_b \tilde u_r \tilde u_s \tilde v_t \beta_r \assoCondB{a}{b}{c}{r}{s}{t} =\tilde u_a \tilde u_b \tilde u_r \tilde u_s \tilde v_t \assoCondC{a}{b}{r}{s}{t} \nonumber \\
=& \sum_{\alpha_t}\sum_{\tilde n_{t^*},\tilde n'_{s^*}}\tilde u_a \tilde u_b \tilde u_r \tilde u_s \tilde v_t\left( w^{ats^*}_{\alpha_a\alpha_t\tilde n'_{s^*}} \right)^*(\Omega^{s})^{-1}_{\tilde n'_{s^*}\alpha_s}\left( w^{brt^*}_{\alpha_b\alpha_r\tilde n_{t^*}} \right)^* (\Omega^{t})^{-1}_{\tilde n_{t^*}\alpha_t} \nonumber \\
=&  \sum_{\alpha_t} f_{a\alpha_at \alpha_ts^* \alpha_{s}} f_{b\alpha_br \alpha_rt^*\alpha_{t}}.\label{eq:assoCond}
\end{align}
In the first equality above, we simply rewrite the left hand side using the definition \eqref{eq:fCorrespondence}. The result is then graphically presented via the second equality. The third and fourth equalities are due to Eq. \eqref{eq:colsedGraph}, while the fifth equality due to Eq. \eqref{eq:ortho}.
The last equality is again rewriting using definition \eqref{eq:fCorrespondence}. To prove the non-degeneracy condition \eqref{eq:faNonDegen}, note that since $\sum_{\tilde n_{c^*}} (w^{0bc^{*}}_{0\alpha_b\tilde n_{c^*}})^{*}(\Omega^c)^{-1}_{\tilde n_{c^*}\alpha_c}\propto \delta_{b,c}\delta_{\alpha_b,\alpha_c}$ due to Schur's lemma, $\det(\sum_{\tilde n_{c^*}} (w^{0bc^{*}}_{0\alpha_b\tilde n_{c^*}})^{*}(\Omega^c)^{-1}_{\tilde n_{c^*}\alpha_c})\neq0$\ . 

Therefore, for any finite group $G$ and a subgroup $K\subseteq G$, the set $L_A$ defined in Eq. \eqref{eq:LA} equipped with the multiplication $f$ defined by Eq. \eqref{eq:fCorrespondence} indeed form a Frobenius algebra $A_{G,K} = (L_A,f)_{G,K}$.

Recall that according to Eq. \eqref{eq:LA}, each pair $(s,\alpha_s)$ labels a local $+1$ eigenstate $\ket{\Psi_{s\alpha_s}}$ at vertex $v$ of $\widetilde{\Ab_v^\QD}$, which is a projector. All such eigenstates sharing the same hidden labels span a subspace in the $d_s$-dimensional representation space $V_s$. The dimension $|s|$ of this subspace is the number of pairs $(s,\alpha_s)$ with the same $s$. A salient point of our recognition of an emergent Frobenius algebra out of the Fourier transform of the extended QD model with gauge group $G$ is that it identifies this dimension $|s|$  as the multiplicity $|s|$ of $s$ appearing in $A$ (see above Eq. \eqref{eq:faAssoc}). This identification is intricately related to the mechanism of anyon condensation in topological phases. Here we briefly describe this relation and shall report the detailed studies elsewhere. 

Anyon condensation has been extensively studied recently (see Refs.\cite{Bais2009,Kong2013,Gu2014a,Burnell2018} and references therein). In a topological phase $\C$, certain types of anyons may condense and cause a phase transition that breaks the topological phase to a simpler child topological phase $\U$. In an extreme case, $\U$ may be merely a vacuum, and the original topological phase is said to be completely broken. An alternative and equivalent perspective is that there is a gapped domain wall separating $\C$ and $\U$. In particular in the case where $\U$ is a vacuum, we say certain types of anyons of $\C$ can move to and condense at the gapped boundary between $\C$ and the vacuum. An interesting phenomenon is that certain types of anyons do not condense at the boundary straightforwardly; rather, such an anyon may split into a number of \textit{pieces} at the boundary, and not all of these pieces can and necessarily condense. If an anyon splits into two pieces, it is said to have multiplicity one (two) in the condensate if only one (both) of the two pieces condense. Anyon splitting only occurs to anyons with quantum dimension greater than or equal to $2$. So far, the understanding of anyon splitting is categorical; hence, it would be important to understand such splitting in a concrete lattice model of topological phases with gapped boundaries in terms of the input data of the model. 

For the extended QD model with input gauge group $G$,  the gapped boundary conditions are specified by the subgroups of $K\subseteq\ G$. It is known that $K = \{e \}$ corresponds to condensing all the $G$-charges at the boundary, $K=G$ corresponds to condensing all the $G$-fluxes, and any $K$ in between corresponds to condensing certain types of dyons. Nevertheless, by the subgroup $K$ alone one cannot immediately tell whether a type of condensed anyons should have a multiplicity greater than one in the condensate. In the Fourier-transformed picture, as we now show, anyon splitting and multiplicity becomes lucid. According to our discussion earlier in this subsection, a subgroup $K$ on the boundary of $\Gamma$ gives rise to a Frobenius algebra $A_{G,K} = (L_A,f)_{G,K}$ at the boundary of $\tilde\Gamma$. Consider the case where $K=\{e\}$, i.e., charge condensation at the boundary. Then by Eqs. \eqref{eq:AbProjector} and \eqref{eq:LA}, all irreducible representations must appear in $L_A$. Namely, for any $s\in L_G$, the set of all irreducible representations of $G$, we have $\{(s,\alpha_s)| \alpha = 1,2,\dots, d_s\} \subset L_A$. Since $s$ labels a type of pure charge excitations, and since each pair $(s,\alpha_s)$ is an independent element of $A_{G,K}$, the pure charge $s$ splits into $d_s$ pieces, each of which condenses at the boundary. Thus, the multiplicity of the charge $s$ in the boundary condensate is $d_s = |s|$, the multiplicity of $s$ in the Frobenius algebra $A_{G,K}$. In the case where $K$ is a nontrivial subgroup, we may have for some $s$, only a set $\{(s,\alpha_s)| \alpha = 1,2,\dots, |s|<d_s\} \subset L_A$. That is although the pure charge $s$ splits into $d_s$ pieces, only $|s|$ pieces of them contribute to the boundary condensate. In Section \ref{subsec:exS3}, we shall see both possibilities in a concrete example.

The Frobenius algebra $A_{G,K}$ could be understood more intuitively in the language of group algebras $\mathds C[G]$ and $\mathds{C}[K]$. ( $\mathds{C}[G]$ is spanned by the group elements $\{g\in G\}$, and similarly for $K$.) To see this, the Frobenius algebra $A_{G,\{e\}}$ is in fact identified with the \textit{canonical} Frobenius algebra in $\rep_G$, namely, $A_{\mathrm{canonical}}= \mathds{C[}G]^*=\mathrm{Fun}(G,\mathds{C})$ with multiplication $f\ox f'\rightarrow ff'$ (as a multiplication of functions), and counit $f\mapsto \sum_{g\in G}f(g)$. The irreducible representations $\{\rho^j_{mn}\}$ form a basis of $\mathds{C[}G]^*$. In terms of representations, $\mathds{C[}G]^*$ is a left regular representation and can be decomposed into irreducible representations in the form
\begin{equation}\label{eq:FAcanonical}
A_{\mathrm{canonical}}=\bigoplus_{\mu\in L_G}V_{\mu}^{\oplus d_\mu},
\end{equation}
with the multiplication being the $3j$-symbol  (up to some normalization factor) as defined in Eq. \eqref{eq:fCorrespondence}.

Similarly, the Frobenius algebra $A_{G,K}$ can be identified with $(\mathds{C}[G]/ \mathds{C}[K])^*$---the dual space of the quotient $\mathds{C}[G]/\mathds{C}[K]$---which is the space of all functions $f$ over $G$ such that $f(g)=f(h)$ if $gk=h$ or $kg=h$ for some $K$. The space $(\mathds{C}[G]/\mathds{C}[K])^*$ is a subalgebra of $\mathds{C}[G]^*$ and is also a Frobenius algebra with the algebra structure inherited from $\mathds{C}[G]^*$. By solving the eigen-equation $\rho(k)=\id$, we obtained the new basis $\{\rho^j_{\alpha'\alpha}\}$, and in terms of irreducible representations we have
\begin{equation}\label{eq:FAcanonicalK}
A_{G,K}=\bigoplus_{(s,\alpha_s)\in L_A}V_{s}^{\alpha_s},
\end{equation}
where $V_{s}^{\alpha_s}$ denotes the direct sum $V_s$ that appears in $A_{G,K}$ with the multiplicity index $\alpha_s$. The multiplication is also the $3j$-symbol (up to some normalization factor) as defined in Eq. \eqref{eq:fCorrespondence}. By the definition of $L_A$ in Eq. \eqref{eq:AbProjector}, $\alpha_s$ is given by a basis vector such that $\rho^s(k)\ket{s,\alpha_s}=\ket{s,\alpha_s}$ for all $k\in K$. The space $A_{G,K}$ spanned by $\ket{s,\alpha_s}$ is thus identified with  $(\mathds{C}[G]/\mathds{C}[K])^*$ defined above.

\section{EM duality in the extended QD model}\label{sec:duality}

\subsection{EM duality in the bulk}

The QD model exhibits an EM duality. Consider the QD models with finite Abelian groups $G$. In such a case, all the irreducible representations of $G$ are 1-dimensional and form a group, with the group multiplication defined as the tensor product of representations. For example, the irreducible representations of $G=\Z_n$ form the group $\rep_G=\Z_n$. Let $g\in\{0,1,\dots,n-1\}$ and $g\cdot h =g+h\mod n$. Then the irreducible representations are $j\in\{0,1,\dots,n-1\}$ and $j\cdot k=j+k\mod n$ in the sense that $\rho^j(g)\rho^k(g)=\rho^{j\cdot k}(g)$ for all $g\in G$.

\begin{figure}[h!]
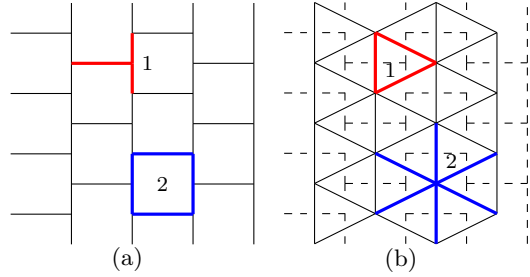

        \centering
        \subfigure[]{\dualGraph\label{fig:dualGraph}}
        \subfigure[]{\dualGraphAA\label{fig:dualGraphAA}}
        \caption{A closed trivalent graph $\Gamma$ (a) and its dual graph $\Gamma^*$ (b). A vertex $v$ (a plaquette $p$) of $\Gamma$ becomes a plaquette $p^*$ (a vertex $v^*$) of $\Gamma^*$.}
\end{figure}

As reviewed in Section \ref{sec:rev}, the QD model on a closed trivalent graph $\Gamma$ consists of vertex and plaquette operators. For example, in Fig. \ref{fig:dualGraph}, we have an operator at vertex $v=1$:
\begin{equation}\label{eq:dualGraphAvBefore}
A_v\ket{g_1,g_2,g_3}=\frac{1}{|G|}\sum_{h\in G}\ket{h\cdot g_1,h\cdot g_2,h\cdot g_3},
\end{equation}
where $g_1,g_2,g_3\in G$ are the group elements on the three edges meeting at $v$. At plaquette $p=2$, we have
\begin{equation}\label{eq:dualGraphBpBefore}
B_p\ket{g'_1,g'_2,g'_3,g'_4,g'_5,g'_6}=\delta_{g'_1\cdot g'_2\cdot g'_3\cdot g'_4\cdot g'_5\cdot g'_6,0}\ket{g'_1,g'_2,g'_3,g'_4,g'_5,g'_6},
\end{equation}
where the $g'$s are the group elements on the edges outlining the plaquette $p$.

We also Fourier transform the basis of Hilbert space $\Hil_\Gamma$ on $\Gamma$ from the group space to the representation space. Hence, the vertex and plaquette operators act on the new basis states as  
\begin{equation}\label{eq:dualGraphAvAfter}
A_v\ket{j_1,j_2,j_3}_{v=1}=\delta_{j_1\cdot j_2\cdot j_3,0}\ket{ j_1 ,j_2, j_3}_{v=1},
\end{equation}
where $j_1,j_2,j_3\in\rep G$ are on the three edges meeting at the vertex $v=1$ in $\Gamma$, and
\begin{equation}\label{eq:dualGraphBpAfter}
B_p\ket{j'_1,j'_2,j'_3,j'_4,j'_5,j'_6}_{p=2}=\frac{1}{|G|}\sum_{k\in \rep_G}\ket{k\cdot j'_1,k\cdot j'_2,k\cdot j'_3,k\cdot j'_4,k\cdot j'_5,k\cdot j'_6}_{p=2},
\end{equation}
where the $j'$'s are graced
on the six edges outlining the plaquette $p=2$ on $\Gamma$. 

On the other hand, we can draw the dual lattice $\Gamma^*$ (Fig. \ref{fig:dualGraphAA}) of $\Gamma$ and place an element of $\Gamma^*$ on each edge of $\Gamma^*$. Since $\rep_G\cong G$, $\Hil_{\Gamma^*}\cong\Hil_\Gamma$. The dual QD model with $\rep_G$ is naturally defined on $\Gamma^*$. The basis states $\ket{j_1,j_2,j_3}_{v=1}$ and $\ket{j'_1,j'_2,j'_3,j'_4,j'_5,j'_6}_{p=2}$ on $\Gamma$ are the same as the basis states $\ket{j_1,j_2,j_3}_{p^*=1}$ and $\ket{j'_1,j'_2,j'_3,j'_4,j'_5,j'_6}_{v^*=2}$ on $\Gamma^*$. Hence, we have
\begin{align}
 B^*_{p^*} \ket{j_1,j_2,j_3}_{p^*=1} &= \delta_{j_1\cdot j_2\cdot j_3,0} \ket{ j_1 ,j_2, j_3}_{p^*=1} , \label{eq:dualBp} \\
 A^*_{v^*}\ket{j'_1,j'_2,j'_3,j'_4,j'_5,j'_6}_{v^*=2} &= \frac{1}{|G|} \sum_{k\in \rep_G}\ket{k\cdot j'_1,k\cdot j'_2,k\cdot j'_3,k\cdot j'_4,k\cdot j'_5,k\cdot j'_6}_{v^*=2}, \label{eq:dualAv}
\end{align}
where $B^*_{p^*}$ and $A^*_{v^*}$ are the plaquette and vertex operators of the dual QD model. Comparing Eqs. \eqref{eq:dualGraphAvAfter} and \eqref{eq:dualGraphBpAfter} to \eqref{eq:dualBp} and \eqref{eq:dualAv}, and since the above discussion applies to any vertex and plaquette on $\Gamma$, we conclude that the QD model with a group $G$ on $\Gamma$ is mapped to the dual QD model with a group $\rep_G$ on $\Gamma^*$, with $\mathcal{H}_{\Gamma}\cong \mathcal{H}_{\Gamma^*}$, and
\begin{equation}\label{eq:dualGraphAABB}
A_v=B^*_{p^*}, B_p = A^*_{v^*}.
\end{equation}

Since $A_v$ measures the gauge charges (which have an electric nature) and $B_p$ measures the gauge fluxes (which have a magnetic nature), the above duality is an electric-magnetic duality: an electric charge/magnetic flux on $\Gamma$ is mapped to a magnetic flux/electric charge on $\Gamma^{*}$.

For non-Abelian cases, $\rep_G$ is no longer a group. In a generalized context using quantum groups (Hopf algebra), we can still define the charges and fluxes of $\rep_G$, and thus are able to study the EM duality\cite{Buerschaper2013} in the QD model on a closed surface. In Ref.\cite{Hu2018}, the authors has also checked the EM duality in the enlarged Hilbert space of the LW model on a closed surface.

\subsection{EM duality on the boundary}

\begin{figure}[h!]
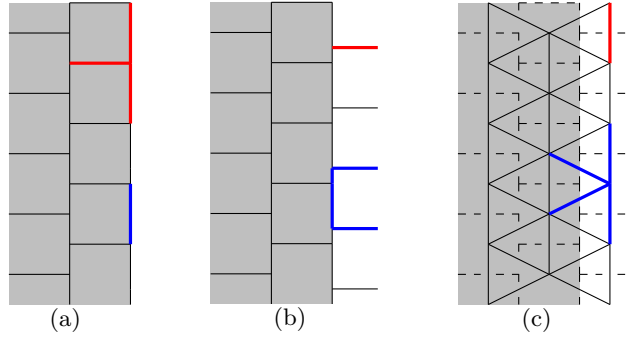

        \centering
        \subfigure[]{\dualGraphBoundaryAA\label{fig:dualGraphBoundaryAA}}\qquad
        \subfigure[]{\dualGraphBoundaryAB\label{fig:dualGraphBoundaryAB}}\qquad
        \subfigure[]{\dualGraphBoundaryAC\label{fig:dualGraphBoundaryAC}}
        \caption{(a) a trivalent lattice $\Gamma$ with a boundary, (b) its extension $\tilde\Gamma$, and (c) the dual lattice $\tilde\Gamma^*$ of $\tilde\Gamma$. Bulk is in grey and to the left of the boundary. In the extended lattice $\tilde\Gamma$, the dangling edges in the bulk (see Fig. \ref{fig:basisTrans}(d)) are neglected because we focus on the boundary. A boundary vertex (edge) in $\Gamma$ is highlighted in red (blue). A boundary vertex (edge) in $\Gamma$ becomes a dangling edge (open plaquette with four edges) in $\tilde\Gamma$ and then becomes a boundary edge (a four-valent boundary vertex) on $\tilde\Gamma^*$.}
\end{figure}

The EM duality in the bulk can be extended to one that is on the boundary. We again consider the Abelian cases first. As reviewed in Section \ref{sec:rev}, on the boundary, given a subgroup $K$, there are two types of operators, $\bar A_v$ acting on the boundary vertices, and $\bar B_p$ on the boundary edges, as illustrated in Fig. \ref{fig:dualGraphBoundaryAA}.

\begin{equation}\label{eq:dualityBoundaryAv}
\bar A_v \ket{g_1,g_2,g_3}=\frac{1}{|K|}\sum_{k\in K}\ket{k\cdot g_1,k\cdot g_2,k\cdot g_3},
\end{equation}
\begin{equation}\label{eq:dualityBoundaryBp}
\bar B_p\ket{g}=\delta_{g\in K}\ket{g}.
\end{equation}
The $\bar A_v$ measures the $K$-charges, while $\bar B_p$ measures the $G/K$-fluxes.

To reveal the EM duality on the boundary, we need to extend the lattice $\Gamma$ in Fig \ref{fig:dualGraphBoundaryAA} to the lattice $\tilde\Gamma$ in \ref{fig:dualGraphBoundaryAB}, following the procedure illustrated in Fig. \ref{fig:basisTrans}. As explained in Section \ref{subsec:HilbertFT}, this extension preserves the Hilbert space of the extended QD model, i.e., $\Hil_\Gamma=\Hil_{\tilde\Gamma}$, and is merely a basis transformation. That is, the basis of the Hilbert space is transformed from the group space to the representation space. 
%
%\begin{figure}[h!]
%        \centering
%        \dualGraphTailExtension
%        \caption{ A trivalent vertex is extended to have an open link. The %new d.o.f. are not independent of the original ones, as they are constraint %by $g_4=g_1\cdot g_2,g_5=g_3\cdot g_4$.}
%\label{fig:latticeTailExtension}
%\end{figure}

 Then according to Eqs. \eqref{eq:qdaOperators} and \eqref{eq:qdbOperators} and the basis transformation, a boundary vertex operator $\bar A_v$ and plaquette operator $\bar B_p$ acts respectively on the highlighted dangling edge and open plaquette on $\tilde\Gamma$ in Fig. \ref{fig:dualGraphBoundaryAB} as
\begin{equation}\label{eq:boundaryAvAfter}
\bar A_v\ket{j}_v=\delta_{j\in L_{A}}\ket{j}_v,
\end{equation}
where $j\in\rep_G$ is graced on the red dangling edge in Fig. \ref{fig:dualGraphBoundaryAB}, and
\begin{equation}\label{eq:boundaryBpAfter}
\bar B_p\ket{j_1,j_2,j_3,j_4}_p=\frac{1}{|L_A|}\sum_{i\in L_A}\ket{i\cdot j_1,i\cdot j_2,i\cdot j_3, i\cdot j_4}_p,
\end{equation}
where $j_1,j_2,j_3,j_4\in \rep_G$ are respectively placed on the four blue edges outlining the open plaquette in Fig. \ref{fig:dualGraphBoundaryAB}, and  $L_A$ is a subgroup of $\rep_G$ defined by
\begin{equation}\label{eq:dualDefLA}
L_A=\{j\in \rep_G|\rho^j(k)=1,\forall k\in K\}.
\end{equation}

We then draw in Fig. \ref{fig:dualGraphBoundaryAC} the dual lattice $\tilde\Gamma^*$ of the extended lattice $\tilde\Gamma$. This dual lattice $\tilde\Gamma^*$ is a triangular lattice with a boundary and naturally defines a dual extended QD model with $\rep_G$ being the gauge group. The dual boundary operators $\Bb^*_{p*}$ and $\Ab^*_{v^*}$ act on the dual boundary edges and vertices as
\begin{align}
\Bb^*_{p*} \ket{j}_{p^*} &= \delta_{j\in L_{A}}\ket{j}, \label{eq:dualBdryBp} \\
\Ab^*_{v^*} \ket{j_1,j_2,j_3,j_4}_{v^*} &= \frac{1}{|L_A|}\sum_{i\in L_A}\ket{i\cdot j_1,i\cdot j_2,i\cdot j_3, i\cdot j_4}_{v^*}. \label{eq:dualBdryAv}
\end{align}
Comparing Eqs. 
\eqref{eq:boundaryAvAfter} and \eqref{eq:boundaryBpAfter} to \eqref{eq:dualBdryBp} and \eqref{eq:dualBdryAv}, we conclude that the extended QD model with a group $G$ on $\Gamma$ is mapped to the dual extended QD model with the group $\rep_G$ on $\tilde\Gamma^*$, with $\mathcal{H}_{\Gamma}\cong \mathcal{H}_{\tilde\Gamma^*}$, and
\begin{equation}\label{eq:dualGraphBoundaryAABB}
\Ab_v=\Bb^*_{p^*}, \Bb_p=\Ab^{*}_{v^*}.
\end{equation}
The $\bar A_v$ measuring $K$-charges is mapped to $\Bb^*_{p^*}$ measuring $(\rep_G/L_A)$-fluxes, and $\bar B_p$ measuring $G/K$-fluxes to $\Ab^{*}_{v^*}$ measuring $L_A$-charges. Although we consider the Abelian examples here, the duality \eqref{eq:dualGraphBoundaryAABB} can be examined for an arbitrary finite group, in the language of Frobenius algebras. After the Fourier transformation, Eqs. \eqref{eq:boundaryAvAfter} and \eqref{eq:boundaryBpAfter} become Eqs. \eqref{eq:AbProjector} and \eqref{eq:bdBpMatrixNoCharge}, expressed in terms of the Frobenius algebra $A$. They can be viewed as the operators that measure the $(\rep_G/L_A)$-fluxes and $L_A$-charges in a more generalized context.

\section{Mapping the Fourier-transformed extended QD model to the extended LW model} \label{sec:mapToLW}

The Fourier transformed extended QD model with a finite gauge group $G$ can actually be regarded as an extended LW model with a input unitary fusion category whose simple objects are the irreducible representations of $G$, commonly denoted by $\rep_G$ too. 
\begin{figure}[!ht]
\centering
\lwReviewTail
\caption{A portion of the lattice $\tilde\Gamma$ on which the basis states of the Fourier-transformed Hilbert space. See also Fig. \ref{fig:basisTrans}(d) for reminiscence.  }
\label{fig:LWlattice}
\end{figure}
According to the results of Section \ref{sec:FTeQD}, the Fourier -transformed Hilbert space of the extended QD model is spanned by the basis states defined on a trivalent lattice $\tilde\Gamma$, each vertex of which has an associated dangling edge, as shown in Fig. \ref{fig:LWlattice}. We have denoted by $L_G$ the set of all irreducible representations of $G$. Each edge (an internal one or dangling one) of the lattice carries an element of $L_G$.  The open end of each dangling edge also carries an extra degree of freedom ($\tilde m_{s_i}$ or $\tilde m_{q_i}$ in the figure), which is analogously the $z$-component of the irreducible representation on the edge ($s_i$ or $q_i$ in the figure). Note that as defined in Eq. \eqref{eq:AbProjector}, here we choose the indices $\tilde m_{s_i}$ and $\tilde m_{q_i}$ that diagonalize the vertex operators $A_v$ and $\Ab_v$.

The original LW model\cite{Levin2004} is known to have a Hilbert space smaller than that of the QD model\cite{Buerschaper2009} because the Hilbert space of the original LW model does not contain excited states with charges or dyons\cite{Hu2018}. In Ref.\cite{Hu2018}, the authors generalized the original LW model by adding a tail to each vertex in the trivalent lattice of the model. The generalized LW model has a larger Hilbert space encompassing a full dyon spectrum due to the tails, which play the same role as the dangling edges in Fig. \ref{fig:LWlattice}. Therefore, to map the Fourier-transformed extended QD model to the extended LW model with an enlarged Hilbert space, there is no need of reducing the Hilbert space of the extended QD model, in contrary to what was done in Ref.\cite{Buerschaper2009}.

What acts on the basis states in Fig. \ref{fig:LWlattice} is the Fourier transformed version of the Hamiltonian of the extended QD model. Namely,
\be\label{eq:lwHamiltonian}
H_{L_G,A_{G,K}}^{\QD,\tilde\Gamma} =\widetilde{ H^{\QD,\tilde\Gamma}_{G,K} } = -\sum_{v\in \tilde\Gamma\setminus \partial\tilde\Gamma} \widetilde{A_v^\QD} -\sum_{p\in \tilde\Gamma\setminus \partial\tilde\Gamma} \widetilde{B_p^\QD} -\sum_{v\in\partial\tilde\Gamma} \widetilde{ \Ab_{v}^{\QD} } - \sum_{p\in\partial\tilde\Gamma} \widetilde{ \Bb_{p}^{\QD} }.
\ee
 In Eqs.  \eqref{eq:AbProjector} and \eqref{eq:bpOperator}, we have already obtained the action of $\widetilde{ \Ab_{v}^{\QD} }$ on boundary vertices and that of $\widetilde{ \Bb_{p}^{\QD} }$ on boundary plaquettes in the lattice states in Fig. \ref{fig:LWlattice}. To see how the bulk vertex operators $\widetilde{ A_v^\QD }$ on the lattice states in Fig. \ref{fig:LWlattice}, note that according to Section \ref{subsec:AvFT}, the action of $\widetilde{ A_v^\QD }$ is the same as that of $\widetilde{ \Ab_v^\QD }$ when $K=G$. Hence, by Eq. \eqref{eq:AbProjector}, we have
\be
\widetilde{A_v^{QD}}\ket{\Psi_{s\tilde m_s}}=P^s_G \ket{\Psi_{s \tilde m_s}} = \delta_{s,0} \ket{\Psi_{s \tilde m_s}},
\ee
where use is made of $P^s_G =\tfrac{1}{|G|}\sum_g D^s_{\tilde m_s \tilde m'_s}(g) = \delta_{s,0}$, with $0$ being the trivial representation of $G$. Clearly, when $s=0$, $m_s$ is not a degree of freedom and can be omitted. As to a bulk $\widetilde{B_p^\QD}$ operator, its action on a bulk plaquette would be straightforwardly obtained and would be similar to that of the boundary $\widetilde{\Bb_p^\QD}$ but with seven more $6j$-symbols because a bulk plaquette has seven more vertices on its perimeter. Besides, there is no restriction to $L_A$ in a bulk $\widetilde{B_p^\QD}$ operator. 

By comparing  to the boundary vertex and plaquette operators in the extended LW model reviewed in Appendix \ref{appd:eLW} or in Ref.\cite{Hu2017,Hu2017a} and the bulk operators in the enlarged LW model in Ref.\cite{Hu2018}, we can actually identify the model $H_{L_G,A_{G,K}}^{\QD,\tilde\Gamma}$ defined on $\tilde\Gamma$ as the extended LW model $H_{\rep_G,A}^{\LW,\tilde\Gamma}$ with $A=A_{G,K}$ defined on the same lattice $\tilde\Gamma$. That is,
\be
H^{\QD,\Gamma}_{G,K} \xrightarrow{\text{FT \& basis rewriting}} H_{L_G,A_{G,K}}^{\QD,\tilde\Gamma} = H_{\rep_G,A}^{\LW,\tilde\Gamma},\quad A=A_{G,K}.
\ee
The two models have the same Hilbert space and Hamiltonians term by term. 
\section{Examples}\label{sec:examples}
In this section we offer two explicit examples, one for $G$ being Abelian and one for $G=S_3$, to aid the understanding of the results. 
\subsection{Abelian groups}\label{subsec:exZn}
Since an Abelian group is always isomorphic to the product of cyclic groups, let us consider for example $G=\mathbb{Z}_{N_{1}}\times\mathbb{Z}_{N_2}...$. The group elements are denoted by tuples $g=(g_1, g_2, ...)$ with $g_1=0, 1, ...N_{1}-1$, $g_2=0, 1 ,... N_2-1,$... All the irreducible representations of $G$ are $1$-dimensional; they form a set $L=\{(\mu_1, \mu_2,...), \mu_1=0,1,2,...N_1-1, \mu_2=0,1,2,...N_2-1,...\}$. We denote   element  $(\mu_1, \mu_2,...)$ in the set for short by $\mu$. The $3j$-symbols are simply $w^{\mu\nu\lambda}_{111}=\delta_{\mu\nu\lambda}$, which are $1$ iff $\mu_l+\nu_l+\lambda_l=0$ mod $N_l$ for all $l$ in $\mu$. The dual irreducible representations are defined by $\nu=\mu^*$ iff $\nu_l+\mu_l=0$ mod $N_l$ for all $l$ in $\mu$.

Since the irreducible representations are $1$-dimensional, the representation matrices are merely complex numbers; hence, all the matrix indices can be removed. The Fourier transform of the local group-basis states on an edge takes the simple form:
\be\label{eq:fTAbelian}
\ket{\mu}=\frac{1}{|G|} \sum_{g \in G}D^\mu(g)\ket{g},
\ee
where
\be\label{eq:oneDRep}
D^\mu(g)=e^{2\pi i(\frac{ \mu_1 g_1}{N_1}+\frac{\mu_2 g_2}{N_2}+...)}.
\ee
The rep-basis local states \eqref{eq:basisLT} of the vertex operators are now denoted simply by $\ket{\Psi_s}$. The fusion of three irreducible representations $\mu,\nu$, and $\lambda$ is determined by the delta function $\delta_{\mu\nu\lambda}$ just introduced. Note that in Eq. \eqref{eq:oneDRep}, the group elements and irreducible representations are on equal footing, the fusion of irreducible representation is the same as the multiplication of group elements. Indeed, for an Abelian group $G$, $\rep_G$ has a group structure and is isomorphic to $G$ itself. This fact results in the  self duality under Fourier transforming the extended QD model with an Abelian gauge group $G$. This self duality is actually the EM-duality in Eq. \eqref{eq:dualGraphBoundaryAABB}. 

To elucidate the self duality in this example, let us study how an $\widetilde{\Ab_v^{QD}}$ and $\widetilde{\Bb_p^{QD}}$ act in the rep-basis.  Following Eq. \eqref{eq:avFourierBasis}, we have\be
\begin{aligned}\label{eq:avFourierBasisAbel}
&\widetilde{\Ab_v^{QD}}\ket{\Psi_{s}}=\sum_{g,h,l\in G}\sum_{\substack{\mu',\nu',\gamma'\\ \lambda',s'}} \ket{ \Psi'_{s'}} \bra{ \Psi'_{s'}}\Ab_v^{QD}\ket{g,h,l}\bra{g,h,l}\Psi^{}_{s}\rangle\\ =&\sum_{\substack{\mu',\nu',\gamma'\\ \lambda',s'}} \ket{ \Psi'_{s'}} \bra{ \Psi'_{s'}} \sum_{g,h,l\in G}\sum_{k\in K}\frac{1}{|K|}\vert k{g,kh,l}\bar k\rangle\bra{g,h,l}\Psi_{sm_s}\rangle
\\
=&  \frac{1}{|K|}\sum_{k\in K}D^{s}(k)\ket{\Psi_{s}}=\frac{1}{|K|}\sum_{k \in K}e^{2\pi i(\frac{ s_1 k_1}{N_1}+\frac{s_2 k_2}{N_2}+...)}\ket{\Psi_{s}}.
\end{aligned}\ee
As in Eq. \eqref{eq:AbProjector}, we define
\be\label{eq:subset}
\delta_{s \in L_A}:= \frac{1}{|K|} \sum_{k \in K}e^{2\pi i(\frac{ s_1 k_1}{N_1}+\frac{s_2 k_2}{N_2}+...)}.
\ee
Then, following Eq. \eqref{eq:bpOperator}, we have
\be\begin{aligned}\label{eq:bpOperatorAbel}
&\langle\Psi^{\eta'\lambda'}_{r';s'} \vert\widetilde{\Bb_p^{QD}}\vert \ \Psi^{\lambda\eta}_{r;s}\rangle
\\
=&\!\sum_{g,h,l,x,y}\! \langle\Psi^{\lambda'\eta'}_{r';s'} \vert\Bb^{QD}_p\vert g,h,l,x,y\rangle\langle g,h,l,x,y| \Psi^{\lambda\eta}_{r;s}\rangle 
\\
=&\sum_{g,h,l,x,y}
\sum_{t \in L_A}\frac{|K|}{|G|}D^{t}(l)\langle\Psi^{\lambda'\eta'}_{r';s'} \vert g,h,l,x,y\rangle\langle g,h,l,x,y|\Psi^{\lambda\eta}_{r;s}\rangle \\
=&\sum_{t\in L_A}\frac{|K|}{|G|}\delta_{t^*s'^*s}\delta_{tr'^*r}\delta_{\rho^*\eta s^*}\delta_{\eta t^* \eta'^{*}}\delta_{\rho^* \eta' s'^*}\delta_{\kappa \lambda \eta^*}\delta_{\lambda t^* \lambda'^*}\delta_{\kappa \lambda' \eta'^*}\delta_{\gamma r^* \lambda^*}\delta_{\gamma r'^* \lambda'^*}        .
\end{aligned}\ee
This result can be understood in the following way. Since $G\cong \rep_G$ in this case, the constraint $\delta_{t^*s'^*s}=1$ indicates the equality $s'=s\cdot t^*$, where the dot is the group multiplication. Then, a  $\widetilde{\Bb_p^{QD}}$ acts on its local basis states as
\be\begin{aligned}\label{eq:rewriteBpAbelian}
\widetilde{\Bb_p^{QD}}\vert \ \Psi^{\lambda\eta}_{r;s}\rangle=&\sum_{t\in L_A}\frac{|K|}{|G|}\delta_{\rho^*\eta s^*}\delta_{\rho^* \eta'^* s'^*}\delta_{\kappa \lambda \eta^*}\delta_{\kappa \lambda' \eta'^*}\delta_{\gamma r^* \lambda^*}\delta_{\gamma r'^* \lambda'^*}        \ket{\Psi^{\lambda \cdot t^*,\eta \cdot t^*}_{t \cdot r;s \cdot t^*}},
\\
=&\sum_{t\in L_A}\frac{|K|}{|G|}\ket{\Psi^{\lambda \cdot t^*,\eta \cdot t^*}_{t \cdot r;s \cdot t^*}},
\end{aligned}\ee
where we omit the delta functions that are implied in the definition of $\ket{\Psi^{\lambda\eta}_{r;s}}$. (All vertices in figure in Eq. \eqref{eq:bpBasis} must satisfy the fusion rule.) The results here exemplify the EM-duality in Eq. \eqref{eq:dualGraphBoundaryAABB}. That is, the operators $\widetilde{\Ab_v^{QD}}$ in the rep-basis act on $\tilde\Gamma$ as the operators $\Bb_p^{QD}$ in the corresponding group-basis act on the dual lattice $\tilde\Gamma^*$, while the operators $\widetilde{\Bb_p^{QD}}$ in the rep-basis act on $\tilde\Gamma$ as the operators $\Ab_v^{QD}$ in the group-basis act on $\tilde\Gamma^*$.  

On the other hand,  $\rep_G$ has the $6j$ symbols   $G^{\mu \nu \lambda}_{\gamma \kappa \rho}=\delta_{\mu\nu\lambda}\delta_{\gamma \kappa\lambda^*}\delta_{\nu\gamma\rho^*}\delta_{\mu\rho\kappa}$ and Frobenius algebra multiplication rule $f_{abc}=\delta_{abc}$. Take $\rep_G$ as the input data of the extended LW model and compare it with the boundary operator actions \eqref{eq:avFourierBasis} and \eqref{eq:bpOperator}, one can see that the Fourier-transformed boundary Hamiltonian with $K\subseteq G$ of the extended QD model is exactly the same as the boundary Hamiltonian with $A_{G,K}$ of the extended LW model.  
 
Finally, we study Eq. \eqref{eq:subset} more explicitly.  Let us consider in particular $G=\mathbb{Z}_m$ with a subgroup $K=\mathbb{Z}_n$, such that $n|m$. The subgroup elements are labeled by a  set of integers $\{0, \frac{m}{n}, \frac{2m}{n}, ..., \frac{m(n-1)}{n}\}$. According to Eq. \eqref{eq:subset}, we have $\delta_{s \in L_A}=\frac{1}{n}(1+e^{2\pi i\frac{s}{n}}+e^{4\pi i\frac{s}{n}}+...+e^{2\pi i \frac{s(n-1)}{n}})=\frac{1}{n}\frac{1-e^{2\pi i s}}{1-e^{2\pi i\frac{s}{n}}}$. Thus, only when $s=jn$, $j\in \mathbb{N}$, we have $\delta_{s \in L_A}=1$; hence, $L_A=\{0, n, 2n, ... m-n\}$, and $A_{\Z_m,\Z_n}= (L_A,\delta_{abc})\cong \Z_{m/n}$. Clearly, the larger (smaller) the $\mathbb{Z}_n$,  the smaller (larger) the $A_{\Z_m,\Z_n}$. Since $n|m$, $\Z_n$ is a normal subgroup of $\Z_m$, and $\Z_{m/n} = \Z_m/\Z_n$. This is a special case of the general EM duality $A_{G,K} = \mathds{(C}[G]/\mathds{C}[K])^*$ on the boundary of the extended QD model.  

\subsection{$G=S_3$}\label{subsec:exS3}
When $G=S_3$, the output of the extended QD model is a topological phase described by $D(S_3)$, the quantum double of $S_3$. The anyons carry representations of $([c],\mu)$ of $D(S_3)$, where $[c]$ is a conjugacy class in $S_3$, and $\mu$ is a irreducible representation of the centralizer of $[c]$ in $S_3$. Thus, the anyons can be divided into three types: $S_3$ charges, fluxes, and dyons, as shown in Table \ref{tab:anyonS3}. 
\begin{table}[!h]
\centering
\begin{tabular}{|c|c|c|c|}
                \hline
Anyon types&$A$, $B$, $C$ &$D$, $E$ & $F$, $G$, $H$                
                \\
                \hline
           Conjugacy classes & $\{e\}$ & $\{r,rs,r^2s\}$ & $\{r,r^2\}$
\\
\hline
Centralizer & $S_3$ & $\mathbb{Z}_2$ & $\mathbb{Z}_3$
\\
\hline 
Irrep of centralizer & $1$, $sign$, $\pi$ & $1$, $-1$ & $1$, $\omega$, $\omega^{*}$
\\
\hline
Quantum dimension & $1$, $1$, $2$ & $3$, $3$ & $2$, $2$, $2$
\\
\hline
Twist & $1$, $1$, $1$ & $1$, $-1$ & $1$, $e^{2\pi i/3}$, $e^{-2\pi i/3}$
\\
\hline
\end{tabular}
\caption{Information of anyons of the $D(S_3)$ model. Anyons of types $A,B$, and $C$ are pure charges, $D$ and $F$ are pure fluxes, and $E,G,H$ are dyons.}
\label{tab:anyonS3}
\end{table}

For notation simplicity, we rename the three irreducible representations of $S_3$ by $0,1$, and $2$, respectively corresponding to the $1,sign$, and $\pi$ in the table above. One set of irreducible representations matrices is listed as follows. 
\begin{align}\label{eq:repS3}
&D^0(e)=D^0(r)=D^0(r^{2})=D^0(sr)=D^0(s)=D^0(sr^{2})=1,\notag
\\
&D^1(e)=D^1(r)=D^1(r^{2})=1, \quad D^1(sr)=D^1(s)=D^1(sr^{2})=-1, \notag
\\
&D_{m_2m_2'}^2(e)= \begin{pmatrix}1 & 0 \\
0 & 1 \\
\end{pmatrix}_{m_2m_2'},D_{m_2m_2'}^2(r)=\begin{pmatrix}-\frac{1}{2} & -\frac{\sqrt{3}}{2} \\
\frac{\sqrt{3}}{2} & -\frac{1}{2} \\
\end{pmatrix}_{m_2m_2'}, \notag \\ &D_{m_2m_2'}^2(r^2)=\begin{pmatrix}-\frac{1}{2} & \frac{\sqrt{3}}{2} \\
-\frac{\sqrt{3}}{2} & -\frac{1}{2} \\
\end{pmatrix}_{m_2m_2'}, D_{m_2m_2'}^2(sr)=\begin{pmatrix}\frac{1}{2} & \frac{\sqrt{3}}{2} \\
\frac{\sqrt{3}}{2} & -\frac{1}{2} \\
\end{pmatrix}_{m_2m_2'},  \notag 
\\
&D_{m_2m_2'}^2(s)=\begin{pmatrix}-1 & 0 \\
0 & 1 \\
\end{pmatrix}_{m_2m_2'}, D_{m_2m_2'}^2(sr^2)=\begin{pmatrix}\frac{1}{2} & -\frac{\sqrt{3}}{2} \\
-\frac{\sqrt{3}}{2} & -\frac{1}{2} \\
\end{pmatrix}_{m_2m_2'}.
\end{align}
The nonvanishing $3j$-symbols of $S_3$ are 
\be\begin{aligned}\label{eq:threejSymbolSThree}
&w^{000}_{111}=w^{011}_{111}=1,
\\
&w^{022}_{1m_1m_2}=\begin{pmatrix}\frac{1}{\sqrt2} & 0 \\
0 & \frac{1}{\sqrt2} \\
\end{pmatrix}_{m_1m_2},
\\
&w^{122}_{1m_1m_2}=\begin{pmatrix}0 & \frac{i}{\sqrt{2}} \\
 -\frac{i}{\sqrt{2}} & 0 \\
\end{pmatrix}_{m_1m_2},
\\
&w^{222}_{m_1m_2m_3}=\begin{pmatrix}\{0,\frac{1}{2}\} & \{\frac{1}{2},0\} \\
\{\frac{1}{2},0\} & \{0,-\frac{1}{2}\} \\
\end{pmatrix}_{m_1m_2m_3}.
\end{aligned}\ee
\begin{table}[!h]
\small
\centering
\begin{tabular}{|c|c|c|}
                \hline
Subgroups $K$   & $L_A = \{(s,\alpha_s)\}$\ & $f_{a\alpha_ab\alpha_bc\alpha_c}:=u_au_bu_c(w^{abc}_{\alpha_a\alpha_b\alpha_c})^*$\\&&$f_{abc}=f_{cab}$, $f_{aa^*0}=1$              
                \\
                \hline
               $S_3$  & $\{(0,1)\}$  & $f_{0_{1}0_10_1}=1$
               \\ \hline $\mathbb{Z}_3$ & $\{(0,1),(1,1)\}$ & $f_{0_{1}1_11_1}=1$
               \\ \hline $\mathbb{Z}_2$  & $\{(0,1),(2,1)\}$ & $f_{2_22_22_2}=-2^{-\frac{1}{4}}$, $f_{0_{1}2_22_2}=1$
               \\ \hline $\{e\}$ &$\{(0,1),(1,1),(2,1),(2,2)\}$ & $f_{2_{2}1_12_1}=-i$, $f_{2_{2}2_11_1}=i$, \\ 
&  & $f_{2_{2}2_12_1}=2^{-\frac{1}{4}}$, $f_{2_{2}2_22_2}=-2^{-\frac{1}{4}}$
\\
\hline 
\end{tabular}
\caption{Four emergent Frobenius algebras $A_{S_3,K} = (L_A,f)_{S_3,K}$ of the Fourier-transformed extended QD model with $G=S_3$. In the third column, only the non-vanishing symbols $f$ are shown.}
\label{tab:checkSthree}
\end{table}
\begin{table}[!h]
\small
\centering
\begin{tabular}{|c|}
\hline
Frobenius algebra (modulo Morita equivalence) objects in $\text{Rep}_{S_3}$\\ $f_{abc}=f_{cab}$, $f_{aa^*0}=1$\\
\hline
$A_1=0$ (Morita equivalent to $A_1'=0\oplus1\oplus2$, $f_{122}=\pm1$) \\ $A_2=0\oplus1$ \\
$A_3=0\oplus2$, $f_{222}=-2^{-\frac{1}{4}}$  \\
$A_4=0\oplus1\oplus2_1\oplus2_2$, $f_{110}=1$, $f_{2_{2}12_{1}}=-i$, $f_{2_22_11}=i$,\ $f_{2_22_12_1}=2^{-\frac{1}{4}}$, $f_{2_22_22_2}=-2^{-\frac{1}{4}}$  
\\
\hline 
\end{tabular}
\caption{Inequivalent Frobenius algebra objects in the UFC $\rep_{S_3}$ obtained by solving the defining conditions \eqref{eq:faCondi}. Frobenius algebra objects $A_1=0$ and $A_1'=0\oplus 1\oplus 2$ are Morita equivalent\cite{Hu2017a}. One thus can forget about $A'_1$.}
\label{tab:Z23sols}
\end{table}
Plugging the $3j$-symbols \eqref{eq:threejSymbolSThree} into Eq. \eqref{eq:fCorrespondence}, we can obtain four emergent Frobenius algebras at the boundary of the Fourier-transformed extended QD model, corresponding to the four distinct subgroups of $S_3$, as recorded in Table \ref{tab:checkSthree}. The four emergent Frobenius algebras can indeed be identified with the four Frobenius algebra objects in Table \ref{tab:Z23sols}
obtained by solving the defining conditions \eqref{eq:faCondi} of a Frobenius algebra object in a UFC $\F$, which in this case is $\rep_{S_3}$. 

Here we can try to understand explicitly the relation between the emergent Frobenius algebras at the boundary and anyon condensation, in particular anyon splitting and multiplicity.

\begin{table}[!ht]
\small
\centering
\begin{tabular}{|c|c|c|c|}
\hline
Condensation type&Boundary condensate&Boundary $A_{S_3,K}$ & Boundary $K$ \\
\hline
Charge  & $\mathcal{A}_1=A\oplus F\oplus D$ & $A_1$ & $S_3$
\\
\hline
Dyon  & $\mathcal{A}_2=A\oplus B\oplus 2F$ & $A_2$ & $\mathbb{Z}_3$
\\
\hline 
Dyon  & $\mathcal{A}_3=A\oplus C\oplus D$ & $A_3$ & $\mathbb{Z}_2$
\\
\hline
Flux  & $\mathcal{A}_4=A\oplus B\oplus 2C$ & $A_4$ & ${\{e\}}$
\\
\hline
\end{tabular}
\caption{Correspondence between (column 1) anyon condensates of  the $D(S_3)$ model, emergent Frobenius algebras (column 2), and subgroups of $S_3$ (column 3).}
\label{tab:S3corrd}
\end{table}

In the current example, there are four different gapped boundary conditions respectively characterized by the four subgroups\footnote{$S_3$ actually has three $\Z_2$ subgroups, which however specifies the same boundary condition and corresponds to the same Frobenius algebra.} of $S_3$, imposed by the boundary plaquette operators $\Bb_p^\QD$ of the model. These four gapped boundary conditions respectively correspond to four different boundary condensates composed of different types of anyons condensed at the boundary. These correspondences are shown in Table \ref{tab:S3corrd}. One can see in the table that the type $C$ anyons, which are pure charges $([e],2)$, appear with multiplicity $1$ in the condensate $\A_3$ but multiplicity $2$ in the condensate $\A_4$. 

Condensate $\A_3$ corresponds to the emergent Frobenius algebra $A_{3} = (0,1) + (2,1)$. The element $s=2$ in the pair $(2,1)$ in $A_3$ actually is the $2$-dimensional irreducible representation of $S_3$ and thus can be directly identified with the pure charge $C$ in the condensate $\A_3$. Note that for each $K\subseteq S_3$ in Table \ref{tab:checkSthree}, the pairs $(s,\alpha_s)$ in the corresponding row each labels a $+1$ local eigenstate $\ket{\Psi_{s\alpha_s}}$ of the $\widetilde{\Ab_v^\QD}$ operator acting on the relevant vertex. In the case with $K=\Z_2$ and $s=2$, seen in Eq. \eqref{eq:avFourierBasis}, an operator $\widetilde{\Ab_v^\QD}$ is represented in the $2$-dimensional local space $V_s$ by the projector
\be \label{eq:Ps2Z2}
P^{s=2}_{K=\Z_2}=\sum_{k\in\Z_2}\frac{1}{|\Z_2|}D^{2}_{m_2m'_2}(k)=\frac{1}{4}\begin{pmatrix}1 & \sqrt{3} \\
\sqrt{3} & 3 \\
\end{pmatrix},
\ee 
where we didn't differentiate a matrix and its matrix elements labeled by $m_s m'_s$ to emphasize that the operator is not diagonalized yet. We can diagonalize the projector in the basis states $e_{\tilde m_s}$ labeled by $\tilde m_s$ 
and find the $+1$ eigenstate $e_1=(\sqrt{3},1)^T$ and $0$ eigenstate $e_2 = (-\frac{1}{\sqrt{3}},1)^T$. Hence, only $\alpha_{s=2}:= \tilde m_s=1$ is allowed. That is, the pair $(2, 1)\in L_A$ labels the only $+1$ local eigenstate $\ket{\Psi_{21}}$ of the $\widetilde{ \Ab_v^\QD}$. Hence, only half of the $2$-dimensional space $V_s$ or half of the charge-$2$ condenses at the boundary. As such, we can say that the charge-$2$ splits: $2 = (2,1) \oplus (2,2)$ but only the half $(2,1)$ condenses. That is why only $(2,1)$ appears in the $A_3$ in Table \ref{tab:checkSthree}, and type $C$ anyons appear with unit multiplicity in $\A_3$ in Table \ref{tab:S3corrd}. If we denote $(2,1)$ by $2_1$ and $(2,2)$ by $2_2$, we found that the Frobenius algebra object $A_3$ in Table \ref{tab:Z23sols} should be rigorously written as $A_3 = 0 \oplus 2_2$; however, one often just write $A_{3}=0\oplus2$ customarily.

Now in the case with $\A_4$ corresponding to $A_4$, the subgroup is $K=\{e\}$. Then, for $s=2$, an operator $\widetilde{\Ab_v^\QD}$ is represented in the $2$-dimensional local space $V_s$ by the matrix  $P^{s=2}_{K=\{e\}} = D^{s}_{m_sm'_s}(e) = \delta_{m_s m'_s} = \delta_{\tilde m_s \tilde m'_s}$, which is automatically diagonal and in fact a $2\x 2$ identity matrix. Thus both local eigenstates of $\widetilde{\Ab_v^\QD}$ in $V_s$ are $+1$ eigenstates, which can be labeled by $(2,1)$ and $(2,2)$. Note that these two eigenstates are not those of the operator in Eq. \eqref{eq:Ps2Z2}. As such, the charge $2$ splits as $2 = (2,1) \oplus (2,2)$, but both pieces condense at the boundary. Therefore, in the boundary condensate, the type $C$ anyons count twice, rendering $\A_4 =  A\oplus B\oplus 2C$. If we write $(2,1)$ and $(2,2)$ as $2_1$ and $2_2$, we can identify the emergent Frobenius algebra $A_4$ with the Frobenius algebra object $A_4=0\oplus1\oplus2_1\oplus2_2$ in Table \ref{tab:Z23sols}.  

As a further corroboration of the mapping between the extended QD model and the extended LW model, we can compute the GSDs of both models on a cylinder, whose two boundaries may not necessarily possess the same boundary conditions. Using 
Eqs. \eqref{eq:qdGSD} and \eqref{eq:lwGSDProj}, we obtain respectively Table \ref{tab:GSDQD} and Table \ref{tab:GSDLW}. Matching the GSDs in these two tables futher confirms the correspondence between the gapped boundary conditions specified by the subgroups of $S_3$ in the extended QD model and those specified by the Morita inequivalent Frobenius algebra objects in the extended LW model.
\begin{table}[!h]
\small
\centering
\begin{tabular}{|c||c|c|c|c|} 
\hline
Subgroups& $\{e\}$ & $S_3$ & $\mathbb{Z}_3$ & $\mathbb{Z}_2$
\\
\hline\hline 
$\{e\}$ & $6$ & $1$ & $2$& $3$
\\
\hline
$S_3$ & $1$ & $3$ & $3$ & $2$ 
\\
\hline
$\mathbb{Z}_3$ & $2$ & $3$ & $6$ & $1$ 
\\
\hline
$\mathbb{Z}_2$ & $3$ & $2$ & $1$ & $3$ 
\\
\hline    
\end{tabular}
\caption{The GSDs of the extended QD model with $G=S_3$ on a cylinder, whose two boundaries are specified by two subgroups of $S_3$. }
\label{tab:GSDQD}
\end{table} 
\begin{table}[!h]
\small
\centering
\begin{tabular}{|c||c|c|c|c|}
\hline
Frobenius algebras& $A_4$ & $A_1$ & $A_2$ & $A_3$ 
\\
\hline\hline 
$A_4$ & $6$ & $1$ & $2$ & $3$
\\
\hline
$A_1$ & $1$ & $3$ & $3$ & $2$
\\
\hline
$A_2$ & $2$ & $3$ & $6$ & $1$ 
\\
\hline
$A_3$ & $3$ & $2$ & $1$ & $3$
\\
\hline
\end{tabular}
\caption{The GSDs of the extended LW model with $\rep_{S_3}$ on a cylinder, whose boundaries are specified by two Frobenius algebra objects of $\rep_{S_3}$.}
\label{tab:GSDLW}
\end{table}

For completeness of the data in this example, we collect in Table \ref{tab:Mod}
the Modules of the Frobenius algebras in $\rep_{S_3}$. 
\begin{table}[!h]
\tiny
\centering
\resizebox{\textwidth}{31mm}{
\begin{tabular}{|c|c|}
\hline
Frobenius algebras of $\text{Rep}_{S_3}$ LW model & Modules                \\ $f_{abc}=f_{cab}$, $f_{aa^*0}=1$ &
                \\
                \hline
                 $A_1=0$  & $M_{A_1}^1=0$, $\rho_{00}^0=1$; 
                 
       $M_{A_1}^2=1$, $\rho_{11}^0=1$;\\ & $M_{A_1}^3=2$, $\rho_{22}^0=1$
                 \\
                 \hline $A_2=0\oplus1$ & $M_{A_2}^1=0\oplus1$, $\rho_{j_1j_2}^{a}=f_{aj^*_2j_1}$; \\ & $M_{A_2}^2=2$, $\rho_{22}^0=1$, $\rho_{22}^1=-1$; \\ & $M_{A_2}^3=2$, $\rho_{22}^0=1$, $\rho_{22}^1=1$;
                 \\
                 \hline$A_3=0\oplus2$, $f_{222}=-2^{-\frac{1}{4}}$  & $M_{A_{3}}^1=0\oplus2$, $\rho_{j_1j_2}^{a}=f_{aj^*_2j_1}$; \\ & $M_{A_3}^2=1\oplus2$, $\rho_{11}^0=1$, $\rho_{22}^0=1$, $\rho_{12}^2=(\rho_{21}^2)^{-1}$,\\ & $\rho_{22}^2=-2^{-1/4}$; 
\\
\hline$A_4=0\oplus1\oplus2_1\oplus2_2$, $f_{110}=1$,& $M_{A_{4}}=0\oplus1\oplus2_1\oplus2_2$, $\rho^{a\lambda_a}_{j_1\lambda_{j_1},j_{2}\lambda_{j_2}}=f_{a\lambda_a,j^*_2\lambda_{j_2},j_1\lambda_{j_1}}$ \\ $f_{2_{2}12_{1}}=-i$, $f_{2_22_11}=i$,\ $f_{2_22_12_1}=2^{-\frac{1}{4}}$, $f_{2_22_22_2}=-2^{\frac{1}{4}}$ &
\\
\hline 
\end{tabular}}
\caption{Modules of the Frobenius algebra objects in $\rep_{S_3}$.}
\label{tab:Mod}
\end{table}

\acknowledgments
The authors thank Perimeter Institute for hospitality during their visit, where this project was largely done and completed. We thank Zichang Huang for helpful discussions. YW is supported by the NSF grant No. 11875109. 

\appendix

\section{Some properties of fusion categories}\label{appd:6j}
Simple objects in a UFC $\F$ obey a set of fusion rules:
\be\label{eq:fusionRules}
i\ox j=\sum_kN_{ij}^k k,
\ee
where $N_{ij}^k$ are the fusion coefficients satisfying
\be\begin{aligned}\label{eq:fusionMultiplicity}
        &N_{0i}^j=N_{i0}^j=\delta_{i,j},
        \\
        &N_{ij}^0=\delta_{i,j^*},
        \\
        &\sum_{m}N_{ij}^mN_{mk}^l=\sum_n N_{jk}^nN_{in}^l.
\end{aligned}\ee
Here, $j^*$ is the dual of $j$ defined by the second row in the equation above. Each object $j$ also has a nonzero characteristic number $\tilde{d_j}$, called quantum dimension of $j$, such that $\tilde{d_j}=\tilde{d_{j^*}}$ and 
\be\label{eq:quantumdim}
\tilde d_i\tilde d_j=\sum_k N^k_{ij}\tilde d_k. 
\ee
In particular, $\tilde d_0=1$. Let $\beta_j=\mathrm{sgn}(\tilde d_j)$, then $\beta_i\beta_j\beta_k=1$ is satisfied if $N_{ij}^{k}>0$. When $\F=\rep_G$ for a finite group $G$, we have $\tilde d_j = \beta_j d_j$, and $\beta_j$ is identified with the FS indicator of the representation $j$ of $G$. Now we can define  tetrahedral-symmetric unitary $6j$-symbols as the coefficients of the map $G: L^6\rightarrow\mathbb{C}$ that satisfies the following conditions
\be\begin{aligned}\label{eq:sjSymbol}
        &G^{ijm}_{kln}=G^{mij}_{nk^*l^*}=G^{klm^*}_{ijn^*}=\beta_m\beta_{n} (G^{j^*i^*m^*}_{l^*k^*n})^*,
        \\
        &\sum_n\tilde d_nG^{mlq}_{kp^*n}G^{jip}_{mns^*}G^{js^*n}_{lkr^*}=G^{jip}_{q^*kr^*}G^{riq^*}_{mls^*},
        \\
        &\sum_n\tilde d_nG^{mlq}_{kp^*n}G^{l^*m^*i^*}_{pk^*n}=\frac{\delta_{iq}}{\tilde d_i}\delta_{mlq}\delta_{k^*ip}.
\end{aligned}\ee

\section{Extended LW model}\label{appd:eLW}
\begin{figure}[h!]
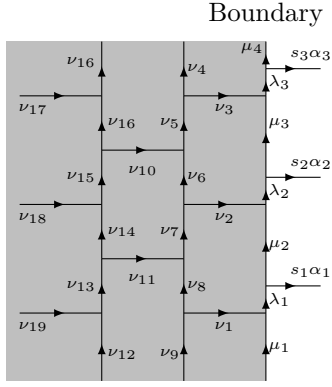

\centering
\lwReview
\caption{A portion of the oriented trivalent lattice $\tilde\Gamma'$ on which the extended LW model is defined. The bulk (grey region) is to the left of the boundary (consisting of the dangling edges).}
\label{fig:lwbdGraph}
\end{figure}

In this section, we review the basic ingredients of extended LW model. The extended LW model is  defined on an oriented trivalent lattice $\tilde\Gamma'$, part of which is depicted in Fig. \ref{fig:lwbdGraph}.

The Hilbert space is spanned by all configurations of assigning of the simple objects of a UFC $\F$ on the oriented edges of $\tilde\Gamma'$. Conventionally, we call these simple objects string types and label them by integers in $L=\{0,1,...,N\}$. Each string types $j$ has a dual $j^*$, which is also an element of $L$ and satisfies $j^{**}=j$. If we reverse the orientation of the edge and conjugate the string type $j\mapsto j^*$ associated with the edge at the same time, the state remains the same. There is always a trivial (unit)\ element $0\in L$ satisfying $0^*=0$.

The set $\{N_{ij}^k, \tilde d_j, G^{ijk}_{lmn}\}$ is the input data of the extended LW model, which can be derived from representation theory of a finite group or quantum group. In this work, we take the string types $\mu$ to be the irreducible representation of a finite group $G$ and label them with Greek letters. The trivial representation is $\mu=0$. The set $L$ is identified with $L_G$---the set of irreducible representations of $G$. The fusion coefficients $N_{\mu\nu}^\lambda$ are the numbers of trivial representations appearing in the decomposition of $\mu\otimes \nu \otimes \lambda$. In such cases, as mentioned earlier, $\tilde d_\mu=\beta_\mu d_\mu$ with $\beta_\mu$ the FS indicator. The dual representation $\mu^*$ of $\mu$ satisfies $\tilde d_{\mu^*}=\tilde d_{\mu}$. The $6j$-symbols $G^{\mu\nu\lambda}_{\kappa\eta\gamma}$ are the same as the symmetrized Racah $6j$-symbols of $G$. Similar to the extended QD models, there are also two types of local operators in the bulk Hamiltonian of the extended LW model, namely
\be\label{eq:lwHam}
H^{\mathrm{LW}}_{\F}=-\sum_{v\in \tilde\Gamma'\setminus \partial \tilde\Gamma'} A_v^\LW-\sum_{v\in \tilde\Gamma'\setminus \partial \tilde\Gamma'} B_p^{\LW},
\ee
where $A_v^{\LW}$ and $B_p^{LW}$ are respectively the vertex and plaquette operators. The action of $A_v^{\LW}$ on a local trivalent state as 
\be\label{eq:lwAvop}
A_v^{LW}\bket{\lwAvop{\mu_1}{\mu_2}{\mu_3}}=\delta_{ijk}\bket{\lwAvop{\mu_1}{\mu_2}{\mu_3}},
\ee
where $\delta_{ijk}$ is delta function that is equal to $1$ if $N_{ij}^k>0$ and otherwise $0$. The action of $B_p^{\LW}$ is more involved and reads
\be\begin{aligned}\label{eq:lwBpop}
B_p^{LW}=&\frac{1}{D}\sum_{s\in L}\tilde d_sB_p^{LW}(s),
\\
B_p^{LW}(s)\bket{\lwBpop{\mu_1}{\mu_2}{\mu_3}{\mu_4}{\mu_5}{\mu_6}} = &\sum_{\mu'_1,\mu'_2,\mu'_3} \prod_{i=1}^6 \tilde v_{\mu_i}\tilde v_{\mu'_i} G^{\nu_1\mu_1^*\mu_6}_{s\mu'_6\mu_1'^*} G^{\nu_2\mu_2^*\mu_1}_{s\mu_1'\mu_2'^*} G^{\nu_3\mu_3^*\mu_2}_{s\mu_2'\mu_{3}'^*}
\\ 
&\times  G^{\nu_4\mu_4^*\mu_3}_{s\mu'_3\mu_4'^*}G^{\nu_5\mu_5^*\mu_4}_{s\mu_4'\mu_5'^*}G^{\nu_6\mu_6^*\mu_5}_{s\mu_5'\mu_6'^*}\bket{\lwBpop{\mu_1'}{\mu_2'}{\mu_3'}{\mu'_4}{\mu'_5}{\mu'_6}}.
\end{aligned}\ee
The bulk Hamiltonian is exactly solvable because all the operators therein are commuting projectors. 

On the boundary $\partial \tilde \Gamma'$ of $\tilde\Gamma'$ however, we need an extra input data for the extended LW to be well-defined. This extra input data is the maximal Frobenius algebra object in $\rep_G$. This maximal Frobenius algebra object can be identified with the canonical Frobenius algebra defined in Eq. \eqref{eq:FAcanonical}. But to comply with the definition of boundary vertex operators in Refs.\cite{Hu2017,Hu2017a}, where the extended LW model was constructed. We rewrite this canonical Frobenius algebra as
\be
A_\mathrm{canonical} = \bigoplus_{s\in \mathrm{Irrep_G}}\bigoplus_{\alpha_s=1}^{d_s} s\alpha_s.
\ee 
As in the main text, here $d_s$ counts the multiplicity of $s$. Note that Refs.\cite{Hu2017,Hu2017a} actually did not consider the cases with multiplicities in a Frobenius algebra and thus did not require an extra input data on the boundary. In this review, we offer a generalization to such cases. We emphasize that in an extended LW model, the boundary $\partial \tilde\Gamma'$ consists of the dangling edges only. As far as boundary degrees of freedom are concerned, we regard $A_\mathrm{canonical}$ as a set and allocate to each tail on the boundary an element $s\alpha_s$. The boundary Hamiltonian of the model is then defined on the boundary lattice $\partial \tilde \Gamma'$ of $\tilde\Gamma'$ by  
\be\label{eq:qdbdHamiltonian}
H_{A}^{\LW}=-\sum_{n\in \partial \tilde\Gamma'} \Ab_{n}^{\LW} - \sum_{p\in \partial \tilde\Gamma'} \Bb_{p}^{\LW},
\ee
where the $A_n^{\LW}$ operator acts on the tail $n$ on the boundary and projects it to a Frobenius algebra object $A\subseteq A_{\mathrm{cananical}}$ of $\rep_G$:
\be\label{eq:lwbdAvop}
\Ab_n^{\LW}\bket{\lwbdAvop{\mu}{\nu}{s_n\alpha_{s_n}}}=\delta_{s_n\alpha_{s_{n}} \in  A}\bket{\lwbdAvop{\mu}{\nu}{s_n\alpha_{s_n}}},
\ee
and $\overline B_p^{\LW}$ operator is a operator comprised of $\overline B_p^{\LW}(t)$:
\be\label{eq:lwbdBpop}
\Bb_p^{\LW}=\frac{1}{d_A} \sum_{t\alpha_t \in L_A} \tilde v_t \Bb_p^{\LW}(t\alpha_t), 
\ee
where $d_A=\sum_{t\alpha_t \in L_A}d_t$, and $\Bb_p^{\LW}(t\alpha_t)$ acts on the open boundary $p$ between two nearest neighbouring tails as:
\begin{align}
\Bb_p^{\LW}(t\alpha_t) \bket{\lwbdBpop{\mu}{\nu}{\gamma}{r\alpha_r}{\lambda}{\kappa}{\eta}{s\alpha_s}{\rho}}&=\sum_{s'\alpha_{s'},\eta',\lambda',r'\alpha_{r'}} f_{t^{*}\alpha_{t} s'^*\alpha_{s'} s\alpha_s}f_{r\alpha_r t\alpha_t r'^*\alpha_{r'}}\tilde u_r \tilde u_s \tilde u_{r'} \tilde u_{s'}\times
\nonumber \\
&G^{\rho^* \eta s^*}_{t^*s'^*\eta'}G^{\kappa \lambda \eta^*}_{t^*  \eta'^* \lambda'}G^{\gamma \lambda^* r^*}_{t^* r'^* \lambda'^*}\tilde v_{\lambda}\tilde v_{\eta} \tilde v_{\lambda'} \tilde v_{\eta'} \bket{\lwbdBpop{\mu}{\nu}{\gamma}{r'\alpha_{r'}}{\lambda'}{\kappa}{\eta'}{s'\alpha_{s'}}{\rho}}, \label{eq:lwbdBptop}
\end{align}
if one of the tail $r\alpha_r$ or $s\alpha_s \notin A$, then $f_{t^{*}\alpha_{t} s'^*\alpha_{s'} s\alpha_s}f_{r\alpha_r t\alpha_t r'^*\alpha_{r'}}=0$, and hence $\Bb_p^{\LW}(t\alpha_t)=0$. Each $A$ specifies a gapped boundary condition. The total Hamiltonian of the extended LW model then is the sum
\be\label{eq:lwBdryHamiltonian}
H^{\LW}_{\F, A}= H_{\F}^{\LW} + H_A^{\LW}.
\ee
The GSD formula on a cylinder takes the form
\be\label{eq:lwGSDProj}
\mathrm{GSD} = \mathrm{Tr} P^0_{cyl}=\prod_{v\in\tilde\Gamma'\setminus \partial \tilde\Gamma'}A_v^{\LW}\prod_{n \in\partial \tilde\Gamma'}\Ab_{n}^{\LW}\prod_{p\in\tilde \Gamma'\setminus \partial \tilde\Gamma'}B_p^{\LW}\prod_{p'\in\partial \tilde\Gamma'}\Bb^{\LW}_{p'}.
\ee
Note that on a cyliner, $\partial \tilde\Gamma'$ contains two components, on which the operators $\Ab_n^{\LW}$ may project the boundary degrees of freedom into respectively two (not necessarily) different Frobenius algebras.
\section{Some proofs}\label{appd:proof}
Here we prove that the rep-basis defined in Eq. \eqref{eq:basisLT} is orthonormal and complete.
\begin{align}
&\langle\Psi'_{s'm_{s'}}\vert\Psi_{sm_{s}}\rangle \nonumber
\\
&=\frac{v_{\mu}v_{\nu}v_{\lambda}v_{\gamma}v_sv_{\mu'}v_{\nu'}v_{\lambda'}v_{\gamma'}v_{s'}}{|G|^3}\sum_{\substack{g,h,l\\x,y,z}}\repBasisB\left( \repBasisAA \right)^*\langle x,y,z\vert g,h,l\rangle \nonumber
\\
&=\frac{v_{\mu}v_{\nu}v_{\lambda}v_{\gamma}v_sv_{\mu'}v_{\nu'}v_{\lambda'}v_{\gamma'}v_{s'}}{|G|^3}\sum_{\substack{g,h,l}}\repBasisA\left( \repBasisAA \right)^* \nonumber
\\
&=\frac{v_{\mu}v_{\nu}v_{\lambda}v_{\gamma}v_sv_{\mu'}v_{\nu'}v_{\lambda'} v_{\gamma'}v_{s'}}{|G|^3}\sum_{\substack{g,h,l}}\repBasisC  \nonumber
\\
&=\frac{v_{\mu}v_{\nu}v_{\lambda}v_{\gamma} v_sv_{\mu'}v_{\nu'} v_{\lambda'}v_{\gamma'} v_{s'}}{d_{\nu}d_{\mu}d_{\lambda}}\repBasisD \nonumber
\\
&=\delta_{\mu,\mu'}\delta_{\nu,\nu'}\delta_{\gamma,\gamma'}\delta_{\lambda,\lambda'} \delta_{s,s'}\delta_{m_s,m_{s'}} \delta_{n_\mu,n_{\mu'}} \delta_{n_\nu,n_{\nu'}} \delta_{n_\lambda,n_{\lambda'}}, \label{eq:orthogonal}
\end{align}
where the second line uses Eq. \eqref{eq:basisLT} and last equality uses Eq. \eqref{eq:schursLemma}. To prove the completeness, we check that
\begin{align*}\label{eq:complete}
&\sum_{\substack{\mu,\nu,\gamma\\ \lambda,s,m_{s'}\\n_{\mu},n_{\nu},n_{\lambda}}}\sum_{\substack{g,h,l\\x,y,z}}\ket{g,h,l}\langle g,h,l\ket{\Psi_{sm_s}}\bra{\Psi_{sm_s}}x,y,z\rangle\bra{x,y,z}
\\
=&\sum_{\substack{\mu,\nu,\gamma\\ \lambda,s,m_{s}\\n_{\mu},n_{\nu},n_{\lambda}}}\sum_{\substack{g,h,l\\x,y,z}}\frac{d_{\mu}d_{\nu}d_{\lambda}d_{s}d_{\gamma}}{|G|^3}\ket{g,h,l}\bra{x,y,z}\repBasisA\left( \repBasisB \right)^*
\\
=&\sum_{\substack{\mu,\nu,\gamma\\ \lambda,s,m_{s}\\n_{\mu},n_{\nu},n_{\lambda}}}\sum_{\substack{g,h,l\\x,y,z}}\frac{d_{\mu}d_{\nu}d_{\lambda}d_{s}d_{\gamma}}{|G|^3}\ket{g,h,l}\bra{x,y,z}\repBasisE
\\
=&\sum_{\substack{\mu,\nu,\gamma\\ \lambda,s\\}}\sum_{\substack{g,h,l\\x,y,z}}\frac{d_{\mu}d_{\nu}d_{\lambda}d_{s}d_{\gamma}}{|G|^3}\ket{g,h,l}\bra{x,y,z}\repBasisF
\\
=&\sum_{\substack{\mu,\nu,\gamma\\ \lambda,s\\}}\sum_{\substack{g,h,l\\x,y,z,g'}}\frac{d_{\mu}d_{\nu}d_{\lambda}d_{s}d_{\gamma}\beta_\gamma}{|G|^4}\ket{g,h,l}\bra{x,y,z}\repBasisG
\\
=&\sum_{\substack{\mu,\nu,\gamma\\ \lambda,s\\}}\sum_{\substack{g,h,l\\x,y,z,g'}}\frac{d_{\mu}d_{\nu}d_{\lambda}d_{s}d_{\gamma}\beta_\gamma}{|G|^4}\ket{g,h,l}\bra{x,y,z}\traceA \traceB \traceC
\\
=&\sum_{\substack{\mu,\nu,\gamma\\ \lambda,s\\}}\sum_{\substack{g,h,l\\x,y,z,g',g''}}\frac{d_{\mu}d_{\nu}d_{\lambda}d_{s}d_{\gamma}}{|G|^5}\ket{g,h,l}\bra{x,y,z}\traceA \traceB \traceD\traceE\traceF
\\
=&\sum_{\substack{g,h,l\\x,y,z,g',g''}}\ket{g,h,l}\bra{x,y,z}\delta_{g'',e}\delta_{g'',g'}\delta_{\bar yh,g'}\delta_{g \bar x,g'}\delta_{\bar z l,g''}
\\
=&\sum_{\substack{g,h,l\\x,y,z}}\ket{g,h,l}\bra{x,y,z}\delta_{g,x}\delta_{h,y}\delta_{l,z}=\sum_{g,h,l}\ket{g,h,l}\bra{g,h,l},
\end{align*}
where we use the great orthogonality theorem \eqref{eq:greatOrTheom}
\be\label{eq:eq:greatOrTheomII}
\sum_{\mu}\frac{d_{\mu}}{|G|}\greatOrTheomC=\sum_{\mu}\frac{d_{\mu}}{|G|}\greatOrTheomD=\delta_{g,h}.
\ee
Thus, the rep-basis is orthonormal and complete.

Details of the action of the operators $\widetilde{\Ab_v^{QD}}$ in rep-basis is as follows.
\begin{align*}
&\widetilde{\Ab_v^\QD}\ket{\Psi_{sm_s}} 
\\
=&\sum_{\substack{\mu',\nu',\gamma'\\ \lambda',s',m_{s'}\\n_{\mu'},n_{\nu'},n_{\lambda'}}} \ket{ \Psi'_{s'm_{s'}}} \bra{ \Psi'_{s'm_{s'}}} \sum_{g,h,l\in G}\Ab_v^{QD}\ket{g,h,l}\bra{g,h,l}\Psi_{sm_s}\rangle\\
\\ 
=&\sum_{\substack{\mu',\nu',\gamma'\\ \lambda',s',m_{s'}\\n_{\mu'},n_{\nu'},n_{\lambda'}}}\sum_{g,h,l\in G}\sum_{k\in K}\frac{1}{|K|}\ket{\Psi'_{s'm_{s'}}}\bra{\Psi'_{s'm_{s'}}}k{g,kh,l}\bar k\rangle\bra{g,h,l}\Psi_{sm_s}\rangle
\\
=&\sum_{\substack{\mu',\nu',\gamma'\\ \lambda',s',m_{s'}\\n_{\mu'},n_{\nu'},n_{\lambda'}}}\sum_{g,h,l\in G}\sum_{k\in K}\frac{v_{\mu}v_{\nu}v_{\lambda}v_sv_{\gamma}v_{\mu'}v_{\nu'}v_{\lambda'}v_{s'}v_{\gamma'}}{|G|^3|K|}
\left( \avOperatorA \right)^*\contractA\ket{\Psi'_{s'm_{s'}}}
\end{align*}
\be\begin{aligned}\label{eq:avFourierBasisDetail}
=&\sum_{\substack{\mu',\nu',\gamma'\\ \lambda',s',m_{s'}\\n_{\mu'},n_{\nu'},n_{\lambda'}}}\sum_{g,h,l\in G}\sum_{k\in K}\frac{v_{\mu}v_{\nu}v_{\lambda}v_sv_{\gamma}v_{\mu'}v_{\nu'}v_{\lambda'}v_{s'}v_{\gamma'}}{|G|^3|K|}\contractAvOperator\ket{\Psi'_{s'm_{s'}}}
\\
=&\!\sum_{\substack{\mu',\gamma'\\ \lambda',s',m_{s'}\\n_{\mu'},n_{\lambda'}}}\sum_{k\in K}\frac{v_{s}v_{\gamma}v_{s'}v_{\gamma'}}{|K|}\avOperatorB\ket{\Psi'_{s'm_{s'}}}
\\
=&\!\sum_{\substack{s',m_{s'}\\ \gamma'}}\sum_{k\in K}\frac{v_{s}v_{\gamma}v_{s'}v_{\gamma'}}{|K|}\avOperatorC\ket{\Psi^{\gamma'}_{s'm_{s'}}}=\sum_{m'_s}\sum_{k\in K}\frac{1}{|K|}D^{s}_{m_sm'_s}(k)\ket{\Psi_{sm'_s}}.
\end{aligned}\ee

Detailed action of the operators $\widetilde{\Bb_p^{\QD}}$ in the rep-basis reads as follows
\begin{align*}
&\langle\Psi^{\eta'\lambda'}_{r'\tilde{m}_{r'};s'\tilde{m}_{s'}}\vert\widetilde{\Bb^{QD}_p}\vert\ \Psi^{\eta\lambda}_{r\tilde{m}_{r};s\tilde{m}_{s}}\rangle
\\
=&\!\sum_{g,h,l,x,y}\!\langle\Psi^{\eta'\lambda'}_{r'\tilde{m}_{r'};s'\tilde{m}_{s'}}\vert\Bb^{QD}_p\vert g,h,l,x,y\rangle\langle g,h,l,x,y|\Psi^{\eta\lambda}_{r\tilde{m}_{r};s\tilde{m}_{s}}\rangle
\\
=&\sum_{g,h,l,x,y}
\sum_{(t,\alpha_t)\in L_A}\frac{|K|}{|G|}d_{t}D^{t}(l)_{\alpha_t\alpha_t}\langle\Psi^{\eta'\lambda'}_{r'\tilde{m}_{r'};s'\tilde{m}_{s'}}\vert g,h,l,x,y\rangle\langle g,h,l,x,y|\Psi^{\eta\lambda}_{r\tilde{m}_{r};s\tilde{m}_{s}}\rangle \\
=&\sum_{g,h,l,x,y}\frac{v_{\eta}v_{s}v_{r}d_{\mu}d_{\nu}v_{\lambda}v_{\lambda'}d_{\kappa}d_{\beta}v_{ \gamma}v_{\eta'}v_{s'}v_{r'}v_{\gamma'}}{|G|^5}\\&\times\sum_{(t,\alpha_t)\in L_A}d_t\frac{|K|}{|G|}\left( \bpOperatorA{\mu}{\nu}{\gamma'}{r'}{\lambda'}{\kappa}{\eta'}{\rho}{s'} \right)^*\bpOperatorC\bpOperatorAA{\mu}{\nu}{\gamma}{r}{\lambda}{\kappa}{\eta}{\rho}{s}
\\
=&\sum_{g,h,x,y,l_{1},l_2,l_3,l_{4}}\sum_{(t,\alpha_t)\in L_A}\frac{|K|}{|G|}\frac{v_{\eta}v_{s}v_{r}d_{\mu}d_{\nu}v_{\lambda}v_{\lambda'}d_{\kappa}d_{\beta}v_{ \gamma}v_{\eta'}v_{s'}v_{r'}v_{\gamma'}}{|G|^8}d_t
\\&\times\left( \bpOperatorG{\mu}{\nu}{\gamma'}{r'}{\lambda'}{\kappa}{\eta'}{\rho}{s'} \right)^*\bpOperatorD\bpOperatorGG{\mu}{\nu}{\gamma}{r}{\lambda}{\kappa}{\eta}{\rho}{s}
\\
=&\sum_{g',h',x',y',l_{1},l_2,l_3,l_{4}}\sum_{(t,\alpha_t)\in L_A}\frac{|K|}{|G|}\frac{v_{\eta}v_{s}v_{r}d_{\mu}d_{\nu}v_{\lambda}v_{\lambda'}d_{\kappa}d_{\beta}v_{ \gamma}v_{\eta'}v_{s'}v_{r'}v_{\gamma'}}{|G|^8}d_t
\\&\times\left( \bpOperatorB{\mu}{\nu}{\gamma'}{r'}{\lambda'}{\kappa}{\eta'}{\rho}{s'} \right)^*\bpOperatorD\bpOperatorBB{\mu}{\nu}{\gamma}{r}{\lambda}{\kappa}{\eta}{\rho}{s}
\\
=&\sum_{l_{1},l_2,l_3,l_{4}}\sum_{(t,\alpha_t)\in L_A}\frac{|K|}{|G|}\frac{v_{\eta}v_{s}v_{r}v_{\lambda}v_{\gamma}v_{\lambda'}v_{\eta'}v_{s'}v_{r'}v_{\gamma'}}{|G|^4}d_t\bpOperatorE
\\
\end{align*}

\be\begin{aligned}\label{eq:bpOperatorDetail}
=&\sum_{(t,\alpha_t)\in L_A}\frac{|K|}{|G|}\beta_s \beta_\eta \beta_\lambda \beta_{r'} \beta_{\gamma} d_tv_{\eta}v_{s}v_{r}v_{\lambda}v_{\lambda'}v_{\eta'}v_{s'}v_{r'}\bpOperatorF,
\\
=&\sum_{(t,\alpha_t)\in L_A}\sum_{\tilde n_{r'^*},\tilde n_{t^*},\tilde n_{s'^*}}\frac{1}{d_A}\tilde d_t \tilde v_\eta \tilde v_s \tilde v_r \tilde v_\lambda \tilde v_{\lambda'} \tilde v_{\eta'} \tilde v_{s'} \tilde v_{r'} (w^{rtr'^*}_{\tilde m_{r}\alpha_t \tilde n_{r'^*}})^*(\Omega^{r'})^{-1}_{\tilde n_{r'^*}\tilde m_{r'}}
\\
&\times(w^{t^*s'^*s}_{\tilde n_{t^*}\tilde n_{s'^*}\tilde m_s})^*(\Omega^{s'^*})^{-1}_{\tilde m_{s'}\tilde n_{s'^*}}(\Omega^{t^*})^{-1}_{\alpha_{t}\tilde n_{t^*}}G^{\rho^*\eta s^*}_{t^*s'^*\eta'}G^{\kappa\lambda\eta^*}_{t^*\eta'^*\lambda'}G^{\gamma r^*\lambda^*}_{t^*\lambda'^* r'^*}.
\end{aligned}\ee

We prove $d_A:=\sum_{t\alpha_t \in L_A}d_{t}=\frac{|G|}{|K|}$. Using $P^t_K=\frac{1}{|K|}\sum_{k \in K}D^{t}(k)_{\tilde m_t \tilde m_t}=\delta_{(t, \tilde m_t)\in L_A}$ and great orthogonality theorem, we have
\be\label{eq:proveDim}
\frac{1}{|K|}\sum_{t\in L,\tilde m_t} \sum_{k \in K}d_{t}D^{t}(k)_{\tilde m_t \tilde m_t}=\frac{|G|}{|K|}=\sum_{t \in L, \tilde m_t}\delta_{(t, \tilde m_t)\in L_A}d_t.
\ee

%%%%%%%%%%%%%%%%%%%
%%%%%%%%%%%%%%%%%%%
%%%%%%%%%%%%%%%%%%%

\bibliographystyle{apsrev4-1}
\bibliography{StringNet}
\end{document}